\documentclass[footinbib,a4paper,aps,pra,reprint,twocolumn,preprintnumbers,amsmath,amssymb,10pt, superscriptaddress]{revtex4-1}
\usepackage{verbatim}
\usepackage{color}
\usepackage{tikz}
\usepackage{xr}
\usepackage[colorlinks=true,citecolor=blue,linkcolor=blue,urlcolor=blue]{hyperref}
\usepackage{braket}
\usepackage[normalem]{ulem}
\usepackage{upgreek}
\usepackage[percent]{overpic}
\usepackage{graphicx,xcolor}
\usepackage{dcolumn}
\usepackage{bm}
\usepackage[shortlabels]{enumitem}

\definecolor{forestgreen}{rgb}{0.1, 0.6, 0.2}

\makeatletter
\renewcommand{\fnum@figure}{FIG. \thefigure}
\makeatother

\newcommand{\bracket}[2]{\langle{#1}|{#2}\rangle}

\begin{document}

\title{Critical light-matter entanglement at cavity mediated phase transition}
\author{Giuliano Chiriacò}
\email{gchiriac@ictp.it}
\affiliation{The Abdus Salam International Centre for Theoretical Physics (ICTP), Strada Costiera 11, 34151 Trieste,
Italy}
\affiliation{SISSA — International School of Advanced Studies, via Bonomea 265, 34136 Trieste, Italy}
\author{Marcello Dalmonte}
\email{mdalmont@ictp.it}
\affiliation{The Abdus Salam International Centre for Theoretical Physics (ICTP), Strada Costiera 11, 34151 Trieste,
Italy}
\affiliation{SISSA — International School of Advanced Studies, via Bonomea 265, 34136 Trieste, Italy}
\author{Titas Chanda}
\email{tchanda@ictp.it}
\affiliation{The Abdus Salam International Centre for Theoretical Physics (ICTP), Strada Costiera 11, 34151 Trieste,
Italy}

\date{\today}

\begin{abstract}
We consider a model of a light-matter system, in which a system of fermions (or bosons) is coupled to a photonic mode that drives a phase transitions in the matter degrees of freedom. Starting from a simplified analytical model, we show that the entanglement between light and matter vanishes at small and large coupling strength, and shows a peak in the proximity of the transition. We perform numerical simulations for a specific model (relevant to both solid state and cold atom platforms), and show that the entanglement displays critical behavior at the transition, and features maximum susceptibility, as demonstrated by a maximal entanglement capacity. Remarkably, light-matter entanglement provides direct access to critical exponents, suggesting a novel approach to measure universal properties without direct matter probes. 
\end{abstract}

\maketitle

\section{Introduction}\label{Intro}
A many-body ``matter'' system coupled to a photonic mode experiences a variety of effects depending on the coupling strength. In both solid state and cold atom communities~\cite{Rev:Schlawin2022,Rev:Garcia-Vidal2021,New:Kurizki2015,Ritsch13, Dicke:Landig2016, Cosme18, Zupancic19, Georges18, Mivehvar2021}, a lot of attention has been given to the (ultra) strong coupling regime~\cite{Exp:Bayer2017} where the interaction strength is comparable with the relevant energy scales of cavity and matter system. In such regime, the strong hybridization between light and matter induces shared hybrid properties absent in the decoupled components of the system, such as entangled polaritonic states and coherence in the photon properties \cite{Exc:Schachenmayer2015a,MB:Hagenmuller2018b,MB:Bartolo2018b,Exc:Latini2019}.

Several experiments on solid state platforms, such as molecules \cite{Mol:Zhong2017a,Mol:Feist2018,Mol:Flick2018,Mol:Ribeiro2018,Mol:Rozenman2018b,Mol:Kena-Cohen2019a,Mol:Thomas2019b,Mol:Liu2020b,Mol:Takahashi2020b,Mol:Wellnitz2020a} and semiconductors \cite{Exc:Feist2015,Exc:Orgiu2015,Exc:Wei2019b,Exc:Lenk2020b}, have shown that the cavity mediated long-range interactions dramatically change the properties of the system, for example by enhancing its transport properties \cite{MB:Hagenmuller2017b,MB:Du2018b,MB:Hagenmuller2018b,TRANSPORT:Rokaj2022,TRANSPORT:Eckhardt2022}. Successive theoretical studies investigated the appearance of new phases of matter in quantum materials embedded in cavities \cite{New:Kiffner2019b,New:Kiffner2019d,New:Kiffner2019c,New:Ashida2020}, including superconductivity \cite{SC:Sentef2018, New:Schlawin2019a,SC:Chakraborty2021}, topological phases \cite{Chiral:Kollath2016, Mivehvar2017,Topo:Wang2019, ENT:Mendez2020TopologyCavity, Chanda2021,Topo:Dmytruk2022, Chanda2022,Topo:Appugliese2022,Topo:Rokaj2022}, many-body localized phases~\cite{Sierant_Scipost_2019, Kubala_PRB_2021}, and quantum spin liquids \cite{New:Chiocchetta2021}. Other works predicted a superradiant phase transition \cite{Super:Mazza2019,Super:Nataf2019,Super:Guerci2020,Super:Stokes2020,Super:Ferri2021,IsingSuperRad:Rohn2020}: above a critical value of the coupling strength the system is ordered and the cavity photons spontaneously condense into a state with a non-zero electric field. These works sparked a debate over the correct way of modeling the light-matter coupling and whether no-go theorems forbid spontaneous photon condensation  \cite{Gauge:Li2020b,Gauge:Dmytruk2021b,NoGo:Andolina2019,NoGo:Nataf2010,NoGo:Andolina2020}.

The aforementioned phenomena clearly signal that hybrid systems are governed by strong quantum effects. An earlier work \cite{ENT:Lambert2004EntCavity} found critical quantum correlations in the Dicke model (where the matter degrees of freedom are decoupled), while others studied the intra-matter entanglement in hybrid systems \cite{ENT:Mendez2020TopologyCavity,ENT:Sharma2022EntanglementCavity}. However, the extent to which the most basic form of quantum correlation - entanglement - between light and matter plays a role in genuine many-body hybrid systems is presently unresolved. The exotic structure of the coupling (global between matter and light, and local within the former) represents a completely different scenario with respect to the traditional - and long-studied - entanglement structure of locally interacting many-body systems~\cite{Amico2008}. Understanding the role of entanglement in hybrid many-body systems can provide a new window to understand their functioning, and particularly pave new avenues for their manipulation and probing.

\begin{figure}[t]
    \centering
    \includegraphics[width=0.9\columnwidth]{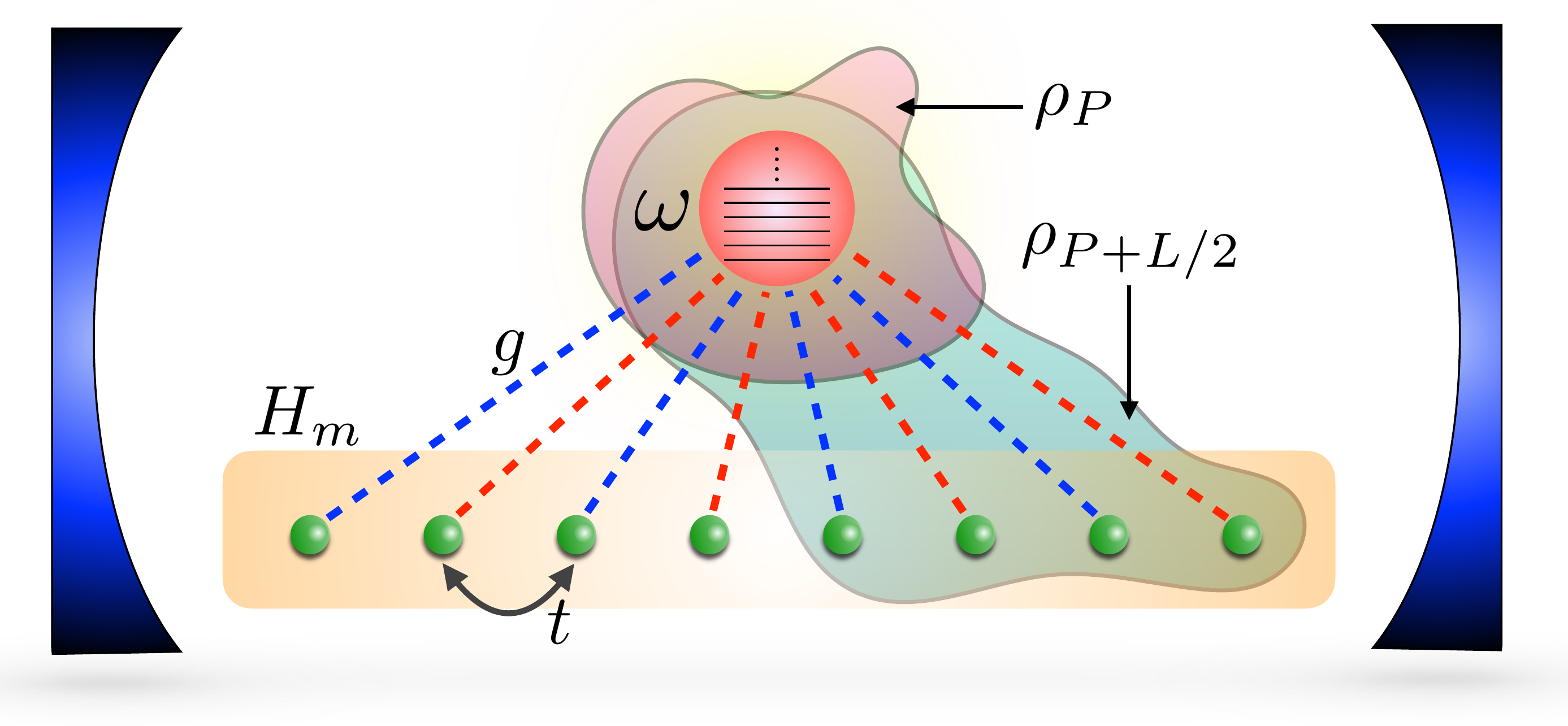}
    \caption{Schematic depiction of the system. The matter degrees of freedom (green spheres) are confined inside a cavity and governed by a Hamiltonian $H_m$. The cavity hosts a single photon mode (circle)  of frequency $\omega$, that couples to the entire matter system (dashed lines) with strength $g$. We characterize light-matter entanglement via the density matrix of the photon ($\rho_P$), and of a hybrid partition including photon and half of the matter sites ($\rho_{P+L/2}$).}
    \label{fig:sketch}
\end{figure}

In this work we aim to fill this gap and investigate the entanglement between light and matter in hybrid systems. We consider a model in which the cavity photons drive a phase transition in a suitable order parameter of the matter, and analyze the behavior of the entanglement across the phase transition~\cite{ENT:Osterloh2002,ENT:Osborne2002,ENT:Vidal2003}. We first study a simplified, yet very general, analytical model to get a qualitative physical insight, and then perform numerical calculations to characterize the critical behavior of the entanglement and of the matter order parameter.

We find that matter and light are unentangled in the disordered phase and deep in the ordered phase, while they become highly entangled near the phase transition. There, a large amount of quantum correlations is shared between matter and the photons, and is accompanied by a maximal entanglement capacity - analogous to specific heat for quantum correlations. Surprisingly, we find that the photon entanglement with the matter component displays a critical behavior at the transition, and that its critical exponents are in remarkable agreement with those of the order parameter. This opens up the possibility to extract information about the matter component by observing the photon state, and to probe the transition through a quantity that is experimentally more convenient. 

The paper is organized as follows. In Section \ref{Analytical} we present a simple yet general analytical model to estimate the light-matter entanglement. In Section \ref{Numerical} we focus on a specific model and perform numerical calculations to extract the critical behavior at the transition. In Section \ref{Experiment} we briefly discuss possible experimental implications of our work, while in Section \ref{Conclusion} we present our conclusions.

\section{Analytical model}\label{Analytical} 
We consider a generic hybrid system, consisting of a matter part (electrons, fermions, spins, etc.) described by the Hamiltonian $\hat H_m$ coupled to a light part, i.e., one bosonic degree of freedom (such as photons or phonons) characterized by the creation operator $\hat a^{\dagger}$. The photon \footnote{For convenience, we use ``photon'' to indicate the light part even when we do not specify its nature.} couples with all the matter degrees of freedom through the electric field $\hat E\equiv \hat a + \hat a^{\dagger}$. The coupling drives a phase transition in a suitable order parameter of the matter system, which we assume to be the expectation value of the operator $\hat\Delta$. The Hamiltonian of the system is
\begin{equation}\label{Eq:simpleH}
\hat H = \hat H_m+\omega \hat a^{\dagger} \hat a- g \hat E\hat\Delta.
\end{equation}

We calculate the entanglement of the system using the reduced density matrix $\rho_{P+A}$, where $P$ indicates the photon, $A$ is a partition of the matter system (with $\bar A$ its complement) and $\rho_{P+A}=\text{Tr}_{\bar A}\ket{\psi}\bra{\psi}$, where $\ket{\psi}$ is the ground state of the system. We focus on $A=\varnothing$ (i.e., trace out the matter) and $A=L/2$ (i.e., take a bipartition of the matter), see Fig. \ref{fig:sketch}. We employ as entanglement witness the 2$^{\text{nd}}$ Renyi entropy 
\begin{equation}
S_{P+A}=-\ln\text{Tr}\rho^2_{P+A},
\end{equation} 
the von Neumann entropy 
\begin{equation}
S_{P+A}^v=-\text{Tr}(\rho_{P+A}\ln\rho_{P+A}),
\end{equation}
or the entanglement capacity~\cite{Yao2010,Schliemann2011,Boer2019,ENT:Nandy2021EntanglementCapacity}
\begin{equation}
C_E=\text{Tr}(\rho_{P+A}(\ln\rho_{P+A})^2)-(\text{Tr}(\rho_{P+A}\ln\rho_{P+A}))^2.
\label{eq:entCap}
\end{equation}
See Appendix~\ref{app:entCap} for a brief discussion about the entanglement capacity.

In order to carry out the analytical calculations and gain an intuitive insight about the entanglement physics, we make a very simplistic assumption: all matter degrees of freedom can be collapsed into two states, a disordered state $\hat\Delta\ket{0}_m=0$ and an ordered state $\hat\Delta\ket{1}_m=\ket{1}_m$ \footnote{The exact value of $\braket{\hat\Delta}$ in the ordered phase is not important since it can be rescaled. Here we assume it to be unity.}. This approximation is true far away from the transition, but quite crude close to the critical point. Nonetheless, it can capture the essential features of the entanglement even at the critical point, as we show later for a specific system.
We then introduce a variational ansatz for the quantum state:
\begin{equation}\label{Eq:VarAns}
\ket{\psi}=\sqrt{1-\Delta}\ket{0}_m\ket{\Psi_0}+\sqrt{\Delta}\ket{1}_m\ket{\Psi_1},
\end{equation}
such that $0 \leq \bra{\psi}\hat\Delta\ket{\psi}=\Delta \leq 1$. For simplicity we neglect for now any degeneracy in the ordered state.
We substitute into Eq. \eqref{Eq:simpleH} to minimize the energy
\begin{equation}\label{Eq:VarEn}
\braket{\hat H}=F(\Delta)+(1-\Delta)\omega\braket{\hat a^{\dagger}\hat a}_0+\Delta\braket{\omega \hat a^{\dagger} \hat a-g\hat E}_1,
\end{equation}
where $F(\Delta)\equiv\bra{\psi}\hat H_m\ket{\psi}$ and $\braket{}_{0/1}$ is the expectation value on $\ket{\Psi_{0/1}}$. The second term is minimized by choosing a state with no photons in it: $\ket{\Psi_0}=\ket{0}$; the third term is minimized by choosing $\ket{\Psi_1}$ to be the ground state of the Hamiltonian
\begin{equation}\label{Eq:Hph}
\hat H_{ph}=\omega \hat a^{\dagger}\hat a-g(\hat a+\hat a^{\dagger}) =\omega \hat{\tilde{a}}^{\dagger}\hat{\tilde a}-\frac{g^2}{\omega};\quad\, \hat{\tilde a} \equiv \hat a-\frac g{\omega},
\end{equation}
meaning $\hat{\tilde a}\ket{\Psi_1}=0$ or $\ket{\Psi_1}$ is a coherent state with parameter $g/\omega$~\footnote{The value of $\Delta$ can then be determined by minimizing the model-dependent free energy $\braket{\hat H}=F(\Delta)-g^2\Delta/\omega$.}:
\begin{equation}
\ket{\Psi_1}=e^{-g^2/2\omega^2}\sum_n(g/\omega)^n\ket{n}/\sqrt{n!}.
\end{equation}
Upon tracing out the matter degrees of freedom, we obtain the photon reduced density matrix
\begin{equation}\label{Eq:rhoP}
\rho_P=(1-\Delta)\ket{0}\bra{0}+\Delta e^{-\frac{g^2}{\omega^2}}\sum_{n,m}\left(\frac{g}{\omega}\right)^{n+m}\frac{\ket{n}\bra{m}}{\sqrt{n!m!}}.
\end{equation}

We calculate analytically Renyi entropy $S_P=-\ln\text{Tr}\rho_P^2$, using $\bracket{\alpha}{\beta}=e^{-|\alpha|^2/2-|\beta|^2/2+\alpha^*\beta}$ for coherent states $\alpha$ and $\beta$, as
\begin{equation}\label{Eq:Renyi}
\text{Tr}\rho_P^2=(1-\Delta)^2+\Delta^2+2\Delta(1-\Delta)e^{-g^2/\omega^2}.
\end{equation}
The entropy vanishes deep into the disordered ($\Delta\rightarrow0$) and ordered ($\Delta\rightarrow1$) phases, while is maximized for $\Delta\sim1/2$. In other words, the light-matter entanglement is peaked around the phase transition in $\Delta$ and can be in principle considered to be an indicator of the transition.

\subsection*{Spontaneous symmetry breaking} In the presence of spontaneous symmetry breaking, the ordered state is degenerate. For $\mathbb Z_2$ symmetry (Ising transition), the ordered state is a quantum superposition even for $|\Delta| \rightarrow1$, since the matter can be in a third state: $\hat\Delta\ket{-1}_m=-\ket{-1}_m$, so that we need to replace $\ket{1}_m\ket{\Psi_1}$ in Eq.~\eqref{Eq:VarAns} by $c_1\ket{1}_m\ket{\Psi_1}+c_{-1}\ket{-1}_m\ket{\Psi_{-1}}$ with $|c_1|^2+|c_{-1}|^2=1$.

The expectation value of the matter Hamiltonian is symmetric in $|\Delta|$, so we find
\begin{eqnarray}
\notag\braket{\hat H}&=&F(\Delta)+(1-|\Delta|)\omega\braket{\hat a^{\dagger}\hat a}_0+|\Delta||c_1|^2\braket{\omega \hat a^{\dagger}\hat a-g\hat E}_1\\
\label{Eq:VarEn2}&+&|\Delta||c_{-1}|^2\braket{\omega \hat a^{\dagger}\hat a+g\hat E}_{-1}.
\end{eqnarray}

Minimizing separately we find again that $\ket{\Psi_0}$ is the vacuum state, $\ket{\Psi_1}$ is the coherent state with parameter $g/\omega$, and $\ket{\Psi_{-1}}$ is the coherent state with parameter $-g/\omega$. The coefficients $c_1$ and $c_{-1}$ may be determined by introducing a phenomenological tunneling term between the degenerate states, which in the $\{\ket{1}_m,\ket{-1}_m\}$ basis is written very generally as $-t_{\perp}\begin{pmatrix}0&1\\1&0\end{pmatrix}$ (with $t_{\perp}\rightarrow0^+$). Such mixing selects the symmetric combination $c_1=c_2=1/\sqrt2$, such that even in the limit $|\Delta| \rightarrow 1$ the photon density matrix is a mixed state of two coherent states with opposite electric fields:
\begin{equation}\label{Eq:rhoP3}
\rho_P=\frac{e^{-g^2/\omega^2}}2\sum_{n,m}\left(\frac{g}{\omega}\right)^{n+m}(1+(-1)^{n+m})\frac{\ket{n}\bra{m}}{\sqrt{n!m!}},
\end{equation}
corresponding to a finite Renyi entropy $S_P=\ln2-\ln\left(1+e^{-2g^2/\omega^2}\right)
$, which saturates to $\ln2$, i.e. the entanglement of a superposition of the two degenerate states (Fig. \ref{fig:collapse_Om1}(f)).

For an ordered phase with $\mathbb{Z}_q$ degeneracy, the density matrix elements are of the type $(\rho_P)_{mn}\sim\sum_kz_k^{m+n}$ where $z_k^q=1$, which vanish unless $m+n\equiv0\text{ (mod }q)$ -- the density matrix shows the same symmetry as the underlying phase transition, as expected. Introducing a symmetry breaking term we select only one ordered state, ending up with the result of Eq.~\eqref{Eq:Renyi}.

We remark that the model \eqref{Eq:simpleH} and its consequences are very general, and apply to any system where a many-body matter component couples to a bosonic mode, including scenarios where the transition is already present even in the absence of coupling to the photon \cite{ENT:Mendez2020TopologyCavity}.

\section{Numerical results}\label{Numerical}

We now consider a specific model, describing spinless free fermions coupled to the photon through staggered density \footnote{Our analysis can be applied to any kind of coupling, including the usual Peierls substitution (which has the advantage of making gauge invariance manifest).}:
\begin{equation}\label{Eq:Hcdw}
\hat H_m=-t\sum_{\braket{ij}}\hat c^{\dagger}_i \hat c_j+\text{h.c.};\qquad \hat\Delta=\sum_j(-1)^j \hat c^{\dagger}_j \hat c_j.
\end{equation}

We add a small degeneracy-lifting term ($g_2 (\hat c_1^{\dagger} \hat c_1 - \hat c_L^{\dagger} \hat c_L)$ with $g_2=10^{-3}$ and $L$ the system size, see Appendix~\ref{app:symmetrybroken}) and study the system at half-filling, where it presents an instability to a gapped phase with a non-zero density wave for strong enough coupling $g$. 

The model \eqref{Eq:Hcdw} can describe both cold atoms coupled to an optical cavity mode with periodicity twice that of the optical lattice~\cite{Habibian13,Ritsch13, Caballero15, Caballero16, Elliott16}, or solid state electrons exhibiting a Fermi surface nesting and coupled to a lattice distortion (modeled by a dominant phonon mode) that drives a charge density wave (CDW) transition \cite{CDW:Grunwe1988,CDW:gorkov2012,CDW:monceau1985electronic,CDW:Zhu2015,CDW:Chiriaco2018}. Indeed, the connection between metal to CDW transition and the superradiant transition in cavity systems has been recently investigated \cite{SuperCDW:Chen2014,SuperCDW:Keeling2014,SuperCDW:Piazza2014,SuperCDW:Rylands2020}. In addition, the (hard-core) bosonic version of the model can be realized in circuit QED settings~\cite{viehmann2013observing,dalmonte2015realizing}, considering 3D cavities that are quasi-resonant with the superconducting qubit frequencies.

We study the dependence of the order parameter $O_D\equiv\braket{\hat\Delta}/L$ and of the light-matter entanglement (characterized through $S_P$) on $g$ and on $\omega$. We perform numerical simulations using exact diagonalization (ED)~\cite{Sandvik2010} and density matrix renormalization group (DMRG)~\cite{white_prl_1992,white_prb_1993,schollwock_aop_2011, Orus_aop_2014, Singh_PRA_2010,Singh_PRB_2011, Halati20a, Halati20b, Halati21} calculations (see Appendix~\ref{app:numerics} for details). The maximal bond dimension for the matrix-product state (MPS) representation in the DMRG calculations is set to $\chi=600$, enough to ensure convergence for simulations with system sizes up to $L=100$.

\begin{figure*}[!ht]
    \centering
    \includegraphics[width=0.9\textwidth]{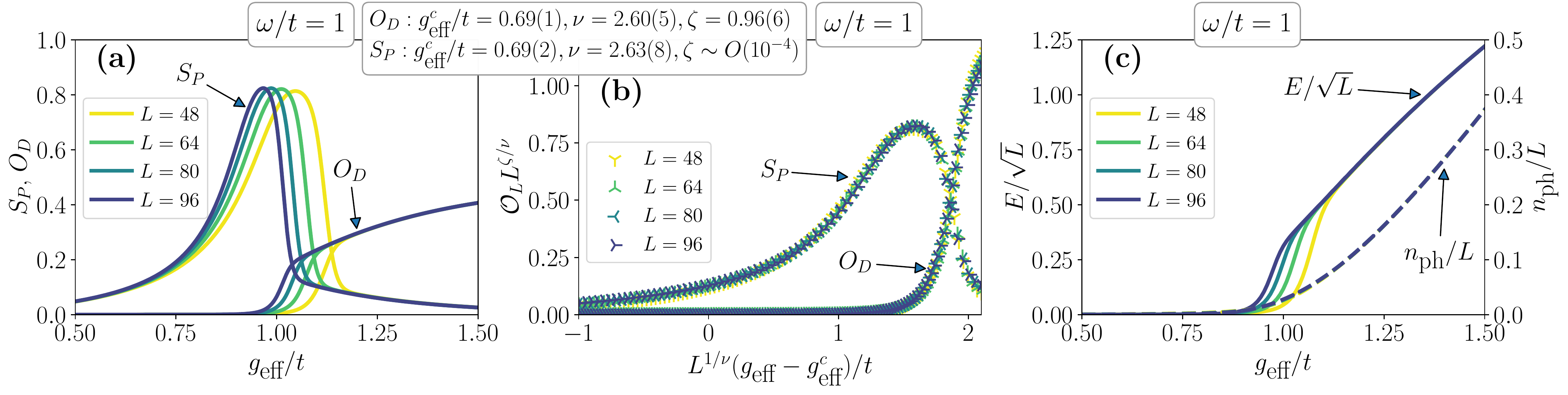}\\
    \includegraphics[width=0.9\textwidth]{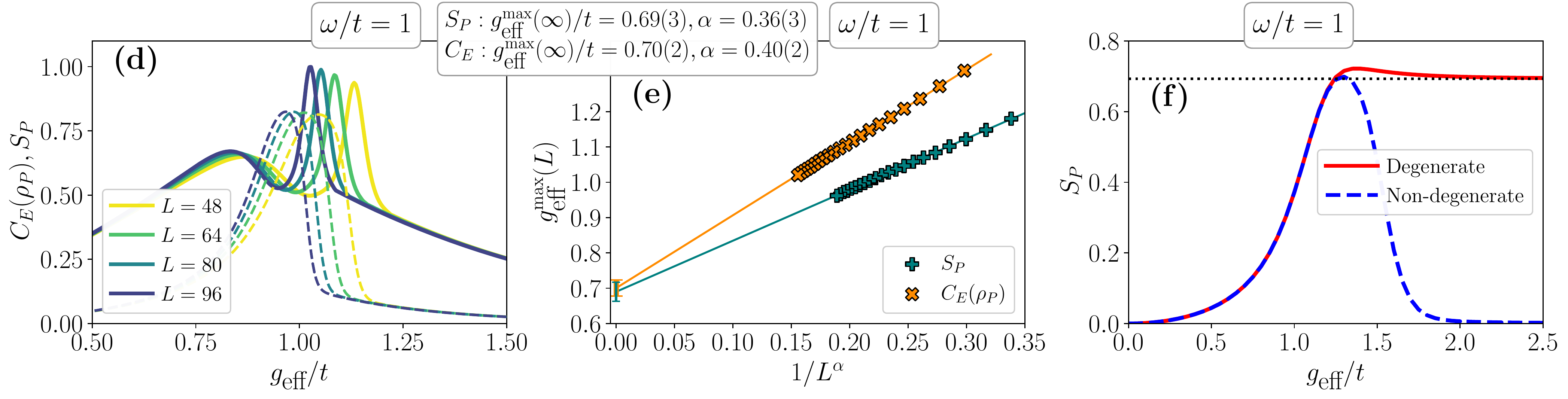}
    \caption{Phase transition analysis at $\omega/t=1$. (a) The photon entanglement entropy $S_P$ and the order parameter $O_D$ as functions of the effective coupling $g_{\textrm{eff}}$. (b) Collapse of the curves using  scaling hypothesis near the transition point. We consider system-sizes in the range $L \in [48, 100]$ for the scaling analysis, and plot only for $L=48, 64, 80, 96$ for clarity. The values of the critical coupling and of the critical exponents for the two transitions are reported in the box. (c) Rescaled expectation value of the electric field $E/\sqrt{L}$ and photon number $n_{\textrm{ph}}/L$ as functions of $g_{\textrm{eff}}$. (d) Entanglement capacity $C_E(\rho_P)$ (solid lines) and photon entropy $S_P$ (dashed).
    (e) Scaling of the locations $g_{\textrm{eff}}^{\max}(L)$ of the peak values with respect to the system size $L$ for both $S_P$ and $C_E(\rho_P)$. For both quantities, $g_{\textrm{eff}}^{\max}$ coincides with the critical point $g_{\textrm{eff}}^c$ in the thermodynamic limit as $g_{\textrm{eff}}^{\max}(L)/t - g_{\textrm{eff}}^c/t \propto L^{-\alpha}$. The data for (a)-(e) have been obtained with DMRG simulations. (f) Plot of $S_P$ obtained with ED ($L=16$) for the degenerate ($g_2=0$) and non-degenerate case ($g_2=0.001$). The degenerate entropy saturates to $\ln2$ (black dotted line). }
    \label{fig:collapse_Om1}
\end{figure*}

Our main results are reported in Fig. \ref{fig:collapse_Om1} for $\omega/t=1$ (see Appendix~\ref{app:otherOmega} for other values of $\omega/t$). We plot $S_P$ and $O_D$ in terms of the effective coupling strength $g_{\textrm{eff}}=g\sqrt{L}$. The rescaling ensures that all energy contributions are extensive in size \cite{New:Kiffner2019b, Habibian13,Ritsch13, Caballero15, Caballero16, Elliott16}: the internal fermion energy scales as $\sim L$, while the photon contribution is $\sim g^2L^2/\omega$ (because the photon mode couples to each fermion, mediating an all-to-all interaction). An effective coupling $g_{\textrm{eff}}^2=g^2L$ guarantees that the scaling properties with $L$ are not dominated by the electric field, and also means that the resulting critical exponents differ from those obtained within a mean field analysis.

The entropy $S_P$ goes to zero for small and large $g_{\textrm{eff}}$, and exhibits a peak in correspondence to the transition as signaled by $O_D$ (Fig.~\ref{fig:collapse_Om1}(a)). This agrees with our analytical prediction, see Eq.~\eqref{Eq:Renyi}, and confirms that the photon entanglement is an indicator of phase transitions in hybrid light-matter systems. Moreover it implies that integrating out the photon is fundamentally non-trivial near the transition, given its large entanglement with the matter.

The precise values of the critical coupling $g_{\textrm{eff}}^c$ and the critical exponent $\nu$ of the transition
can be determined by performing a finite size scaling analysis, based on the critical scaling hypothesis $\mathcal{O}(L)=L^{-\zeta/\nu}f(L^{1/\nu}(g_{\textrm{eff}}-g_{\textrm{eff}}^c))$ with $\mathcal{O}$ any observable. 
The values of $g_{\textrm{eff}}^c$ and $\nu$ for the transitions in $S_P$ and in $O_D$ are equal within error bars (Fig.~\ref{fig:collapse_Om1}(b)), signaling a common origin of the critical behavior, and suggesting that it is possible to study the transition in the order parameter by looking at the criticality of the photon entanglement $S_P$. This is a striking feature, since it allows to probe the transition of a many-body system by simply looking at the {\it single} degree of freedom of the photon - likely due to the inherently non-local coupling between light and matter.

The quality of the scaling analysis is proven by the collapse of the curves in the vicinity of the transition (Fig. \ref{fig:collapse_Om1}(b)), which fall onto each other almost perfectly. We note that there is no fundamental reason for the exponent $\zeta$ to coincide for the two different quantities. Indeed, it differs for $S_P$ and $O_D$, and changes with $\omega$ (see Appendix~\ref{app:otherOmega}).

We observe that deep in the ordered phase, the electric field $E = |\braket{\hat a + \hat a^{\dagger}}|$ (which can be directly measured by the cavity output~\cite{Dicke:Baumann2011, Dicke:Landig2016, Hruby2018}) and the average number of photons $n_{\textrm{ph}} = \braket{\hat a^{\dagger} \hat a}$ are non-zero, similarly to what happens in a superradiant condensed phase; despite the small amount of entanglement shared between the photon and the fermions, this occurs because the photon is in a coherent state, see Eq. \eqref{Eq:rhoP}. In particular, we note that $n_{\textrm{ph}}$ is proportional to $L$, so that the rescaled $n_{\textrm{ph}}/L$ is \textit{scale-invariant} even near the transition (Fig.~\ref{fig:collapse_Om1}(c)). On the other hand, $E$ scales as $\sqrt{L}$ as expected in the CDW phase, but the rescaled $E/\sqrt{L}$ shows a critical finite-size dependence similar to that of the order parameter $O_D$ (Fig.~\ref{fig:collapse_Om1}(c)), see Appendix~\ref{app:otherOmega}.

In Fig.~\ref{fig:collapse_Om1}(d) we plot the entanglement capacity $C_E(\rho_P)$, that interestingly shows a local dip followed by a global maximum. The locations of the local minima roughly coincide with those of the peaks in $S_P$. Moreover, the locations of the global peak in $C_E(\rho_P)$ and in $S_P$ both converge to the same value in the thermodynamics limit, which also corresponds to the critical coupling $g_{\textrm{eff}}^c$, i.e., $g_{\textrm{eff}}^{\max}(L)-g_{\textrm{eff}}^c  \sim L ^{-\alpha}$, see Fig.~\ref{fig:collapse_Om1}(e). Therefore, at the critical point light-matter entanglement is not only maximal, but is also easy to extract, as shown by maximal entanglement capacity.

We also show the results of ED calculations (Fig. \ref{fig:collapse_Om1}(f)) in the absence (degenerate) and presence (non-degenerate) of the $g_2$ symmetry breaking term. The entropy coincides before the transition and goes to zero in the non-degenerate case, but saturates to $\ln2$ in the degenerate case, as predicted by Eq. \eqref{Eq:rhoP3}. The presence of a small $g_2$ term is helpful computationally: while ED can always access the superposed quantum state (it being the true ground state), DMRG calculations may get stuck in one of the two degenerate states when $g_{\textrm{eff}}/t$ or $L$ become too large. In fact DMRG favors the minimally entangled state between states exponentially close in energy, even if it is not the true ground state, leading to sudden jumps in $S_P$ and unphysical results.

\begin{figure}
    \centering
    \includegraphics[width=\linewidth]{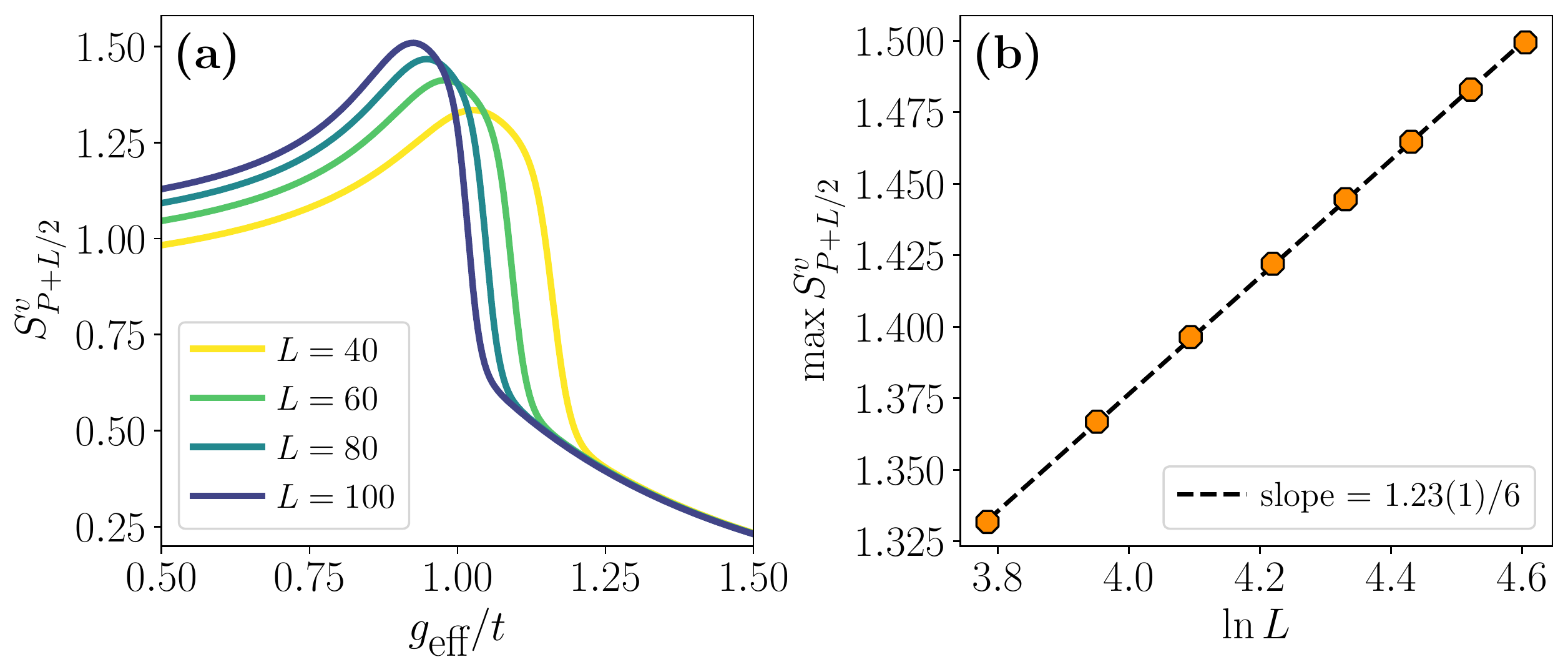}
    \caption{
    (a) The von Neumann entropy $S^v_{P+L/2}$ of the photon and half of the fermionic chain as function of $g_{\textrm{eff}}$, obtained with DMRG for $\omega/t=1$. (b) The variation of the peak value of $S^v_{P+L/2}$ with respect to $L$. We can observe a logarithmic divergence akin to the critical theory in homogeneous systems.}
    \label{fig:vonNeu_scaling_L}
\end{figure}

In Fig.~\ref{fig:vonNeu_scaling_L} we show the behavior of the von Neumann entanglement entropy of the photon plus half of the fermionic chain. This quantity does not vanish for $g_{\textrm{eff}}=0$ since it also account for correlations between the fermions, but still exhibits a peak at the transition (Fig. \ref{fig:vonNeu_scaling_L}(a)). The peak value of the entropy scales logarithmically with $L$ (Fig. \ref{fig:vonNeu_scaling_L}(b)) with a slope $\sim 1.23/6$. This behavior is very similar to the critical scaling of entanglement entropy \cite{callan_geometric_1994, ENT:Vidal2003,calabrese_entanglement_2004, ENT:Calabrese2009}, but an exact parallel is hard to make given the inhomogeneity of the partition due to the presence of the photon.

\section{Experimental relevance}\label{Experiment}

Coupling itinerant fermions (or hard-core bosons) to cavities can be realized experimentally on many different platforms \cite{Exp:Ballarini2019,Exp:Forn-Diaz2019}, such as microcavities \cite{Exp:Kavokin2008,Exp:Schneider2018}, split-ring resonators \cite{Exp:Halbhuber2020,Exp:Maissen2014,Exp:Scalari2012}, semiconductor heterostructures \cite{Exp:Li2018,Exp:Zhang2016}, THz Fabry-Perot cavities \cite{Exp:Jarc2022} and other solid state settings \cite{Exp:Bylinkin2020,Exp:Epstein2020,Exp:Sivarajah2019}. The models we investigate host a light-matter coupling inspired by these experiments. Indeed, our model can be concretely applied to a recent experiment measuring the properties of the superradiant transition \cite{SuperEXP:Zhang2021}.

Demonstrating critical properties form entanglement is very challenging, even for the most remarkable and experimentally demonstrated measurement schemes~\cite{PROBING:Elben2018,PROBING:Brydges2019probing,PROBING:elben2020mixed,PROBING:neven2021symmetry,PROBING:Vitale2022}. The key advantage of our proposal is that the fundamental collective properties can actually be determined by measuring entanglement between matter and light. Measuring the corresponding entropy is scalable, and the corresponding toolbox has already been experimentally demonstrated in several architectures~\cite{EXP:Wang2011,EXP:eichler2011observation,EXP:brennecke2013real,ENT:Mlynek2012Dicke}.

\section{Conclusions and outlook}\label{Conclusion}

We have shown how entanglement plays a key role at quantum critical points in hybrid many-body light-matter systems. Within a general analytical model, we found that the entanglement between light and matter peaks at the transition of the matter order parameter. We confirmed such results numerically in a model of spinless free fermions with an instability to a charge density wave phase driven by the photon. Importantly the entropy displays the same critical behavior as the order parameter of the density wave transition. This is a remarkable result, showing that in a system with a photon-driven transition a single degree of freedom (that of the photon) can be used to investigate a many body phase transition and probe its critical properties. 

\acknowledgments

We thank J. P. Brantut, D. Fausti, D. Guerci, R. Kraus, G. Morigi, M. Schiró and J. Zakrzewski for insightful discussions and correspondence. We are particularly grateful to F. Piazza for his comments and correspondence.
The work of G. C. and M. D. was partly supported by the ERC under grant number 758329 (AGEnTh), by the MIUR Programme FARE (MEPH), and by the European Union's Horizon 2020 research and innovation programme under grant agreement No 817482 (Pasquans).

\appendix

\section{Entanglement capacity}
\label{app:entCap}
The entanglement capacity $C_{E}$ \cite{Yao2010,Schliemann2011,Boer2019,ENT:Nandy2021EntanglementCapacity} that we define in Eq.~\eqref{eq:entCap} is a measure of how susceptible the entanglement stored in the system is upon its extraction. It can be interpreted as the quantum information theoretic counterpart of the thermodynamic heat capacity. It is obtained starting from the entanglement Hamiltonian $\mathcal{K}_A\equiv-\ln\rho_A$, with $\rho_A$ being the density matrix after tracing-out rest of the system in a given bipartition $A$.  The entanglement capacity is defined as:
\begin{align}
C_E(\rho_A) &=\text{Tr}(\rho_A\mathcal{K}_A^2)-\text{Tr}(\rho_A\mathcal{K}_A)^2 \nonumber \\
&=\text{Tr}(\rho_{A}(\ln\rho_{A})^2)-(\text{Tr}(\rho_{A}\ln\rho_{A}))^2.
\label{EqA:EntCap}
\end{align}

In particular the entanglement capacity can be related to the Renyi entropies $S^{(m)}_A=\frac1{1-m}\ln\text{Tr}\rho_A^m$ by
\begin{equation}\label{EqA:EntCapRenyi}
C_E(\rho_A)=\lim_{m\rightarrow1}m^2\partial_m^2((1-m)S^{(m)}_A).
\end{equation}

\section{Details about Numerical Simulations}
\label{app:numerics}

In order to study the criticality of the light-matter system, we perform numerical simulations using exact diagonalization (ED)~\cite{Sandvik2010} and density matrix renormalization group (DMRG)~\cite{white_prl_1992,white_prb_1993, schollwock_aop_2011, Orus_aop_2014} calculations. 

Within the ED simulations, the Hamiltonian of the system is written in a sparse matrix form and a Lanczos algorithm is used to find the ground state and the first few excited states. We only consider the system at half-filling so that the Hilbert space dimension for the matter part is $L!/(L/2)!^2$, corresponding to $\sim10^5$ for $L=20$. In principle, the photon Hilbert space is infinite dimensional, but in practice it can be truncated up to a maximum photon number $n^{\max}_{\textrm{ph}}$. The total dimension of the system Hilbert space makes ED calculations viable only for small system sizes ($L\lesssim20$) with limited maximum number of photons $n^{\max}_{\textrm{ph}}\sim 5$. It is to be noted that the average photon occupancy $n_{\textrm{ph}} = \braket{\hat a^{\dagger} \hat a}$ increases linearly with system-size in the symmetry-broken charge-density wave (CDW) phase (see Fig.~\ref{fig:collapse_Om1}(c)); thus due to the small value of $L$, the photon population is not too large and typically saturates to $n_{\textrm{ph}}\lesssim 3$. Despite the large computation cost, ED calculations have the advantage of giving access to the full quantum state, allowing the study of any degeneracy of the ground state.

\begin{figure}[htb!]
    \centering
    \includegraphics[width=\linewidth]{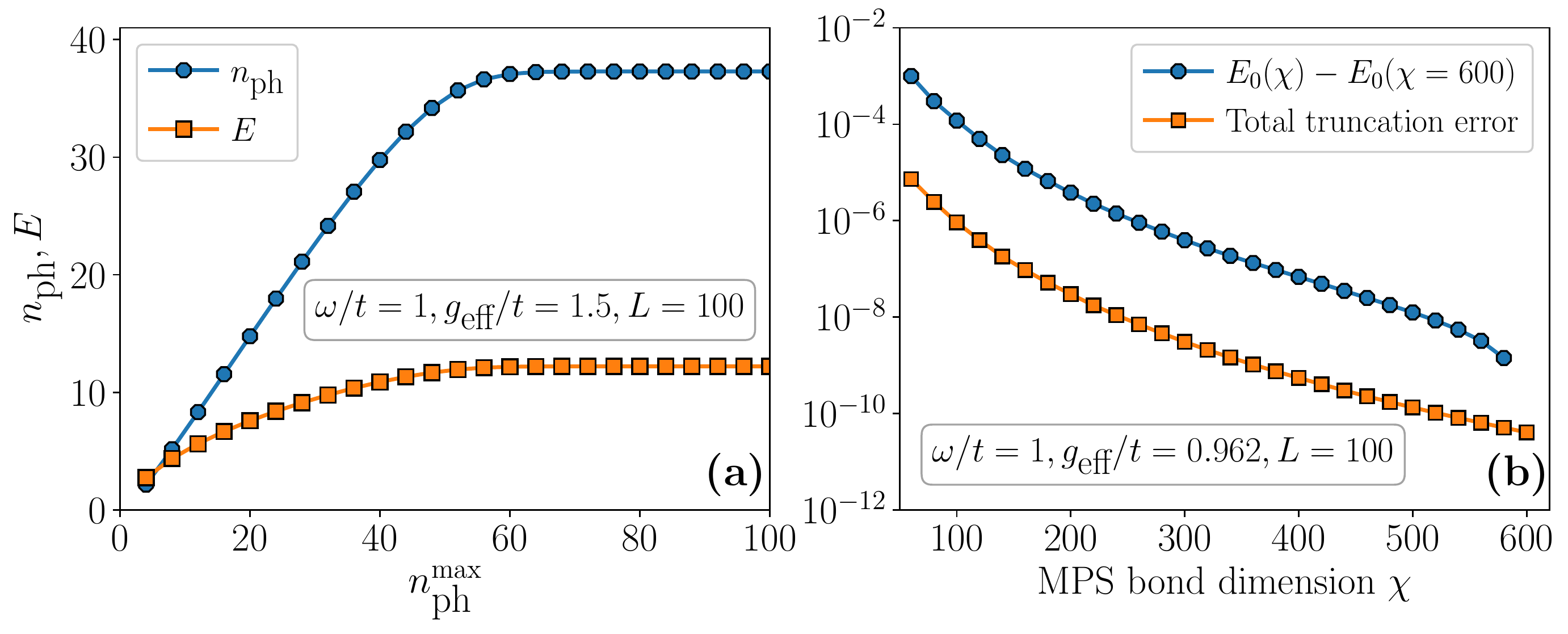}
    \caption{(a) The average photon occupancy $n_{\textrm{ph}}$ and the electric field $E = |\braket{\hat a + \hat a^{\dagger}}|$ as functions of the maximum photon cut-off $n^{\max}_{\textrm{ph}}$ in the symmetry broken CDW phase ($g_{\textrm{eff}}/t=1.5$) for $\omega/t=1$ and system-size $L=100$. Both $n_{\textrm{ph}}$ and $E$ saturates to converged values for $n^{\max}_{\textrm{ph}} \gtrsim 70$.
    (b) The ground state energy difference $E_0(\chi) - E_0(\chi=600)$ with respect to the ground state energy obtained with MPS bond dimension $\chi=600$ and the total DMRG truncation error with varying MPS bond dimension $\chi$. Here, we set $\omega/t=1$ and $L=100$, and consider $g_{\textrm{eff}}/t=0.962$ where the photon entanglement $S_P$ is highest.
    }
    \label{fig:conv}
\end{figure}

The DMRG calculations are based on a matrix-product states (MPS)~\cite{schollwock_aop_2011, Orus_aop_2014} ansatz for a finite system size $L$ with open boundary condition and for truncated photon space. Compared to ED simulations, DMRG calculations allow us to access larger system sizes ($L \sim 100$) with a higher cut-off in the photonic Hilbert-space ($n^{\max}_{\textrm{ph}} \sim 100$), enough to ensure convergence with respect to the truncated photon space (see Fig.~\ref{fig:conv}(a)). Thus, our MPS ansatz consists of $L+1$ sites where the first MPS site corresponds to the photon with local dimension $d_{\textrm{ph}} = n^{\max}_{\textrm{ph}}+1 = 101$, while the other sites are for the fermions with local dimension of $2$. Moreover, to keep the fermion number fixed, we utilize $U(1)$ symmetry-preserving tensors \cite{Singh_PRA_2010,Singh_PRB_2011} at the fermionic MPS sites. Such an MPS setup has been proven to be very successful in similar hybrid systems~\cite{Halati20a, Halati20b, Halati21}. The maximal bond dimension for the MPS representation is set to $\chi=600$, which is sufficient to keep the total discarded weight of the singular values below $10^{-10}$ ($\sim 10^{-12}$ per MPS bond) and ensure convergence for simulations with system sizes up to $L=100$ (see Fig.~\ref{fig:conv}(b)).

\begin{figure}[t]
    \centering
    \includegraphics[width=\linewidth]{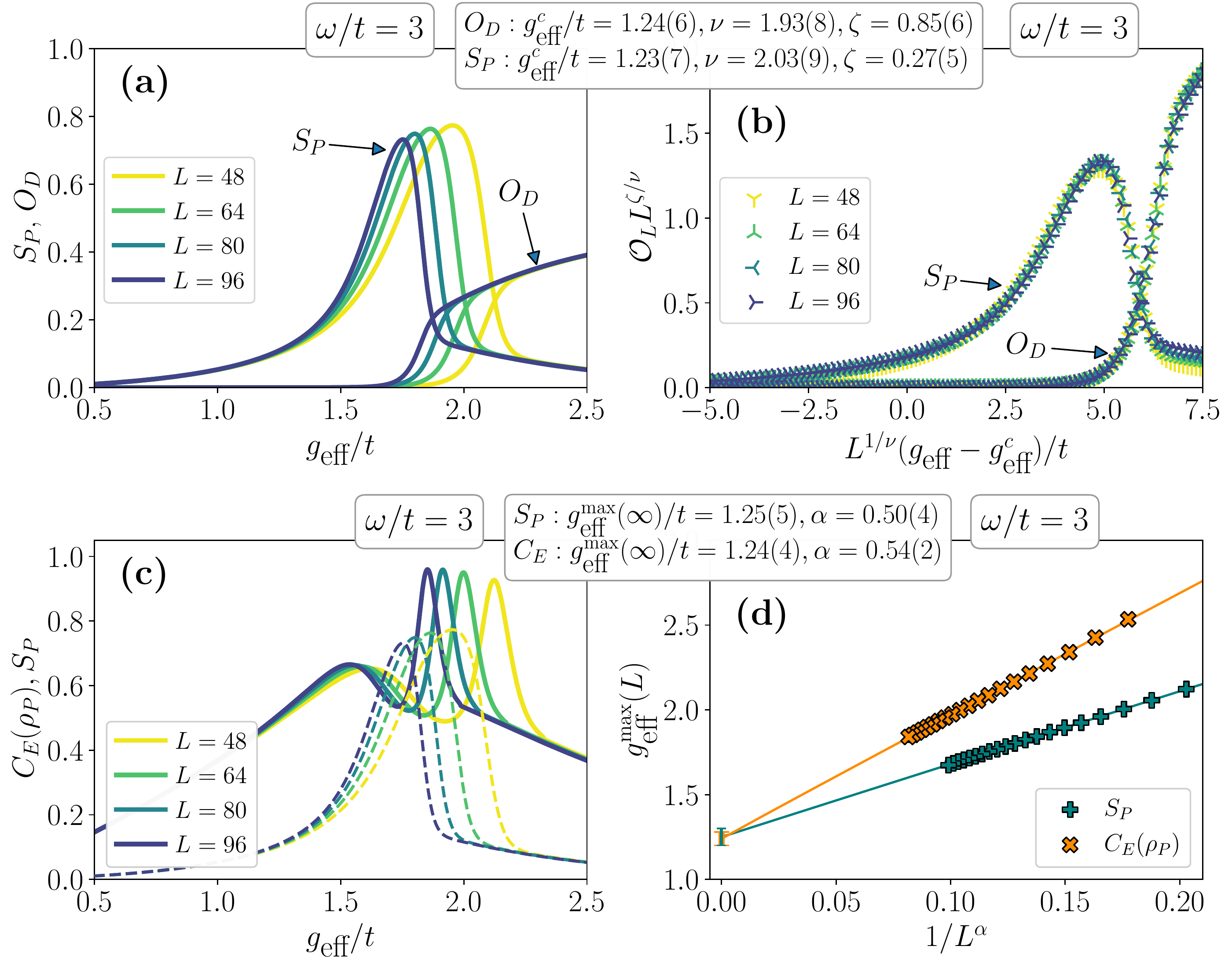}
    \caption{Phase transition analysis at $\omega/t=3$ for the data obtained from DMRG simulations. (a) The photon entanglement entropy $S_P$ and the order parameter $O_D$ as functions of the effective coupling $g_{\textrm{eff}}$. (b) Collapse of the curves using the scaling hypothesis near the transition point. As in the main text, we consider system sizes in the range $L \in [48, 100]$ for the scaling analysis, and plot the data only for $L=48, 64, 80, 96$ for clarity. The values of the critical coupling and of the critical exponents for the two transitions are reported in the box; we observe that the values of $g^c_{\textrm{eff}}$ and $\nu$ agree within error. (c) Entanglement capacity $C_E(\rho_P)$ (solid lines) and photon entropy $S_P$ (dashed). (d) Scaling of the locations $g_{\textrm{eff}}^{\max}(L)$ of the peak values with respect to the system size $L$ for both $S_P$ and $C_E(\rho_P)$. For both quantities, $g_{\textrm{eff}}^{\max}$ approaches the critical point $g_{\textrm{eff}}^c$ in the thermodynamic limit with a power law behavior.}
    \label{figS:collapse_Om3}
\end{figure}

\section{Choice of the symmetry breaking term}
\label{app:symmetrybroken}

We have chosen a symmetry breaking term of the form $g_2(\hat c^{\dagger}_1\hat c_1-\hat c^{\dagger}_L\hat c_L)$, which selects only one of the possible charge density wave states. There is some degree of arbitrariness in this choice: for example one could choose a term that breaks the degeneracy in the electric field, such as $g_2(\hat a+\hat a^{\dagger})$. Indeed, we have performed calculations employing such symmetry breaking term and found results qualitatively similar to those presented in the rest of the work. However, the main drawback of such choice is that the expectation value of the electric field $\hat E=\hat a+\hat a^{\dagger}$ does not depend on the system size $L$ in the disordered phase and scales like $\sqrt{L}$ in the ordered phase. This implies that the symmetry breaking term scales differently with $L$ depending on the value of $g$, making a finite size scaling analysis imprecise. Indeed, we have compared the collapses obtained from the scaling fit and observed that the ones obtained from $g_2(\hat c^{\dagger}_1\hat c_1-\hat c^{\dagger}_L\hat c_L)$ are significantly better.

\section{Results for $\omega/t=3$ and $\omega/t=10$}
\label{app:otherOmega}

\begin{figure}[t]
    \centering
    \includegraphics[width=\linewidth]{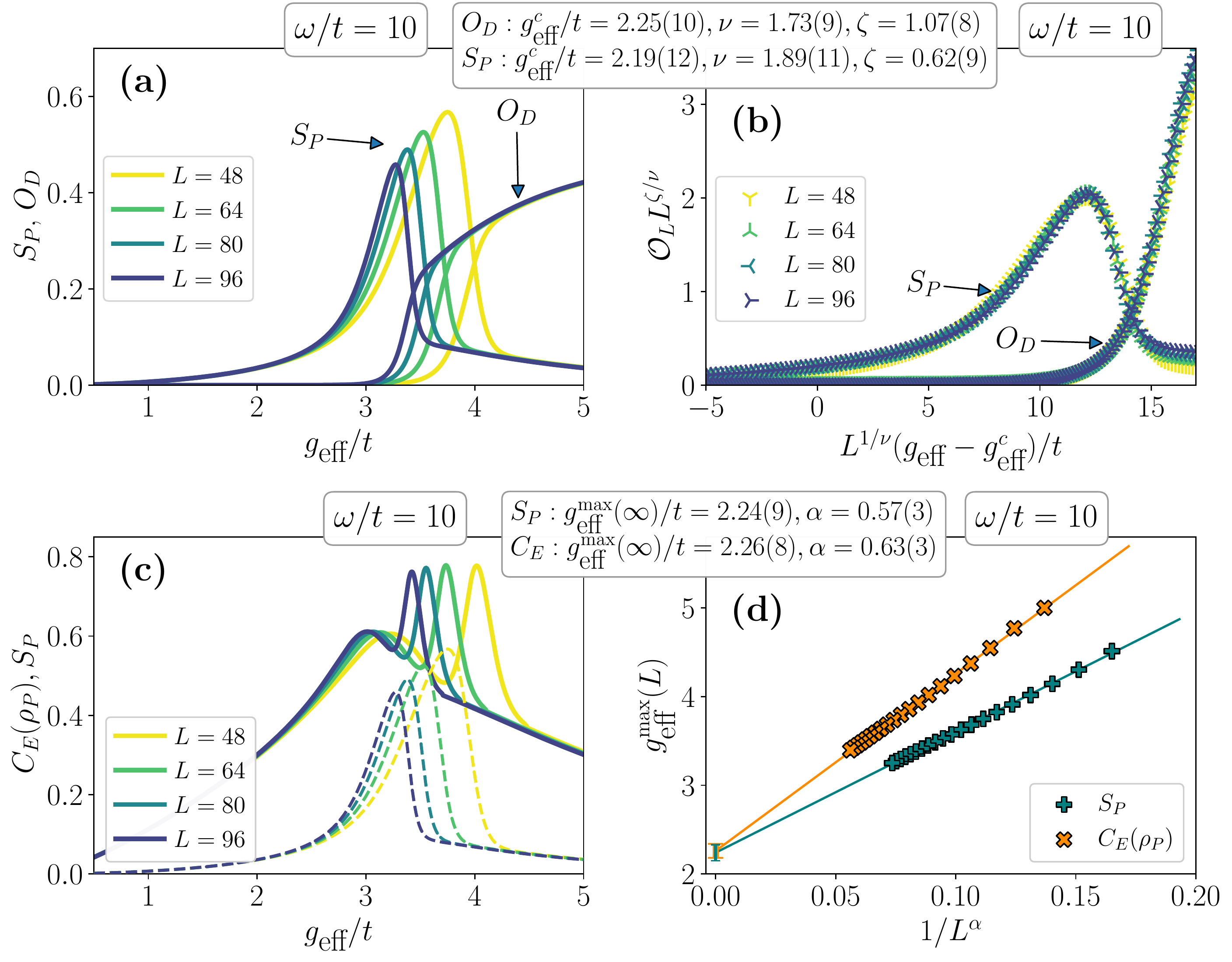}
    \caption{Phase transition analysis at $\omega/t=10$ for the data obtained from DMRG simulations. (a) The photon entanglement entropy $S_P$ and the order parameter $O_D$ as functions of the effective coupling $g_{\textrm{eff}}$. (b) Collapse of the curves using the scaling hypothesis near the transition point. Again we consider system sizes in the range $L \in [48, 100]$ for the scaling analysis, and plot the data only for $L=48, 64, 80, 96$. The values of $g^c_{\textrm{eff}}$ and $\nu$ still agree within uncertainty, but the difference is now larger compared to the $\omega/t=1$ and $\omega/t=3$ cases; the uncertainties on the fit parameters are also larger, signaling a worse quality of the collapses. (c) Entanglement capacity $C_E(\rho_P)$ (solid lines) and photon entropy $S_P$ (dashed). (d) Scaling of the locations $g_{\textrm{eff}}^{\max}(L)$ of the peak values with respect to the system size $L$ for both $S_P$ and $C_E(\rho_P)$. The asymptotic values of $g_{\textrm{eff}}^{\max}$ are again compatible with $g_{\textrm{eff}}^c$.}
    \label{figS:collapse_Om10}
\end{figure}

In the main text, we have presented our analysis for $\omega/t=1$ only. To supplement those results and for the purpose of completeness,
in this section, we present the same analysis for $\omega/t=3$ (Fig. \ref{figS:collapse_Om3}) and $\omega/t=10$ (Fig. \ref{figS:collapse_Om10}).

\begin{figure}[ht!]
    \centering
    \includegraphics[width=\linewidth]{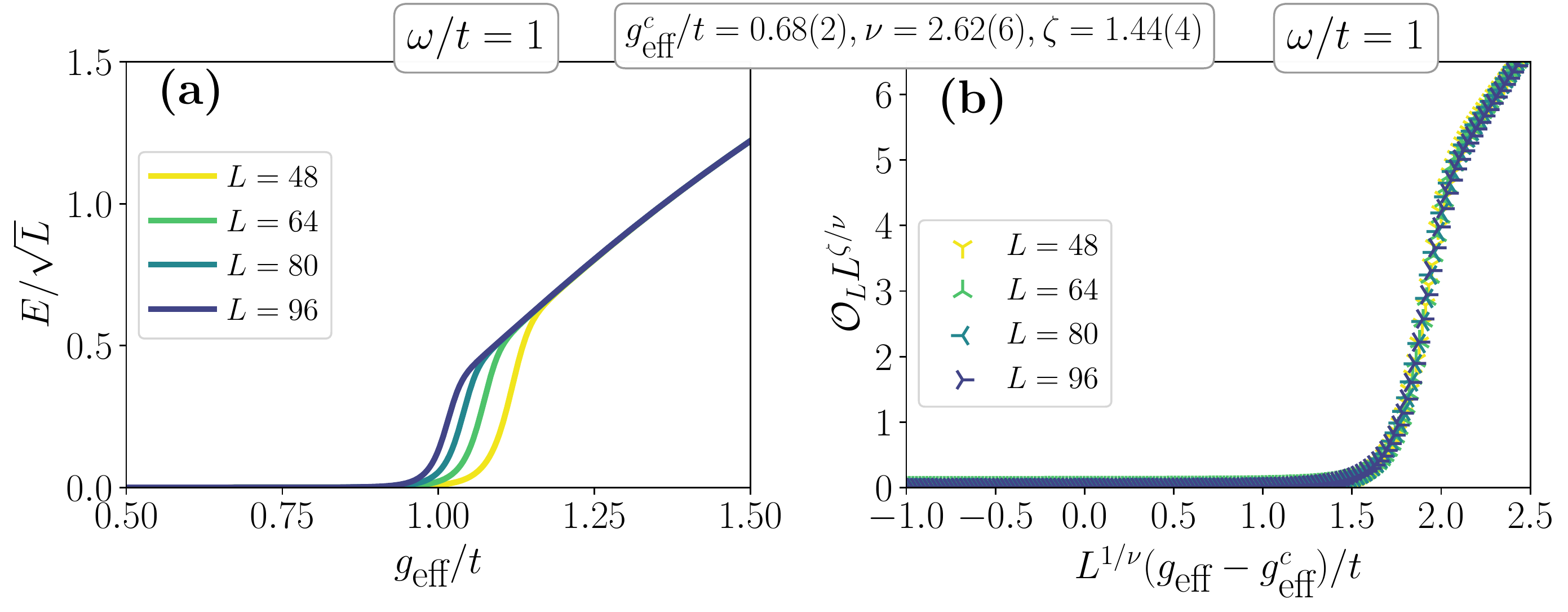}
    \caption{Critical behavior of the rescaled electric field $|\langle\hat a+\hat a^{\dagger}\rangle|/\sqrt L$ at $\omega/t=1$ for data obtained from DMRG simulations. (a) Plot of the rescaled electric field as functions of the effective coupling $g_{\textrm{eff}}$ for selected system sizes. (b) Collapse of the curves  obtained from a finite size scaling analysis of system-sizes in the range $L \in [48, 100]$ for the scaling analysis, and are plotted only for $L=48, 64, 80, 96$ for clarity. The values of the critical coupling and of $\nu$ coincide with those obtained for the entanglement entropy and the order parameter.}
    \label{figS:electric field}
\end{figure}

We find that the results for $\omega/t=3$ are qualitatively similar to those obtained for $\omega/t=1$: in particular, the collapse of the finite size scaling analysis is very good and the critical exponents for entanglement and order parameter are in excellent agreement. On the other hand the quality of the collapse is worse for $\omega/t=10$ as seen by the larger errors and by the fact that the differences in the critical exponent $\nu$ and in the critical coupling $g_{\textrm{eff}}^c/t$  extracted from entanglement and order parameter are also larger.

We postulate that this difference is model dependent: one possible explanation is due to the higher photon energy which makes it difficult to excite and entangle with the electrons. We also remark that in the absence of coupling, the spectrum of the electrons has a bandwidth of $4t$, so that no direct excitations can be mediated by the photon when $\omega \gtrsim 4t$. Thus, we may expect qualitative differences when $\omega/t \gtrsim 4$.

We also report in Fig. \ref{figS:electric field} the scaling analysis and the collapsed curves for the rescaled electric field $E/\sqrt{L}$ for $\omega/t=1$. The uncertainty on the critical parameters are very small and the collapses are excellent, signaling the high quality of the fit. We remark that the electric field exhibits the same critical effective coupling and the same value of $\nu$ as the entanglement entropy of the photon and the order parameter, meaning it can also be used to extract the critical exponents experimentally.

\bibliographystyle{apsrev4-1}
\bibliography{Cavity.bbl}

\begin{thebibliography}{129}%
\makeatletter
\providecommand \@ifxundefined [1]{%
 \@ifx{#1\undefined}
}%
\providecommand \@ifnum [1]{%
 \ifnum #1\expandafter \@firstoftwo
 \else \expandafter \@secondoftwo
 \fi
}%
\providecommand \@ifx [1]{%
 \ifx #1\expandafter \@firstoftwo
 \else \expandafter \@secondoftwo
 \fi
}%
\providecommand \natexlab [1]{#1}%
\providecommand \enquote  [1]{``#1''}%
\providecommand \bibnamefont  [1]{#1}%
\providecommand \bibfnamefont [1]{#1}%
\providecommand \citenamefont [1]{#1}%
\providecommand \href@noop [0]{\@secondoftwo}%
\providecommand \href [0]{\begingroup \@sanitize@url \@href}%
\providecommand \@href[1]{\@@startlink{#1}\@@href}%
\providecommand \@@href[1]{\endgroup#1\@@endlink}%
\providecommand \@sanitize@url [0]{\catcode `\\12\catcode `\$12\catcode
  `\&12\catcode `\#12\catcode `\^12\catcode `\_12\catcode `\%12\relax}%
\providecommand \@@startlink[1]{}%
\providecommand \@@endlink[0]{}%
\providecommand \url  [0]{\begingroup\@sanitize@url \@url }%
\providecommand \@url [1]{\endgroup\@href {#1}{\urlprefix }}%
\providecommand \urlprefix  [0]{URL }%
\providecommand \Eprint [0]{\href }%
\providecommand \doibase [0]{http://dx.doi.org/}%
\providecommand \selectlanguage [0]{\@gobble}%
\providecommand \bibinfo  [0]{\@secondoftwo}%
\providecommand \bibfield  [0]{\@secondoftwo}%
\providecommand \translation [1]{[#1]}%
\providecommand \BibitemOpen [0]{}%
\providecommand \bibitemStop [0]{}%
\providecommand \bibitemNoStop [0]{.\EOS\space}%
\providecommand \EOS [0]{\spacefactor3000\relax}%
\providecommand \BibitemShut  [1]{\csname bibitem#1\endcsname}%
\let\auto@bib@innerbib\@empty
\bibitem [{\citenamefont {Schlawin}\ \emph {et~al.}(2022)\citenamefont
  {Schlawin}, \citenamefont {Kennes},\ and\ \citenamefont
  {Sentef}}]{Rev:Schlawin2022}%
  \BibitemOpen
  \bibfield  {author} {\bibinfo {author} {\bibfnamefont {F.}~\bibnamefont
  {Schlawin}}, \bibinfo {author} {\bibfnamefont {D.~M.}\ \bibnamefont
  {Kennes}}, \ and\ \bibinfo {author} {\bibfnamefont {M.~A.}\ \bibnamefont
  {Sentef}},\ }\href {\doibase 10.1063/5.0083825} {\bibfield  {journal}
  {\bibinfo  {journal} {Applied Physics Reviews}\ }\textbf {\bibinfo {volume}
  {9}},\ \bibinfo {pages} {011312} (\bibinfo {year} {2022})}\BibitemShut
  {NoStop}%
\bibitem [{\citenamefont {Garcia-Vidal}\ \emph {et~al.}(2021)\citenamefont
  {Garcia-Vidal}, \citenamefont {Ciuti},\ and\ \citenamefont
  {Ebbesen}}]{Rev:Garcia-Vidal2021}%
  \BibitemOpen
  \bibfield  {author} {\bibinfo {author} {\bibfnamefont {F.~J.}\ \bibnamefont
  {Garcia-Vidal}}, \bibinfo {author} {\bibfnamefont {C.}~\bibnamefont {Ciuti}},
  \ and\ \bibinfo {author} {\bibfnamefont {T.~W.}\ \bibnamefont {Ebbesen}},\
  }\href {\doibase 10.1126/science.abd0336} {\bibfield  {journal} {\bibinfo
  {journal} {Science}\ }\textbf {\bibinfo {volume} {373}},\ \bibinfo {pages}
  {eabd0336} (\bibinfo {year} {2021})}\BibitemShut {NoStop}%
\bibitem [{\citenamefont {Kurizki}\ \emph {et~al.}(2015)\citenamefont
  {Kurizki}, \citenamefont {Bertet}, \citenamefont {Kubo}, \citenamefont
  {M{\o}lmer}, \citenamefont {Petrosyan}, \citenamefont {Rabl},\ and\
  \citenamefont {Schmiedmayer}}]{New:Kurizki2015}%
  \BibitemOpen
  \bibfield  {author} {\bibinfo {author} {\bibfnamefont {G.}~\bibnamefont
  {Kurizki}}, \bibinfo {author} {\bibfnamefont {P.}~\bibnamefont {Bertet}},
  \bibinfo {author} {\bibfnamefont {Y.}~\bibnamefont {Kubo}}, \bibinfo {author}
  {\bibfnamefont {K.}~\bibnamefont {M{\o}lmer}}, \bibinfo {author}
  {\bibfnamefont {D.}~\bibnamefont {Petrosyan}}, \bibinfo {author}
  {\bibfnamefont {P.}~\bibnamefont {Rabl}}, \ and\ \bibinfo {author}
  {\bibfnamefont {J.}~\bibnamefont {Schmiedmayer}},\ }\href {\doibase
  10.1073/pnas.1419326112} {\bibfield  {journal} {\bibinfo  {journal}
  {Proceedings of the National Academy of Sciences}\ }\textbf {\bibinfo
  {volume} {112}},\ \bibinfo {pages} {3866} (\bibinfo {year}
  {2015})}\BibitemShut {NoStop}%
\bibitem [{\citenamefont {Ritsch}\ \emph {et~al.}(2013)\citenamefont {Ritsch},
  \citenamefont {Domokos}, \citenamefont {Brennecke},\ and\ \citenamefont
  {Esslinger}}]{Ritsch13}%
  \BibitemOpen
  \bibfield  {author} {\bibinfo {author} {\bibfnamefont {H.}~\bibnamefont
  {Ritsch}}, \bibinfo {author} {\bibfnamefont {P.}~\bibnamefont {Domokos}},
  \bibinfo {author} {\bibfnamefont {F.}~\bibnamefont {Brennecke}}, \ and\
  \bibinfo {author} {\bibfnamefont {T.}~\bibnamefont {Esslinger}},\ }\href
  {\doibase 10.1103/RevModPhys.85.553} {\bibfield  {journal} {\bibinfo
  {journal} {Rev. Mod. Phys.}\ }\textbf {\bibinfo {volume} {85}},\ \bibinfo
  {pages} {553} (\bibinfo {year} {2013})}\BibitemShut {NoStop}%
\bibitem [{\citenamefont {Landig}\ \emph {et~al.}(2016)\citenamefont {Landig},
  \citenamefont {Hruby}, \citenamefont {Dogra}, \citenamefont {Landini},
  \citenamefont {Mottl}, \citenamefont {Donner},\ and\ \citenamefont
  {Esslinger}}]{Dicke:Landig2016}%
  \BibitemOpen
  \bibfield  {author} {\bibinfo {author} {\bibfnamefont {R.}~\bibnamefont
  {Landig}}, \bibinfo {author} {\bibfnamefont {L.}~\bibnamefont {Hruby}},
  \bibinfo {author} {\bibfnamefont {N.}~\bibnamefont {Dogra}}, \bibinfo
  {author} {\bibfnamefont {M.}~\bibnamefont {Landini}}, \bibinfo {author}
  {\bibfnamefont {R.}~\bibnamefont {Mottl}}, \bibinfo {author} {\bibfnamefont
  {T.}~\bibnamefont {Donner}}, \ and\ \bibinfo {author} {\bibfnamefont
  {T.}~\bibnamefont {Esslinger}},\ }\href {\doibase 10.1038/nature17409}
  {\bibfield  {journal} {\bibinfo  {journal} {Nature}\ }\textbf {\bibinfo
  {volume} {532}},\ \bibinfo {pages} {476} (\bibinfo {year}
  {2016})}\BibitemShut {NoStop}%
\bibitem [{\citenamefont {Cosme}\ \emph {et~al.}(2018)\citenamefont {Cosme},
  \citenamefont {Georges}, \citenamefont {Hemmerich},\ and\ \citenamefont
  {Mathey}}]{Cosme18}%
  \BibitemOpen
  \bibfield  {author} {\bibinfo {author} {\bibfnamefont {J.~G.}\ \bibnamefont
  {Cosme}}, \bibinfo {author} {\bibfnamefont {C.}~\bibnamefont {Georges}},
  \bibinfo {author} {\bibfnamefont {A.}~\bibnamefont {Hemmerich}}, \ and\
  \bibinfo {author} {\bibfnamefont {L.}~\bibnamefont {Mathey}},\ }\href
  {\doibase 10.1103/PhysRevLett.121.153001} {\bibfield  {journal} {\bibinfo
  {journal} {Phys. Rev. Lett.}\ }\textbf {\bibinfo {volume} {121}},\ \bibinfo
  {pages} {153001} (\bibinfo {year} {2018})}\BibitemShut {NoStop}%
\bibitem [{\citenamefont {Zupancic}\ \emph {et~al.}(2019)\citenamefont
  {Zupancic}, \citenamefont {Dreon}, \citenamefont {Li}, \citenamefont
  {Baumg\"artner}, \citenamefont {Morales}, \citenamefont {Zheng},
  \citenamefont {Cooper}, \citenamefont {Esslinger},\ and\ \citenamefont
  {Donner}}]{Zupancic19}%
  \BibitemOpen
  \bibfield  {author} {\bibinfo {author} {\bibfnamefont {P.}~\bibnamefont
  {Zupancic}}, \bibinfo {author} {\bibfnamefont {D.}~\bibnamefont {Dreon}},
  \bibinfo {author} {\bibfnamefont {X.}~\bibnamefont {Li}}, \bibinfo {author}
  {\bibfnamefont {A.}~\bibnamefont {Baumg\"artner}}, \bibinfo {author}
  {\bibfnamefont {A.}~\bibnamefont {Morales}}, \bibinfo {author} {\bibfnamefont
  {W.}~\bibnamefont {Zheng}}, \bibinfo {author} {\bibfnamefont {N.~R.}\
  \bibnamefont {Cooper}}, \bibinfo {author} {\bibfnamefont {T.}~\bibnamefont
  {Esslinger}}, \ and\ \bibinfo {author} {\bibfnamefont {T.}~\bibnamefont
  {Donner}},\ }\href {\doibase 10.1103/PhysRevLett.123.233601} {\bibfield
  {journal} {\bibinfo  {journal} {Phys. Rev. Lett.}\ }\textbf {\bibinfo
  {volume} {123}},\ \bibinfo {pages} {233601} (\bibinfo {year}
  {2019})}\BibitemShut {NoStop}%
\bibitem [{\citenamefont {Georges}\ \emph {et~al.}(2018)\citenamefont
  {Georges}, \citenamefont {Cosme}, \citenamefont {Mathey},\ and\ \citenamefont
  {Hemmerich}}]{Georges18}%
  \BibitemOpen
  \bibfield  {author} {\bibinfo {author} {\bibfnamefont {C.}~\bibnamefont
  {Georges}}, \bibinfo {author} {\bibfnamefont {J.~G.}\ \bibnamefont {Cosme}},
  \bibinfo {author} {\bibfnamefont {L.}~\bibnamefont {Mathey}}, \ and\ \bibinfo
  {author} {\bibfnamefont {A.}~\bibnamefont {Hemmerich}},\ }\href {\doibase
  10.1103/PhysRevLett.121.220405} {\bibfield  {journal} {\bibinfo  {journal}
  {Phys. Rev. Lett.}\ }\textbf {\bibinfo {volume} {121}},\ \bibinfo {pages}
  {220405} (\bibinfo {year} {2018})}\BibitemShut {NoStop}%
\bibitem [{\citenamefont {Mivehvar}\ \emph {et~al.}(2021)\citenamefont
  {Mivehvar}, \citenamefont {Piazza}, \citenamefont {Donner},\ and\
  \citenamefont {Ritsch}}]{Mivehvar2021}%
  \BibitemOpen
  \bibfield  {author} {\bibinfo {author} {\bibfnamefont {F.}~\bibnamefont
  {Mivehvar}}, \bibinfo {author} {\bibfnamefont {F.}~\bibnamefont {Piazza}},
  \bibinfo {author} {\bibfnamefont {T.}~\bibnamefont {Donner}}, \ and\ \bibinfo
  {author} {\bibfnamefont {H.}~\bibnamefont {Ritsch}},\ }\href {\doibase
  10.1080/00018732.2021.1969727} {\bibfield  {journal} {\bibinfo  {journal}
  {Advances in Physics}\ }\textbf {\bibinfo {volume} {70}},\ \bibinfo {pages}
  {1} (\bibinfo {year} {2021})}\BibitemShut {NoStop}%
\bibitem [{\citenamefont {Bayer}\ \emph {et~al.}(2017)\citenamefont {Bayer},
  \citenamefont {Pozimski}, \citenamefont {Schambeck}, \citenamefont {Schuh},
  \citenamefont {Huber}, \citenamefont {Bougeard},\ and\ \citenamefont
  {Lange}}]{Exp:Bayer2017}%
  \BibitemOpen
  \bibfield  {author} {\bibinfo {author} {\bibfnamefont {A.}~\bibnamefont
  {Bayer}}, \bibinfo {author} {\bibfnamefont {M.}~\bibnamefont {Pozimski}},
  \bibinfo {author} {\bibfnamefont {S.}~\bibnamefont {Schambeck}}, \bibinfo
  {author} {\bibfnamefont {D.}~\bibnamefont {Schuh}}, \bibinfo {author}
  {\bibfnamefont {R.}~\bibnamefont {Huber}}, \bibinfo {author} {\bibfnamefont
  {D.}~\bibnamefont {Bougeard}}, \ and\ \bibinfo {author} {\bibfnamefont
  {C.}~\bibnamefont {Lange}},\ }\href {\doibase 10.1021/acs.nanolett.7b03103}
  {\bibfield  {journal} {\bibinfo  {journal} {Nano Letters}\ }\textbf {\bibinfo
  {volume} {17}},\ \bibinfo {pages} {6340} (\bibinfo {year}
  {2017})}\BibitemShut {NoStop}%
\bibitem [{\citenamefont {Schachenmayer}\ \emph {et~al.}(2015)\citenamefont
  {Schachenmayer}, \citenamefont {Genes}, \citenamefont {Tignone},\ and\
  \citenamefont {Pupillo}}]{Exc:Schachenmayer2015a}%
  \BibitemOpen
  \bibfield  {author} {\bibinfo {author} {\bibfnamefont {J.}~\bibnamefont
  {Schachenmayer}}, \bibinfo {author} {\bibfnamefont {C.}~\bibnamefont
  {Genes}}, \bibinfo {author} {\bibfnamefont {E.}~\bibnamefont {Tignone}}, \
  and\ \bibinfo {author} {\bibfnamefont {G.}~\bibnamefont {Pupillo}},\ }\href
  {\doibase 10.1103/PhysRevLett.114.196403} {\bibfield  {journal} {\bibinfo
  {journal} {Phys. Rev. Lett.}\ }\textbf {\bibinfo {volume} {114}},\ \bibinfo
  {pages} {196403} (\bibinfo {year} {2015})}\BibitemShut {NoStop}%
\bibitem [{\citenamefont {Hagenm\"uller}\ \emph {et~al.}(2018)\citenamefont
  {Hagenm\"uller}, \citenamefont {Sch\"utz}, \citenamefont {Schachenmayer},
  \citenamefont {Genes},\ and\ \citenamefont {Pupillo}}]{MB:Hagenmuller2018b}%
  \BibitemOpen
  \bibfield  {author} {\bibinfo {author} {\bibfnamefont {D.}~\bibnamefont
  {Hagenm\"uller}}, \bibinfo {author} {\bibfnamefont {S.}~\bibnamefont
  {Sch\"utz}}, \bibinfo {author} {\bibfnamefont {J.}~\bibnamefont
  {Schachenmayer}}, \bibinfo {author} {\bibfnamefont {C.}~\bibnamefont
  {Genes}}, \ and\ \bibinfo {author} {\bibfnamefont {G.}~\bibnamefont
  {Pupillo}},\ }\href {\doibase 10.1103/PhysRevB.97.205303} {\bibfield
  {journal} {\bibinfo  {journal} {Phys. Rev. B}\ }\textbf {\bibinfo {volume}
  {97}},\ \bibinfo {pages} {205303} (\bibinfo {year} {2018})}\BibitemShut
  {NoStop}%
\bibitem [{\citenamefont {Bartolo}\ and\ \citenamefont
  {Ciuti}(2018)}]{MB:Bartolo2018b}%
  \BibitemOpen
  \bibfield  {author} {\bibinfo {author} {\bibfnamefont {N.}~\bibnamefont
  {Bartolo}}\ and\ \bibinfo {author} {\bibfnamefont {C.}~\bibnamefont
  {Ciuti}},\ }\href {\doibase 10.1103/PhysRevB.98.205301} {\bibfield  {journal}
  {\bibinfo  {journal} {Phys. Rev. B}\ }\textbf {\bibinfo {volume} {98}},\
  \bibinfo {pages} {205301} (\bibinfo {year} {2018})}\BibitemShut {NoStop}%
\bibitem [{\citenamefont {Latini}\ \emph {et~al.}(2019)\citenamefont {Latini},
  \citenamefont {Ronca}, \citenamefont {Giovannini}, \citenamefont
  {H\"{u}bener},\ and\ \citenamefont {Rubio}}]{Exc:Latini2019}%
  \BibitemOpen
  \bibfield  {author} {\bibinfo {author} {\bibfnamefont {S.}~\bibnamefont
  {Latini}}, \bibinfo {author} {\bibfnamefont {E.}~\bibnamefont {Ronca}},
  \bibinfo {author} {\bibfnamefont {U.~D.}\ \bibnamefont {Giovannini}},
  \bibinfo {author} {\bibfnamefont {H.}~\bibnamefont {H\"{u}bener}}, \ and\
  \bibinfo {author} {\bibfnamefont {A.}~\bibnamefont {Rubio}},\ }\href
  {\doibase 10.1021/acs.nanolett.9b00183} {\bibfield  {journal} {\bibinfo
  {journal} {Nano Letters}\ }\textbf {\bibinfo {volume} {19}},\ \bibinfo
  {pages} {3473} (\bibinfo {year} {2019})}\BibitemShut {NoStop}%
\bibitem [{\citenamefont {Zhong}\ \emph {et~al.}(2017)\citenamefont {Zhong},
  \citenamefont {Chervy}, \citenamefont {Zhang}, \citenamefont {Thomas},
  \citenamefont {George}, \citenamefont {Genet}, \citenamefont {Hutchison},\
  and\ \citenamefont {Ebbesen}}]{Mol:Zhong2017a}%
  \BibitemOpen
  \bibfield  {author} {\bibinfo {author} {\bibfnamefont {X.}~\bibnamefont
  {Zhong}}, \bibinfo {author} {\bibfnamefont {T.}~\bibnamefont {Chervy}},
  \bibinfo {author} {\bibfnamefont {L.}~\bibnamefont {Zhang}}, \bibinfo
  {author} {\bibfnamefont {A.}~\bibnamefont {Thomas}}, \bibinfo {author}
  {\bibfnamefont {J.}~\bibnamefont {George}}, \bibinfo {author} {\bibfnamefont
  {C.}~\bibnamefont {Genet}}, \bibinfo {author} {\bibfnamefont {J.~A.}\
  \bibnamefont {Hutchison}}, \ and\ \bibinfo {author} {\bibfnamefont {T.~W.}\
  \bibnamefont {Ebbesen}},\ }\href {\doibase 10.1002/anie.201703539} {\bibfield
   {journal} {\bibinfo  {journal} {Angewandte Chemie International Edition}\
  }\textbf {\bibinfo {volume} {56}},\ \bibinfo {pages} {9034} (\bibinfo {year}
  {2017})}\BibitemShut {NoStop}%
\bibitem [{\citenamefont {Feist}\ \emph {et~al.}(2017)\citenamefont {Feist},
  \citenamefont {Galego},\ and\ \citenamefont {Garcia-Vidal}}]{Mol:Feist2018}%
  \BibitemOpen
  \bibfield  {author} {\bibinfo {author} {\bibfnamefont {J.}~\bibnamefont
  {Feist}}, \bibinfo {author} {\bibfnamefont {J.}~\bibnamefont {Galego}}, \
  and\ \bibinfo {author} {\bibfnamefont {F.~J.}\ \bibnamefont {Garcia-Vidal}},\
  }\href {\doibase 10.1021/acsphotonics.7b00680} {\bibfield  {journal}
  {\bibinfo  {journal} {{ACS} Photonics}\ }\textbf {\bibinfo {volume} {5}},\
  \bibinfo {pages} {205} (\bibinfo {year} {2017})}\BibitemShut {NoStop}%
\bibitem [{\citenamefont {Flick}\ \emph {et~al.}(2018)\citenamefont {Flick},
  \citenamefont {Rivera},\ and\ \citenamefont {Narang}}]{Mol:Flick2018}%
  \BibitemOpen
  \bibfield  {author} {\bibinfo {author} {\bibfnamefont {J.}~\bibnamefont
  {Flick}}, \bibinfo {author} {\bibfnamefont {N.}~\bibnamefont {Rivera}}, \
  and\ \bibinfo {author} {\bibfnamefont {P.}~\bibnamefont {Narang}},\ }\href
  {\doibase 10.1515/nanoph-2018-0067} {\bibfield  {journal} {\bibinfo
  {journal} {Nanophotonics}\ }\textbf {\bibinfo {volume} {7}},\ \bibinfo
  {pages} {1479} (\bibinfo {year} {2018})}\BibitemShut {NoStop}%
\bibitem [{\citenamefont {Ribeiro}\ \emph {et~al.}(2018)\citenamefont
  {Ribeiro}, \citenamefont {Mart{\'{\i}}nez-Mart{\'{\i}}nez}, \citenamefont
  {Du}, \citenamefont {Campos-Gonzalez-Angulo},\ and\ \citenamefont
  {Yuen-Zhou}}]{Mol:Ribeiro2018}%
  \BibitemOpen
  \bibfield  {author} {\bibinfo {author} {\bibfnamefont {R.~F.}\ \bibnamefont
  {Ribeiro}}, \bibinfo {author} {\bibfnamefont {L.~A.}\ \bibnamefont
  {Mart{\'{\i}}nez-Mart{\'{\i}}nez}}, \bibinfo {author} {\bibfnamefont
  {M.}~\bibnamefont {Du}}, \bibinfo {author} {\bibfnamefont {J.}~\bibnamefont
  {Campos-Gonzalez-Angulo}}, \ and\ \bibinfo {author} {\bibfnamefont
  {J.}~\bibnamefont {Yuen-Zhou}},\ }\href {\doibase 10.1039/c8sc01043a}
  {\bibfield  {journal} {\bibinfo  {journal} {{Chemical Science}}\ }\textbf
  {\bibinfo {volume} {9}},\ \bibinfo {pages} {6325} (\bibinfo {year}
  {2018})}\BibitemShut {NoStop}%
\bibitem [{\citenamefont {Rozenman}\ \emph {et~al.}(2017)\citenamefont
  {Rozenman}, \citenamefont {Akulov}, \citenamefont {Golombek},\ and\
  \citenamefont {Schwartz}}]{Mol:Rozenman2018b}%
  \BibitemOpen
  \bibfield  {author} {\bibinfo {author} {\bibfnamefont {G.~G.}\ \bibnamefont
  {Rozenman}}, \bibinfo {author} {\bibfnamefont {K.}~\bibnamefont {Akulov}},
  \bibinfo {author} {\bibfnamefont {A.}~\bibnamefont {Golombek}}, \ and\
  \bibinfo {author} {\bibfnamefont {T.}~\bibnamefont {Schwartz}},\ }\href
  {\doibase 10.1021/acsphotonics.7b01332} {\bibfield  {journal} {\bibinfo
  {journal} {{ACS} Photonics}\ }\textbf {\bibinfo {volume} {5}},\ \bibinfo
  {pages} {105} (\bibinfo {year} {2017})}\BibitemShut {NoStop}%
\bibitem [{\citenamefont {K{\'{e}}na-Cohen}\ and\ \citenamefont
  {Yuen-Zhou}(2019)}]{Mol:Kena-Cohen2019a}%
  \BibitemOpen
  \bibfield  {author} {\bibinfo {author} {\bibfnamefont {S.}~\bibnamefont
  {K{\'{e}}na-Cohen}}\ and\ \bibinfo {author} {\bibfnamefont {J.}~\bibnamefont
  {Yuen-Zhou}},\ }\href {\doibase 10.1021/acscentsci.9b00219} {\bibfield
  {journal} {\bibinfo  {journal} {{ACS} Central Science}\ }\textbf {\bibinfo
  {volume} {5}},\ \bibinfo {pages} {386} (\bibinfo {year} {2019})}\BibitemShut
  {NoStop}%
\bibitem [{\citenamefont {Thomas}\ \emph {et~al.}(2019)\citenamefont {Thomas},
  \citenamefont {Devaux}, \citenamefont {Nagarajan}, \citenamefont {Chervy},
  \citenamefont {Seidel}, \citenamefont {Hagenmüller}, \citenamefont
  {Schütz}, \citenamefont {Schachenmayer}, \citenamefont {Genet},
  \citenamefont {Pupillo},\ and\ \citenamefont {Ebbesen}}]{Mol:Thomas2019b}%
  \BibitemOpen
  \bibfield  {author} {\bibinfo {author} {\bibfnamefont {A.}~\bibnamefont
  {Thomas}}, \bibinfo {author} {\bibfnamefont {E.}~\bibnamefont {Devaux}},
  \bibinfo {author} {\bibfnamefont {K.}~\bibnamefont {Nagarajan}}, \bibinfo
  {author} {\bibfnamefont {T.}~\bibnamefont {Chervy}}, \bibinfo {author}
  {\bibfnamefont {M.}~\bibnamefont {Seidel}}, \bibinfo {author} {\bibfnamefont
  {D.}~\bibnamefont {Hagenmüller}}, \bibinfo {author} {\bibfnamefont
  {S.}~\bibnamefont {Schütz}}, \bibinfo {author} {\bibfnamefont
  {J.}~\bibnamefont {Schachenmayer}}, \bibinfo {author} {\bibfnamefont
  {C.}~\bibnamefont {Genet}}, \bibinfo {author} {\bibfnamefont
  {G.}~\bibnamefont {Pupillo}}, \ and\ \bibinfo {author} {\bibfnamefont
  {T.~W.}\ \bibnamefont {Ebbesen}},\ }\href@noop {} {\  (\bibinfo {year}
  {2019})},\ \Eprint {http://arxiv.org/abs/arXiv:1911.01459} {arXiv:1911.01459}
  \BibitemShut {NoStop}%
\bibitem [{\citenamefont {Liu}\ \emph {et~al.}(2020)\citenamefont {Liu},
  \citenamefont {Zhao},\ and\ \citenamefont {Wu}}]{Mol:Liu2020b}%
  \BibitemOpen
  \bibfield  {author} {\bibinfo {author} {\bibfnamefont {J.}~\bibnamefont
  {Liu}}, \bibinfo {author} {\bibfnamefont {Q.}~\bibnamefont {Zhao}}, \ and\
  \bibinfo {author} {\bibfnamefont {N.}~\bibnamefont {Wu}},\ }\href {\doibase
  10.1063/5.0014820} {\bibfield  {journal} {\bibinfo  {journal} {The Journal of
  Chemical Physics}\ }\textbf {\bibinfo {volume} {153}},\ \bibinfo {pages}
  {074108} (\bibinfo {year} {2020})}\BibitemShut {NoStop}%
\bibitem [{\citenamefont {Takahashi}\ and\ \citenamefont
  {Watanabe}(2020)}]{Mol:Takahashi2020b}%
  \BibitemOpen
  \bibfield  {author} {\bibinfo {author} {\bibfnamefont {S.}~\bibnamefont
  {Takahashi}}\ and\ \bibinfo {author} {\bibfnamefont {K.}~\bibnamefont
  {Watanabe}},\ }\href {\doibase 10.1021/acs.jpclett.9b03789} {\bibfield
  {journal} {\bibinfo  {journal} {The Journal of Physical Chemistry Letters}\
  }\textbf {\bibinfo {volume} {11}},\ \bibinfo {pages} {1349} (\bibinfo {year}
  {2020})}\BibitemShut {NoStop}%
\bibitem [{\citenamefont {Wellnitz}\ \emph {et~al.}(2020)\citenamefont
  {Wellnitz}, \citenamefont {Sch\"utz}, \citenamefont {Whitlock}, \citenamefont
  {Schachenmayer},\ and\ \citenamefont {Pupillo}}]{Mol:Wellnitz2020a}%
  \BibitemOpen
  \bibfield  {author} {\bibinfo {author} {\bibfnamefont {D.}~\bibnamefont
  {Wellnitz}}, \bibinfo {author} {\bibfnamefont {S.}~\bibnamefont {Sch\"utz}},
  \bibinfo {author} {\bibfnamefont {S.}~\bibnamefont {Whitlock}}, \bibinfo
  {author} {\bibfnamefont {J.}~\bibnamefont {Schachenmayer}}, \ and\ \bibinfo
  {author} {\bibfnamefont {G.}~\bibnamefont {Pupillo}},\ }\href {\doibase
  10.1103/PhysRevLett.125.193201} {\bibfield  {journal} {\bibinfo  {journal}
  {Phys. Rev. Lett.}\ }\textbf {\bibinfo {volume} {125}},\ \bibinfo {pages}
  {193201} (\bibinfo {year} {2020})}\BibitemShut {NoStop}%
\bibitem [{\citenamefont {Feist}\ and\ \citenamefont
  {Garcia-Vidal}(2015)}]{Exc:Feist2015}%
  \BibitemOpen
  \bibfield  {author} {\bibinfo {author} {\bibfnamefont {J.}~\bibnamefont
  {Feist}}\ and\ \bibinfo {author} {\bibfnamefont {F.~J.}\ \bibnamefont
  {Garcia-Vidal}},\ }\href {\doibase 10.1103/PhysRevLett.114.196402} {\bibfield
   {journal} {\bibinfo  {journal} {Phys. Rev. Lett.}\ }\textbf {\bibinfo
  {volume} {114}},\ \bibinfo {pages} {196402} (\bibinfo {year}
  {2015})}\BibitemShut {NoStop}%
\bibitem [{\citenamefont {Orgiu}\ \emph {et~al.}(2015)\citenamefont {Orgiu},
  \citenamefont {George}, \citenamefont {Hutchison}, \citenamefont {Devaux},
  \citenamefont {Dayen}, \citenamefont {Doudin}, \citenamefont {Stellacci},
  \citenamefont {Genet}, \citenamefont {Schachenmayer}, \citenamefont {Genes},
  \citenamefont {Pupillo}, \citenamefont {Samor{\`{\i}}},\ and\ \citenamefont
  {Ebbesen}}]{Exc:Orgiu2015}%
  \BibitemOpen
  \bibfield  {author} {\bibinfo {author} {\bibfnamefont {E.}~\bibnamefont
  {Orgiu}}, \bibinfo {author} {\bibfnamefont {J.}~\bibnamefont {George}},
  \bibinfo {author} {\bibfnamefont {J.~A.}\ \bibnamefont {Hutchison}}, \bibinfo
  {author} {\bibfnamefont {E.}~\bibnamefont {Devaux}}, \bibinfo {author}
  {\bibfnamefont {J.~F.}\ \bibnamefont {Dayen}}, \bibinfo {author}
  {\bibfnamefont {B.}~\bibnamefont {Doudin}}, \bibinfo {author} {\bibfnamefont
  {F.}~\bibnamefont {Stellacci}}, \bibinfo {author} {\bibfnamefont
  {C.}~\bibnamefont {Genet}}, \bibinfo {author} {\bibfnamefont
  {J.}~\bibnamefont {Schachenmayer}}, \bibinfo {author} {\bibfnamefont
  {C.}~\bibnamefont {Genes}}, \bibinfo {author} {\bibfnamefont
  {G.}~\bibnamefont {Pupillo}}, \bibinfo {author} {\bibfnamefont
  {P.}~\bibnamefont {Samor{\`{\i}}}}, \ and\ \bibinfo {author} {\bibfnamefont
  {T.~W.}\ \bibnamefont {Ebbesen}},\ }\href {\doibase 10.1038/nmat4392}
  {\bibfield  {journal} {\bibinfo  {journal} {Nature Materials}\ }\textbf
  {\bibinfo {volume} {14}},\ \bibinfo {pages} {1123} (\bibinfo {year}
  {2015})}\BibitemShut {NoStop}%
\bibitem [{\citenamefont {Wei}\ \emph {et~al.}(2019)\citenamefont {Wei},
  \citenamefont {Zhao}, \citenamefont {Liu}, \citenamefont {Zhao},
  \citenamefont {Wu},\ and\ \citenamefont {Xu}}]{Exc:Wei2019b}%
  \BibitemOpen
  \bibfield  {author} {\bibinfo {author} {\bibfnamefont {J.}~\bibnamefont
  {Wei}}, \bibinfo {author} {\bibfnamefont {F.}~\bibnamefont {Zhao}}, \bibinfo
  {author} {\bibfnamefont {J.}~\bibnamefont {Liu}}, \bibinfo {author}
  {\bibfnamefont {Q.}~\bibnamefont {Zhao}}, \bibinfo {author} {\bibfnamefont
  {N.}~\bibnamefont {Wu}}, \ and\ \bibinfo {author} {\bibfnamefont
  {D.}~\bibnamefont {Xu}},\ }\href {\doibase 10.1103/PhysRevE.100.012125}
  {\bibfield  {journal} {\bibinfo  {journal} {Phys. Rev. E}\ }\textbf {\bibinfo
  {volume} {100}},\ \bibinfo {pages} {012125} (\bibinfo {year}
  {2019})}\BibitemShut {NoStop}%
\bibitem [{\citenamefont {Lenk}\ and\ \citenamefont
  {Eckstein}(2020)}]{Exc:Lenk2020b}%
  \BibitemOpen
  \bibfield  {author} {\bibinfo {author} {\bibfnamefont {K.}~\bibnamefont
  {Lenk}}\ and\ \bibinfo {author} {\bibfnamefont {M.}~\bibnamefont
  {Eckstein}},\ }\href {\doibase 10.1103/PhysRevB.102.205129} {\bibfield
  {journal} {\bibinfo  {journal} {Phys. Rev. B}\ }\textbf {\bibinfo {volume}
  {102}},\ \bibinfo {pages} {205129} (\bibinfo {year} {2020})}\BibitemShut
  {NoStop}%
\bibitem [{\citenamefont {Hagenm\"uller}\ \emph {et~al.}(2017)\citenamefont
  {Hagenm\"uller}, \citenamefont {Schachenmayer}, \citenamefont {Sch\"utz},
  \citenamefont {Genes},\ and\ \citenamefont {Pupillo}}]{MB:Hagenmuller2017b}%
  \BibitemOpen
  \bibfield  {author} {\bibinfo {author} {\bibfnamefont {D.}~\bibnamefont
  {Hagenm\"uller}}, \bibinfo {author} {\bibfnamefont {J.}~\bibnamefont
  {Schachenmayer}}, \bibinfo {author} {\bibfnamefont {S.}~\bibnamefont
  {Sch\"utz}}, \bibinfo {author} {\bibfnamefont {C.}~\bibnamefont {Genes}}, \
  and\ \bibinfo {author} {\bibfnamefont {G.}~\bibnamefont {Pupillo}},\ }\href
  {\doibase 10.1103/PhysRevLett.119.223601} {\bibfield  {journal} {\bibinfo
  {journal} {Phys. Rev. Lett.}\ }\textbf {\bibinfo {volume} {119}},\ \bibinfo
  {pages} {223601} (\bibinfo {year} {2017})}\BibitemShut {NoStop}%
\bibitem [{\citenamefont {Du}\ \emph {et~al.}(2018)\citenamefont {Du},
  \citenamefont {Mart{\'{\i}}nez-Mart{\'{\i}}nez}, \citenamefont {Ribeiro},
  \citenamefont {Hu}, \citenamefont {Menon},\ and\ \citenamefont
  {Yuen-Zhou}}]{MB:Du2018b}%
  \BibitemOpen
  \bibfield  {author} {\bibinfo {author} {\bibfnamefont {M.}~\bibnamefont
  {Du}}, \bibinfo {author} {\bibfnamefont {L.~A.}\ \bibnamefont
  {Mart{\'{\i}}nez-Mart{\'{\i}}nez}}, \bibinfo {author} {\bibfnamefont {R.~F.}\
  \bibnamefont {Ribeiro}}, \bibinfo {author} {\bibfnamefont {Z.}~\bibnamefont
  {Hu}}, \bibinfo {author} {\bibfnamefont {V.~M.}\ \bibnamefont {Menon}}, \
  and\ \bibinfo {author} {\bibfnamefont {J.}~\bibnamefont {Yuen-Zhou}},\ }\href
  {\doibase 10.1039/c8sc00171e} {\bibfield  {journal} {\bibinfo  {journal}
  {Chemical Science}\ }\textbf {\bibinfo {volume} {9}},\ \bibinfo {pages}
  {6659} (\bibinfo {year} {2018})}\BibitemShut {NoStop}%
\bibitem [{\citenamefont {Rokaj}\ \emph
  {et~al.}(2022{\natexlab{a}})\citenamefont {Rokaj}, \citenamefont
  {Ruggenthaler}, \citenamefont {Eich},\ and\ \citenamefont
  {Rubio}}]{TRANSPORT:Rokaj2022}%
  \BibitemOpen
  \bibfield  {author} {\bibinfo {author} {\bibfnamefont {V.}~\bibnamefont
  {Rokaj}}, \bibinfo {author} {\bibfnamefont {M.}~\bibnamefont {Ruggenthaler}},
  \bibinfo {author} {\bibfnamefont {F.~G.}\ \bibnamefont {Eich}}, \ and\
  \bibinfo {author} {\bibfnamefont {A.}~\bibnamefont {Rubio}},\ }\href
  {\doibase 10.1103/PhysRevResearch.4.013012} {\bibfield  {journal} {\bibinfo
  {journal} {Phys. Rev. Research}\ }\textbf {\bibinfo {volume} {4}},\ \bibinfo
  {pages} {013012} (\bibinfo {year} {2022}{\natexlab{a}})}\BibitemShut
  {NoStop}%
\bibitem [{\citenamefont {Eckhardt}\ \emph {et~al.}(2022)\citenamefont
  {Eckhardt}, \citenamefont {Passetti}, \citenamefont {Othman}, \citenamefont
  {Karrasch}, \citenamefont {Cavaliere}, \citenamefont {Sentef},\ and\
  \citenamefont {Kennes}}]{TRANSPORT:Eckhardt2022}%
  \BibitemOpen
  \bibfield  {author} {\bibinfo {author} {\bibfnamefont {C.~J.}\ \bibnamefont
  {Eckhardt}}, \bibinfo {author} {\bibfnamefont {G.}~\bibnamefont {Passetti}},
  \bibinfo {author} {\bibfnamefont {M.}~\bibnamefont {Othman}}, \bibinfo
  {author} {\bibfnamefont {C.}~\bibnamefont {Karrasch}}, \bibinfo {author}
  {\bibfnamefont {F.}~\bibnamefont {Cavaliere}}, \bibinfo {author}
  {\bibfnamefont {M.~A.}\ \bibnamefont {Sentef}}, \ and\ \bibinfo {author}
  {\bibfnamefont {D.~M.}\ \bibnamefont {Kennes}},\ }\href {\doibase
  10.1038/s42005-022-00880-9} {\bibfield  {journal} {\bibinfo  {journal}
  {Communications Physics}\ }\textbf {\bibinfo {volume} {5}},\ \bibinfo {pages}
  {122} (\bibinfo {year} {2022})}\BibitemShut {NoStop}%
\bibitem [{\citenamefont {Kiffner}\ \emph
  {et~al.}(2019{\natexlab{a}})\citenamefont {Kiffner}, \citenamefont
  {Coulthard}, \citenamefont {Schlawin}, \citenamefont {Ardavan},\ and\
  \citenamefont {Jaksch}}]{New:Kiffner2019b}%
  \BibitemOpen
  \bibfield  {author} {\bibinfo {author} {\bibfnamefont {M.}~\bibnamefont
  {Kiffner}}, \bibinfo {author} {\bibfnamefont {J.~R.}\ \bibnamefont
  {Coulthard}}, \bibinfo {author} {\bibfnamefont {F.}~\bibnamefont {Schlawin}},
  \bibinfo {author} {\bibfnamefont {A.}~\bibnamefont {Ardavan}}, \ and\
  \bibinfo {author} {\bibfnamefont {D.}~\bibnamefont {Jaksch}},\ }\href
  {\doibase 10.1103/PhysRevB.99.085116} {\bibfield  {journal} {\bibinfo
  {journal} {Phys. Rev. B}\ }\textbf {\bibinfo {volume} {99}},\ \bibinfo
  {pages} {085116} (\bibinfo {year} {2019}{\natexlab{a}})}\BibitemShut
  {NoStop}%
\bibitem [{\citenamefont {Kiffner}\ \emph
  {et~al.}(2019{\natexlab{b}})\citenamefont {Kiffner}, \citenamefont
  {Coulthard}, \citenamefont {Schlawin}, \citenamefont {Ardavan},\ and\
  \citenamefont {Jaksch}}]{New:Kiffner2019d}%
  \BibitemOpen
  \bibfield  {author} {\bibinfo {author} {\bibfnamefont {M.}~\bibnamefont
  {Kiffner}}, \bibinfo {author} {\bibfnamefont {J.~R.}\ \bibnamefont
  {Coulthard}}, \bibinfo {author} {\bibfnamefont {F.}~\bibnamefont {Schlawin}},
  \bibinfo {author} {\bibfnamefont {A.}~\bibnamefont {Ardavan}}, \ and\
  \bibinfo {author} {\bibfnamefont {D.}~\bibnamefont {Jaksch}},\ }\href
  {\doibase 10.1103/PhysRevB.99.099907} {\bibfield  {journal} {\bibinfo
  {journal} {Phys. Rev. B}\ }\textbf {\bibinfo {volume} {99}},\ \bibinfo
  {pages} {099907} (\bibinfo {year} {2019}{\natexlab{b}})}\BibitemShut
  {NoStop}%
\bibitem [{\citenamefont {Kiffner}\ \emph
  {et~al.}(2019{\natexlab{c}})\citenamefont {Kiffner}, \citenamefont
  {Coulthard}, \citenamefont {Schlawin}, \citenamefont {Ardavan},\ and\
  \citenamefont {Jaksch}}]{New:Kiffner2019c}%
  \BibitemOpen
  \bibfield  {author} {\bibinfo {author} {\bibfnamefont {M.}~\bibnamefont
  {Kiffner}}, \bibinfo {author} {\bibfnamefont {J.}~\bibnamefont {Coulthard}},
  \bibinfo {author} {\bibfnamefont {F.}~\bibnamefont {Schlawin}}, \bibinfo
  {author} {\bibfnamefont {A.}~\bibnamefont {Ardavan}}, \ and\ \bibinfo
  {author} {\bibfnamefont {D.}~\bibnamefont {Jaksch}},\ }\href {\doibase
  10.1088/1367-2630/ab31c7} {\bibfield  {journal} {\bibinfo  {journal} {New
  Journal of Physics}\ }\textbf {\bibinfo {volume} {21}},\ \bibinfo {pages}
  {073066} (\bibinfo {year} {2019}{\natexlab{c}})}\BibitemShut {NoStop}%
\bibitem [{\citenamefont {Ashida}\ \emph {et~al.}(2020)\citenamefont {Ashida},
  \citenamefont {{\.{I}}mamo{\u{g}}lu}, \citenamefont {Faist}, \citenamefont
  {Jaksch}, \citenamefont {Cavalleri},\ and\ \citenamefont
  {Demler}}]{New:Ashida2020}%
  \BibitemOpen
  \bibfield  {author} {\bibinfo {author} {\bibfnamefont {Y.}~\bibnamefont
  {Ashida}}, \bibinfo {author} {\bibfnamefont {A.}~\bibnamefont
  {{\.{I}}mamo{\u{g}}lu}}, \bibinfo {author} {\bibfnamefont {J.}~\bibnamefont
  {Faist}}, \bibinfo {author} {\bibfnamefont {D.}~\bibnamefont {Jaksch}},
  \bibinfo {author} {\bibfnamefont {A.}~\bibnamefont {Cavalleri}}, \ and\
  \bibinfo {author} {\bibfnamefont {E.}~\bibnamefont {Demler}},\ }\href
  {\doibase 10.1103/PhysRevX.10.041027} {\bibfield  {journal} {\bibinfo
  {journal} {Phys. Rev. X}\ }\textbf {\bibinfo {volume} {10}},\ \bibinfo
  {pages} {041027} (\bibinfo {year} {2020})}\BibitemShut {NoStop}%
\bibitem [{\citenamefont {Sentef}\ \emph {et~al.}(2018)\citenamefont {Sentef},
  \citenamefont {Ruggenthaler},\ and\ \citenamefont {Rubio}}]{SC:Sentef2018}%
  \BibitemOpen
  \bibfield  {author} {\bibinfo {author} {\bibfnamefont {M.~A.}\ \bibnamefont
  {Sentef}}, \bibinfo {author} {\bibfnamefont {M.}~\bibnamefont
  {Ruggenthaler}}, \ and\ \bibinfo {author} {\bibfnamefont {A.}~\bibnamefont
  {Rubio}},\ }\href {\doibase 10.1126/sciadv.aau6969} {\bibfield  {journal}
  {\bibinfo  {journal} {Science Advances}\ }\textbf {\bibinfo {volume} {4}},\
  \bibinfo {pages} {eaau6969} (\bibinfo {year} {2018})}\BibitemShut {NoStop}%
\bibitem [{\citenamefont {Schlawin}\ \emph {et~al.}(2019)\citenamefont
  {Schlawin}, \citenamefont {Cavalleri},\ and\ \citenamefont
  {Jaksch}}]{New:Schlawin2019a}%
  \BibitemOpen
  \bibfield  {author} {\bibinfo {author} {\bibfnamefont {F.}~\bibnamefont
  {Schlawin}}, \bibinfo {author} {\bibfnamefont {A.}~\bibnamefont {Cavalleri}},
  \ and\ \bibinfo {author} {\bibfnamefont {D.}~\bibnamefont {Jaksch}},\ }\href
  {\doibase 10.1103/PhysRevLett.122.133602} {\bibfield  {journal} {\bibinfo
  {journal} {Phys. Rev. Lett.}\ }\textbf {\bibinfo {volume} {122}},\ \bibinfo
  {pages} {133602} (\bibinfo {year} {2019})}\BibitemShut {NoStop}%
\bibitem [{\citenamefont {Chakraborty}\ and\ \citenamefont
  {Piazza}(2021)}]{SC:Chakraborty2021}%
  \BibitemOpen
  \bibfield  {author} {\bibinfo {author} {\bibfnamefont {A.}~\bibnamefont
  {Chakraborty}}\ and\ \bibinfo {author} {\bibfnamefont {F.}~\bibnamefont
  {Piazza}},\ }\href {\doibase 10.1103/PhysRevLett.127.177002} {\bibfield
  {journal} {\bibinfo  {journal} {Phys. Rev. Lett.}\ }\textbf {\bibinfo
  {volume} {127}},\ \bibinfo {pages} {177002} (\bibinfo {year}
  {2021})}\BibitemShut {NoStop}%
\bibitem [{\citenamefont {Kollath}\ \emph {et~al.}(2016)\citenamefont
  {Kollath}, \citenamefont {Sheikhan}, \citenamefont {Wolff},\ and\
  \citenamefont {Brennecke}}]{Chiral:Kollath2016}%
  \BibitemOpen
  \bibfield  {author} {\bibinfo {author} {\bibfnamefont {C.}~\bibnamefont
  {Kollath}}, \bibinfo {author} {\bibfnamefont {A.}~\bibnamefont {Sheikhan}},
  \bibinfo {author} {\bibfnamefont {S.}~\bibnamefont {Wolff}}, \ and\ \bibinfo
  {author} {\bibfnamefont {F.}~\bibnamefont {Brennecke}},\ }\href {\doibase
  10.1103/PhysRevLett.116.060401} {\bibfield  {journal} {\bibinfo  {journal}
  {Phys. Rev. Lett.}\ }\textbf {\bibinfo {volume} {116}},\ \bibinfo {pages}
  {060401} (\bibinfo {year} {2016})}\BibitemShut {NoStop}%
\bibitem [{\citenamefont {Mivehvar}\ \emph {et~al.}(2017)\citenamefont
  {Mivehvar}, \citenamefont {Ritsch},\ and\ \citenamefont
  {Piazza}}]{Mivehvar2017}%
  \BibitemOpen
  \bibfield  {author} {\bibinfo {author} {\bibfnamefont {F.}~\bibnamefont
  {Mivehvar}}, \bibinfo {author} {\bibfnamefont {H.}~\bibnamefont {Ritsch}}, \
  and\ \bibinfo {author} {\bibfnamefont {F.}~\bibnamefont {Piazza}},\ }\href
  {\doibase 10.1103/PhysRevLett.118.073602} {\bibfield  {journal} {\bibinfo
  {journal} {Phys. Rev. Lett.}\ }\textbf {\bibinfo {volume} {118}},\ \bibinfo
  {pages} {073602} (\bibinfo {year} {2017})}\BibitemShut {NoStop}%
\bibitem [{\citenamefont {Wang}\ \emph {et~al.}(2019)\citenamefont {Wang},
  \citenamefont {Ronca},\ and\ \citenamefont {Sentef}}]{Topo:Wang2019}%
  \BibitemOpen
  \bibfield  {author} {\bibinfo {author} {\bibfnamefont {X.}~\bibnamefont
  {Wang}}, \bibinfo {author} {\bibfnamefont {E.}~\bibnamefont {Ronca}}, \ and\
  \bibinfo {author} {\bibfnamefont {M.~A.}\ \bibnamefont {Sentef}},\ }\href
  {\doibase 10.1103/PhysRevB.99.235156} {\bibfield  {journal} {\bibinfo
  {journal} {Phys. Rev. B}\ }\textbf {\bibinfo {volume} {99}},\ \bibinfo
  {pages} {235156} (\bibinfo {year} {2019})}\BibitemShut {NoStop}%
\bibitem [{\citenamefont {M\'endez-C\'ordoba}\ \emph
  {et~al.}(2020)\citenamefont {M\'endez-C\'ordoba}, \citenamefont
  {Mendoza-Arenas}, \citenamefont {G\'omez-Ruiz}, \citenamefont
  {Rodr\'{\i}guez}, \citenamefont {Tejedor},\ and\ \citenamefont
  {Quiroga}}]{ENT:Mendez2020TopologyCavity}%
  \BibitemOpen
  \bibfield  {author} {\bibinfo {author} {\bibfnamefont {F.~P.~M.}\
  \bibnamefont {M\'endez-C\'ordoba}}, \bibinfo {author} {\bibfnamefont {J.~J.}\
  \bibnamefont {Mendoza-Arenas}}, \bibinfo {author} {\bibfnamefont {F.~J.}\
  \bibnamefont {G\'omez-Ruiz}}, \bibinfo {author} {\bibfnamefont {F.~J.}\
  \bibnamefont {Rodr\'{\i}guez}}, \bibinfo {author} {\bibfnamefont
  {C.}~\bibnamefont {Tejedor}}, \ and\ \bibinfo {author} {\bibfnamefont
  {L.}~\bibnamefont {Quiroga}},\ }\href {\doibase
  10.1103/PhysRevResearch.2.043264} {\bibfield  {journal} {\bibinfo  {journal}
  {Phys. Rev. Research}\ }\textbf {\bibinfo {volume} {2}},\ \bibinfo {pages}
  {043264} (\bibinfo {year} {2020})}\BibitemShut {NoStop}%
\bibitem [{\citenamefont {Chanda}\ \emph {et~al.}(2021)\citenamefont {Chanda},
  \citenamefont {Kraus}, \citenamefont {Morigi},\ and\ \citenamefont
  {Zakrzewski}}]{Chanda2021}%
  \BibitemOpen
  \bibfield  {author} {\bibinfo {author} {\bibfnamefont {T.}~\bibnamefont
  {Chanda}}, \bibinfo {author} {\bibfnamefont {R.}~\bibnamefont {Kraus}},
  \bibinfo {author} {\bibfnamefont {G.}~\bibnamefont {Morigi}}, \ and\ \bibinfo
  {author} {\bibfnamefont {J.}~\bibnamefont {Zakrzewski}},\ }\href {\doibase
  10.22331/q-2021-07-13-501} {\bibfield  {journal} {\bibinfo  {journal}
  {{Quantum}}\ }\textbf {\bibinfo {volume} {5}},\ \bibinfo {pages} {501}
  (\bibinfo {year} {2021})}\BibitemShut {NoStop}%
\bibitem [{\citenamefont {Dmytruk}\ and\ \citenamefont
  {Schirò}(2022)}]{Topo:Dmytruk2022}%
  \BibitemOpen
  \bibfield  {author} {\bibinfo {author} {\bibfnamefont {O.}~\bibnamefont
  {Dmytruk}}\ and\ \bibinfo {author} {\bibfnamefont {M.}~\bibnamefont
  {Schirò}},\ }\href@noop {} {\enquote {\bibinfo {title} {Controlling
  topological phases of matter with quantum light},}\ } (\bibinfo {year}
  {2022}),\ \Eprint {http://arxiv.org/abs/arXiv:2204.05922} {arXiv:2204.05922}
  \BibitemShut {NoStop}%
\bibitem [{\citenamefont {Chanda}\ \emph {et~al.}(2022)\citenamefont {Chanda},
  \citenamefont {Kraus}, \citenamefont {Zakrzewski},\ and\ \citenamefont
  {Morigi}}]{Chanda2022}%
  \BibitemOpen
  \bibfield  {author} {\bibinfo {author} {\bibfnamefont {T.}~\bibnamefont
  {Chanda}}, \bibinfo {author} {\bibfnamefont {R.}~\bibnamefont {Kraus}},
  \bibinfo {author} {\bibfnamefont {J.}~\bibnamefont {Zakrzewski}}, \ and\
  \bibinfo {author} {\bibfnamefont {G.}~\bibnamefont {Morigi}},\ }\href@noop {}
  {\enquote {\bibinfo {title} {Bond order via cavity-mediated interactions},}\
  } (\bibinfo {year} {2022}),\ \Eprint {http://arxiv.org/abs/arXiv:2201.05466}
  {arXiv:2201.05466} \BibitemShut {NoStop}%
\bibitem [{\citenamefont {Appugliese}\ \emph {et~al.}(2022)\citenamefont
  {Appugliese}, \citenamefont {Enkner}, \citenamefont {Paravicini-Bagliani},
  \citenamefont {Beck}, \citenamefont {Reichl}, \citenamefont {Wegscheider},
  \citenamefont {Scalari}, \citenamefont {Ciuti},\ and\ \citenamefont
  {Faist}}]{Topo:Appugliese2022}%
  \BibitemOpen
  \bibfield  {author} {\bibinfo {author} {\bibfnamefont {F.}~\bibnamefont
  {Appugliese}}, \bibinfo {author} {\bibfnamefont {J.}~\bibnamefont {Enkner}},
  \bibinfo {author} {\bibfnamefont {G.~L.}\ \bibnamefont
  {Paravicini-Bagliani}}, \bibinfo {author} {\bibfnamefont {M.}~\bibnamefont
  {Beck}}, \bibinfo {author} {\bibfnamefont {C.}~\bibnamefont {Reichl}},
  \bibinfo {author} {\bibfnamefont {W.}~\bibnamefont {Wegscheider}}, \bibinfo
  {author} {\bibfnamefont {G.}~\bibnamefont {Scalari}}, \bibinfo {author}
  {\bibfnamefont {C.}~\bibnamefont {Ciuti}}, \ and\ \bibinfo {author}
  {\bibfnamefont {J.}~\bibnamefont {Faist}},\ }\href {\doibase
  10.1126/science.abl5818} {\bibfield  {journal} {\bibinfo  {journal}
  {Science}\ }\textbf {\bibinfo {volume} {375}},\ \bibinfo {pages} {1030}
  (\bibinfo {year} {2022})}\BibitemShut {NoStop}%
\bibitem [{\citenamefont {Rokaj}\ \emph
  {et~al.}(2022{\natexlab{b}})\citenamefont {Rokaj}, \citenamefont {Penz},
  \citenamefont {Sentef}, \citenamefont {Ruggenthaler},\ and\ \citenamefont
  {Rubio}}]{Topo:Rokaj2022}%
  \BibitemOpen
  \bibfield  {author} {\bibinfo {author} {\bibfnamefont {V.}~\bibnamefont
  {Rokaj}}, \bibinfo {author} {\bibfnamefont {M.}~\bibnamefont {Penz}},
  \bibinfo {author} {\bibfnamefont {M.~A.}\ \bibnamefont {Sentef}}, \bibinfo
  {author} {\bibfnamefont {M.}~\bibnamefont {Ruggenthaler}}, \ and\ \bibinfo
  {author} {\bibfnamefont {A.}~\bibnamefont {Rubio}},\ }\href {\doibase
  10.1103/PhysRevB.105.205424} {\bibfield  {journal} {\bibinfo  {journal}
  {Phys. Rev. B}\ }\textbf {\bibinfo {volume} {105}},\ \bibinfo {pages}
  {205424} (\bibinfo {year} {2022}{\natexlab{b}})}\BibitemShut {NoStop}%
\bibitem [{\citenamefont {Sierant}\ \emph {et~al.}(2019)\citenamefont
  {Sierant}, \citenamefont {Biedro{\'{n}}}, \citenamefont {Morigi},\ and\
  \citenamefont {Zakrzewski}}]{Sierant_Scipost_2019}%
  \BibitemOpen
  \bibfield  {author} {\bibinfo {author} {\bibfnamefont {P.}~\bibnamefont
  {Sierant}}, \bibinfo {author} {\bibfnamefont {K.}~\bibnamefont
  {Biedro{\'{n}}}}, \bibinfo {author} {\bibfnamefont {G.}~\bibnamefont
  {Morigi}}, \ and\ \bibinfo {author} {\bibfnamefont {J.}~\bibnamefont
  {Zakrzewski}},\ }\href {\doibase 10.21468/scipostphys.7.1.008} {\bibfield
  {journal} {\bibinfo  {journal} {{SciPost} Physics}\ }\textbf {\bibinfo
  {volume} {7}},\ \bibinfo {pages} {008} (\bibinfo {year} {2019})}\BibitemShut
  {NoStop}%
\bibitem [{\citenamefont {Kubala}\ \emph {et~al.}(2021)\citenamefont {Kubala},
  \citenamefont {Sierant}, \citenamefont {Morigi},\ and\ \citenamefont
  {Zakrzewski}}]{Kubala_PRB_2021}%
  \BibitemOpen
  \bibfield  {author} {\bibinfo {author} {\bibfnamefont {P.}~\bibnamefont
  {Kubala}}, \bibinfo {author} {\bibfnamefont {P.}~\bibnamefont {Sierant}},
  \bibinfo {author} {\bibfnamefont {G.}~\bibnamefont {Morigi}}, \ and\ \bibinfo
  {author} {\bibfnamefont {J.}~\bibnamefont {Zakrzewski}},\ }\href {\doibase
  10.1103/PhysRevB.103.174208} {\bibfield  {journal} {\bibinfo  {journal}
  {Phys. Rev. B}\ }\textbf {\bibinfo {volume} {103}},\ \bibinfo {pages}
  {174208} (\bibinfo {year} {2021})}\BibitemShut {NoStop}%
\bibitem [{\citenamefont {Chiocchetta}\ \emph {et~al.}(2021)\citenamefont
  {Chiocchetta}, \citenamefont {Kiese}, \citenamefont {Zelle}, \citenamefont
  {Piazza},\ and\ \citenamefont {Diehl}}]{New:Chiocchetta2021}%
  \BibitemOpen
  \bibfield  {author} {\bibinfo {author} {\bibfnamefont {A.}~\bibnamefont
  {Chiocchetta}}, \bibinfo {author} {\bibfnamefont {D.}~\bibnamefont {Kiese}},
  \bibinfo {author} {\bibfnamefont {C.~P.}\ \bibnamefont {Zelle}}, \bibinfo
  {author} {\bibfnamefont {F.}~\bibnamefont {Piazza}}, \ and\ \bibinfo {author}
  {\bibfnamefont {S.}~\bibnamefont {Diehl}},\ }\href {\doibase
  10.1038/s41467-021-26076-3} {\bibfield  {journal} {\bibinfo  {journal}
  {Nature Communications}\ }\textbf {\bibinfo {volume} {12}},\ \bibinfo {pages}
  {5901} (\bibinfo {year} {2021})}\BibitemShut {NoStop}%
\bibitem [{\citenamefont {Mazza}\ and\ \citenamefont
  {Georges}(2019)}]{Super:Mazza2019}%
  \BibitemOpen
  \bibfield  {author} {\bibinfo {author} {\bibfnamefont {G.}~\bibnamefont
  {Mazza}}\ and\ \bibinfo {author} {\bibfnamefont {A.}~\bibnamefont
  {Georges}},\ }\href {\doibase 10.1103/PhysRevLett.122.017401} {\bibfield
  {journal} {\bibinfo  {journal} {Phys. Rev. Lett.}\ }\textbf {\bibinfo
  {volume} {122}},\ \bibinfo {pages} {017401} (\bibinfo {year}
  {2019})}\BibitemShut {NoStop}%
\bibitem [{\citenamefont {Nataf}\ \emph {et~al.}(2019)\citenamefont {Nataf},
  \citenamefont {Champel}, \citenamefont {Blatter},\ and\ \citenamefont
  {Basko}}]{Super:Nataf2019}%
  \BibitemOpen
  \bibfield  {author} {\bibinfo {author} {\bibfnamefont {P.}~\bibnamefont
  {Nataf}}, \bibinfo {author} {\bibfnamefont {T.}~\bibnamefont {Champel}},
  \bibinfo {author} {\bibfnamefont {G.}~\bibnamefont {Blatter}}, \ and\
  \bibinfo {author} {\bibfnamefont {D.~M.}\ \bibnamefont {Basko}},\ }\href
  {\doibase 10.1103/PhysRevLett.123.207402} {\bibfield  {journal} {\bibinfo
  {journal} {Phys. Rev. Lett.}\ }\textbf {\bibinfo {volume} {123}},\ \bibinfo
  {pages} {207402} (\bibinfo {year} {2019})}\BibitemShut {NoStop}%
\bibitem [{\citenamefont {Guerci}\ \emph {et~al.}(2020)\citenamefont {Guerci},
  \citenamefont {Simon},\ and\ \citenamefont {Mora}}]{Super:Guerci2020}%
  \BibitemOpen
  \bibfield  {author} {\bibinfo {author} {\bibfnamefont {D.}~\bibnamefont
  {Guerci}}, \bibinfo {author} {\bibfnamefont {P.}~\bibnamefont {Simon}}, \
  and\ \bibinfo {author} {\bibfnamefont {C.}~\bibnamefont {Mora}},\ }\href
  {\doibase 10.1103/PhysRevLett.125.257604} {\bibfield  {journal} {\bibinfo
  {journal} {Phys. Rev. Lett.}\ }\textbf {\bibinfo {volume} {125}},\ \bibinfo
  {pages} {257604} (\bibinfo {year} {2020})}\BibitemShut {NoStop}%
\bibitem [{\citenamefont {Stokes}\ and\ \citenamefont
  {Nazir}(2020)}]{Super:Stokes2020}%
  \BibitemOpen
  \bibfield  {author} {\bibinfo {author} {\bibfnamefont {A.}~\bibnamefont
  {Stokes}}\ and\ \bibinfo {author} {\bibfnamefont {A.}~\bibnamefont {Nazir}},\
  }\href {\doibase 10.1103/PhysRevLett.125.143603} {\bibfield  {journal}
  {\bibinfo  {journal} {Phys. Rev. Lett.}\ }\textbf {\bibinfo {volume} {125}},\
  \bibinfo {pages} {143603} (\bibinfo {year} {2020})}\BibitemShut {NoStop}%
\bibitem [{\citenamefont {Ferri}\ \emph {et~al.}(2021)\citenamefont {Ferri},
  \citenamefont {Rosa-Medina}, \citenamefont {Finger}, \citenamefont {Dogra},
  \citenamefont {Soriente}, \citenamefont {Zilberberg}, \citenamefont
  {Donner},\ and\ \citenamefont {Esslinger}}]{Super:Ferri2021}%
  \BibitemOpen
  \bibfield  {author} {\bibinfo {author} {\bibfnamefont {F.}~\bibnamefont
  {Ferri}}, \bibinfo {author} {\bibfnamefont {R.}~\bibnamefont {Rosa-Medina}},
  \bibinfo {author} {\bibfnamefont {F.}~\bibnamefont {Finger}}, \bibinfo
  {author} {\bibfnamefont {N.}~\bibnamefont {Dogra}}, \bibinfo {author}
  {\bibfnamefont {M.}~\bibnamefont {Soriente}}, \bibinfo {author}
  {\bibfnamefont {O.}~\bibnamefont {Zilberberg}}, \bibinfo {author}
  {\bibfnamefont {T.}~\bibnamefont {Donner}}, \ and\ \bibinfo {author}
  {\bibfnamefont {T.}~\bibnamefont {Esslinger}},\ }\href {\doibase
  10.1103/PhysRevX.11.041046} {\bibfield  {journal} {\bibinfo  {journal} {Phys.
  Rev. X}\ }\textbf {\bibinfo {volume} {11}},\ \bibinfo {pages} {041046}
  (\bibinfo {year} {2021})}\BibitemShut {NoStop}%
\bibitem [{\citenamefont {Rohn}\ \emph {et~al.}(2020)\citenamefont {Rohn},
  \citenamefont {H\"ormann}, \citenamefont {Genes},\ and\ \citenamefont
  {Schmidt}}]{IsingSuperRad:Rohn2020}%
  \BibitemOpen
  \bibfield  {author} {\bibinfo {author} {\bibfnamefont {J.}~\bibnamefont
  {Rohn}}, \bibinfo {author} {\bibfnamefont {M.}~\bibnamefont {H\"ormann}},
  \bibinfo {author} {\bibfnamefont {C.}~\bibnamefont {Genes}}, \ and\ \bibinfo
  {author} {\bibfnamefont {K.~P.}\ \bibnamefont {Schmidt}},\ }\href {\doibase
  10.1103/PhysRevResearch.2.023131} {\bibfield  {journal} {\bibinfo  {journal}
  {Phys. Rev. Research}\ }\textbf {\bibinfo {volume} {2}},\ \bibinfo {pages}
  {023131} (\bibinfo {year} {2020})}\BibitemShut {NoStop}%
\bibitem [{\citenamefont {Li}\ \emph {et~al.}(2020)\citenamefont {Li},
  \citenamefont {Golez}, \citenamefont {Mazza}, \citenamefont {Millis},
  \citenamefont {Georges},\ and\ \citenamefont {Eckstein}}]{Gauge:Li2020b}%
  \BibitemOpen
  \bibfield  {author} {\bibinfo {author} {\bibfnamefont {J.}~\bibnamefont
  {Li}}, \bibinfo {author} {\bibfnamefont {D.}~\bibnamefont {Golez}}, \bibinfo
  {author} {\bibfnamefont {G.}~\bibnamefont {Mazza}}, \bibinfo {author}
  {\bibfnamefont {A.~J.}\ \bibnamefont {Millis}}, \bibinfo {author}
  {\bibfnamefont {A.}~\bibnamefont {Georges}}, \ and\ \bibinfo {author}
  {\bibfnamefont {M.}~\bibnamefont {Eckstein}},\ }\href {\doibase
  10.1103/PhysRevB.101.205140} {\bibfield  {journal} {\bibinfo  {journal}
  {Phys. Rev. B}\ }\textbf {\bibinfo {volume} {101}},\ \bibinfo {pages}
  {205140} (\bibinfo {year} {2020})}\BibitemShut {NoStop}%
\bibitem [{\citenamefont {Dmytruk}\ and\ \citenamefont
  {Schir\'o}(2021)}]{Gauge:Dmytruk2021b}%
  \BibitemOpen
  \bibfield  {author} {\bibinfo {author} {\bibfnamefont {O.}~\bibnamefont
  {Dmytruk}}\ and\ \bibinfo {author} {\bibfnamefont {M.}~\bibnamefont
  {Schir\'o}},\ }\href {\doibase 10.1103/PhysRevB.103.075131} {\bibfield
  {journal} {\bibinfo  {journal} {Phys. Rev. B}\ }\textbf {\bibinfo {volume}
  {103}},\ \bibinfo {pages} {075131} (\bibinfo {year} {2021})}\BibitemShut
  {NoStop}%
\bibitem [{\citenamefont {Andolina}\ \emph {et~al.}(2019)\citenamefont
  {Andolina}, \citenamefont {Pellegrino}, \citenamefont {Giovannetti},
  \citenamefont {MacDonald},\ and\ \citenamefont {Polini}}]{NoGo:Andolina2019}%
  \BibitemOpen
  \bibfield  {author} {\bibinfo {author} {\bibfnamefont {G.~M.}\ \bibnamefont
  {Andolina}}, \bibinfo {author} {\bibfnamefont {F.~M.~D.}\ \bibnamefont
  {Pellegrino}}, \bibinfo {author} {\bibfnamefont {V.}~\bibnamefont
  {Giovannetti}}, \bibinfo {author} {\bibfnamefont {A.~H.}\ \bibnamefont
  {MacDonald}}, \ and\ \bibinfo {author} {\bibfnamefont {M.}~\bibnamefont
  {Polini}},\ }\href {\doibase 10.1103/PhysRevB.100.121109} {\bibfield
  {journal} {\bibinfo  {journal} {Phys. Rev. B}\ }\textbf {\bibinfo {volume}
  {100}},\ \bibinfo {pages} {121109} (\bibinfo {year} {2019})}\BibitemShut
  {NoStop}%
\bibitem [{\citenamefont {Nataf}\ and\ \citenamefont
  {Ciuti}(2010)}]{NoGo:Nataf2010}%
  \BibitemOpen
  \bibfield  {author} {\bibinfo {author} {\bibfnamefont {P.}~\bibnamefont
  {Nataf}}\ and\ \bibinfo {author} {\bibfnamefont {C.}~\bibnamefont {Ciuti}},\
  }\href {\doibase 10.1038/ncomms1069} {\bibfield  {journal} {\bibinfo
  {journal} {Nature Communications}\ }\textbf {\bibinfo {volume} {1}},\
  \bibinfo {pages} {72} (\bibinfo {year} {2010})}\BibitemShut {NoStop}%
\bibitem [{\citenamefont {Andolina}\ \emph {et~al.}(2020)\citenamefont
  {Andolina}, \citenamefont {Pellegrino}, \citenamefont {Giovannetti},
  \citenamefont {MacDonald},\ and\ \citenamefont {Polini}}]{NoGo:Andolina2020}%
  \BibitemOpen
  \bibfield  {author} {\bibinfo {author} {\bibfnamefont {G.~M.}\ \bibnamefont
  {Andolina}}, \bibinfo {author} {\bibfnamefont {F.~M.~D.}\ \bibnamefont
  {Pellegrino}}, \bibinfo {author} {\bibfnamefont {V.}~\bibnamefont
  {Giovannetti}}, \bibinfo {author} {\bibfnamefont {A.~H.}\ \bibnamefont
  {MacDonald}}, \ and\ \bibinfo {author} {\bibfnamefont {M.}~\bibnamefont
  {Polini}},\ }\href {\doibase 10.1103/PhysRevB.102.125137} {\bibfield
  {journal} {\bibinfo  {journal} {Phys. Rev. B}\ }\textbf {\bibinfo {volume}
  {102}},\ \bibinfo {pages} {125137} (\bibinfo {year} {2020})}\BibitemShut
  {NoStop}%
\bibitem [{\citenamefont {Lambert}\ \emph {et~al.}(2004)\citenamefont
  {Lambert}, \citenamefont {Emary},\ and\ \citenamefont
  {Brandes}}]{ENT:Lambert2004EntCavity}%
  \BibitemOpen
  \bibfield  {author} {\bibinfo {author} {\bibfnamefont {N.}~\bibnamefont
  {Lambert}}, \bibinfo {author} {\bibfnamefont {C.}~\bibnamefont {Emary}}, \
  and\ \bibinfo {author} {\bibfnamefont {T.}~\bibnamefont {Brandes}},\ }\href
  {\doibase 10.1103/PhysRevLett.92.073602} {\bibfield  {journal} {\bibinfo
  {journal} {Phys. Rev. Lett.}\ }\textbf {\bibinfo {volume} {92}},\ \bibinfo
  {pages} {073602} (\bibinfo {year} {2004})}\BibitemShut {NoStop}%
\bibitem [{\citenamefont {Sharma}\ \emph {et~al.}(2022)\citenamefont {Sharma},
  \citenamefont {Jaeger}, \citenamefont {Kraus}, \citenamefont {Roscilde},\
  and\ \citenamefont {Morigi}}]{ENT:Sharma2022EntanglementCavity}%
  \BibitemOpen
  \bibfield  {author} {\bibinfo {author} {\bibfnamefont {S.}~\bibnamefont
  {Sharma}}, \bibinfo {author} {\bibfnamefont {S.~B.}\ \bibnamefont {Jaeger}},
  \bibinfo {author} {\bibfnamefont {R.}~\bibnamefont {Kraus}}, \bibinfo
  {author} {\bibfnamefont {T.}~\bibnamefont {Roscilde}}, \ and\ \bibinfo
  {author} {\bibfnamefont {G.}~\bibnamefont {Morigi}},\ }\href@noop {}
  {\enquote {\bibinfo {title} {Quantum critical behavior of entanglement in
  lattice bosons with cavity-mediated long-range interactions},}\ } (\bibinfo
  {year} {2022}),\ \Eprint {http://arxiv.org/abs/arXiv:2204.07712}
  {arXiv:2204.07712} \BibitemShut {NoStop}%
\bibitem [{\citenamefont {Amico}\ \emph {et~al.}(2008)\citenamefont {Amico},
  \citenamefont {Fazio}, \citenamefont {Osterloh},\ and\ \citenamefont
  {Vedral}}]{Amico2008}%
  \BibitemOpen
  \bibfield  {author} {\bibinfo {author} {\bibfnamefont {L.}~\bibnamefont
  {Amico}}, \bibinfo {author} {\bibfnamefont {R.}~\bibnamefont {Fazio}},
  \bibinfo {author} {\bibfnamefont {A.}~\bibnamefont {Osterloh}}, \ and\
  \bibinfo {author} {\bibfnamefont {V.}~\bibnamefont {Vedral}},\ }\href
  {\doibase 10.1103/RevModPhys.80.517} {\bibfield  {journal} {\bibinfo
  {journal} {Rev. Mod. Phys.}\ }\textbf {\bibinfo {volume} {80}},\ \bibinfo
  {pages} {517} (\bibinfo {year} {2008})}\BibitemShut {NoStop}%
\bibitem [{\citenamefont {Osterloh}\ \emph {et~al.}(2002)\citenamefont
  {Osterloh}, \citenamefont {Amico}, \citenamefont {Falci},\ and\ \citenamefont
  {Fazio}}]{ENT:Osterloh2002}%
  \BibitemOpen
  \bibfield  {author} {\bibinfo {author} {\bibfnamefont {A.}~\bibnamefont
  {Osterloh}}, \bibinfo {author} {\bibfnamefont {L.}~\bibnamefont {Amico}},
  \bibinfo {author} {\bibfnamefont {G.}~\bibnamefont {Falci}}, \ and\ \bibinfo
  {author} {\bibfnamefont {R.}~\bibnamefont {Fazio}},\ }\href {\doibase
  10.1038/416608a} {\bibfield  {journal} {\bibinfo  {journal} {Nature}\
  }\textbf {\bibinfo {volume} {416}},\ \bibinfo {pages} {608} (\bibinfo {year}
  {2002})}\BibitemShut {NoStop}%
\bibitem [{\citenamefont {Osborne}\ and\ \citenamefont
  {Nielsen}(2002)}]{ENT:Osborne2002}%
  \BibitemOpen
  \bibfield  {author} {\bibinfo {author} {\bibfnamefont {T.~J.}\ \bibnamefont
  {Osborne}}\ and\ \bibinfo {author} {\bibfnamefont {M.~A.}\ \bibnamefont
  {Nielsen}},\ }\href {\doibase 10.1103/PhysRevA.66.032110} {\bibfield
  {journal} {\bibinfo  {journal} {Phys. Rev. A}\ }\textbf {\bibinfo {volume}
  {66}},\ \bibinfo {pages} {032110} (\bibinfo {year} {2002})}\BibitemShut
  {NoStop}%
\bibitem [{\citenamefont {Vidal}\ \emph {et~al.}(2003)\citenamefont {Vidal},
  \citenamefont {Latorre}, \citenamefont {Rico},\ and\ \citenamefont
  {Kitaev}}]{ENT:Vidal2003}%
  \BibitemOpen
  \bibfield  {author} {\bibinfo {author} {\bibfnamefont {G.}~\bibnamefont
  {Vidal}}, \bibinfo {author} {\bibfnamefont {J.~I.}\ \bibnamefont {Latorre}},
  \bibinfo {author} {\bibfnamefont {E.}~\bibnamefont {Rico}}, \ and\ \bibinfo
  {author} {\bibfnamefont {A.}~\bibnamefont {Kitaev}},\ }\href {\doibase
  10.1103/PhysRevLett.90.227902} {\bibfield  {journal} {\bibinfo  {journal}
  {Phys. Rev. Lett.}\ }\textbf {\bibinfo {volume} {90}},\ \bibinfo {pages}
  {227902} (\bibinfo {year} {2003})}\BibitemShut {NoStop}%
\bibitem [{Note1()}]{Note1}%
  \BibitemOpen
  \bibinfo {note} {For convenience, we use ``photon'' to indicate the light
  part even when we do not specify its nature.}\BibitemShut {Stop}%
\bibitem [{\citenamefont {Yao}\ and\ \citenamefont {Qi}(2010)}]{Yao2010}%
  \BibitemOpen
  \bibfield  {author} {\bibinfo {author} {\bibfnamefont {H.}~\bibnamefont
  {Yao}}\ and\ \bibinfo {author} {\bibfnamefont {X.-L.}\ \bibnamefont {Qi}},\
  }\href {\doibase 10.1103/PhysRevLett.105.080501} {\bibfield  {journal}
  {\bibinfo  {journal} {Phys. Rev. Lett.}\ }\textbf {\bibinfo {volume} {105}},\
  \bibinfo {pages} {080501} (\bibinfo {year} {2010})}\BibitemShut {NoStop}%
\bibitem [{\citenamefont {Schliemann}(2011)}]{Schliemann2011}%
  \BibitemOpen
  \bibfield  {author} {\bibinfo {author} {\bibfnamefont {J.}~\bibnamefont
  {Schliemann}},\ }\href {\doibase 10.1103/PhysRevB.83.115322} {\bibfield
  {journal} {\bibinfo  {journal} {Phys. Rev. B}\ }\textbf {\bibinfo {volume}
  {83}},\ \bibinfo {pages} {115322} (\bibinfo {year} {2011})}\BibitemShut
  {NoStop}%
\bibitem [{\citenamefont {de~Boer}\ \emph {et~al.}(2019)\citenamefont
  {de~Boer}, \citenamefont {J\"arvel\"a},\ and\ \citenamefont
  {Keski-Vakkuri}}]{Boer2019}%
  \BibitemOpen
  \bibfield  {author} {\bibinfo {author} {\bibfnamefont {J.}~\bibnamefont
  {de~Boer}}, \bibinfo {author} {\bibfnamefont {J.}~\bibnamefont
  {J\"arvel\"a}}, \ and\ \bibinfo {author} {\bibfnamefont {E.}~\bibnamefont
  {Keski-Vakkuri}},\ }\href {\doibase 10.1103/PhysRevD.99.066012} {\bibfield
  {journal} {\bibinfo  {journal} {Phys. Rev. D}\ }\textbf {\bibinfo {volume}
  {99}},\ \bibinfo {pages} {066012} (\bibinfo {year} {2019})}\BibitemShut
  {NoStop}%
\bibitem [{\citenamefont {Nandy}(2021)}]{ENT:Nandy2021EntanglementCapacity}%
  \BibitemOpen
  \bibfield  {author} {\bibinfo {author} {\bibfnamefont {P.}~\bibnamefont
  {Nandy}},\ }\href {\doibase 10.1007/JHEP07(2021)019} {\bibfield  {journal}
  {\bibinfo  {journal} {Journal of High Energy Physics}\ }\textbf {\bibinfo
  {volume} {2021}},\ \bibinfo {pages} {19} (\bibinfo {year}
  {2021})}\BibitemShut {NoStop}%
\bibitem [{Note2()}]{Note2}%
  \BibitemOpen
  \bibinfo {note} {The exact value of $\mathinner {\langle {\protect \hat
  \Delta }\rangle }$ in the ordered phase is not important since it can be
  rescaled. Here we assume it to be unity.}\BibitemShut {Stop}%
\bibitem [{Note3()}]{Note3}%
  \BibitemOpen
  \bibinfo {note} {The value of $\Delta $ can then be determined by minimizing
  the model-dependent free energy $\mathinner {\langle {\protect \hat H}\rangle
  }=F(\Delta )-g^2\Delta /\omega $.}\BibitemShut {Stop}%
\bibitem [{Note4()}]{Note4}%
  \BibitemOpen
  \bibinfo {note} {Our analysis can be applied to any kind of coupling,
  including the usual Peierls substitution (which has the advantage of making
  gauge invariance manifest).}\BibitemShut {Stop}%
\bibitem [{\citenamefont {Habibian}\ \emph {et~al.}(2013)\citenamefont
  {Habibian}, \citenamefont {Winter}, \citenamefont {Paganelli}, \citenamefont
  {Rieger},\ and\ \citenamefont {Morigi}}]{Habibian13}%
  \BibitemOpen
  \bibfield  {author} {\bibinfo {author} {\bibfnamefont {H.}~\bibnamefont
  {Habibian}}, \bibinfo {author} {\bibfnamefont {A.}~\bibnamefont {Winter}},
  \bibinfo {author} {\bibfnamefont {S.}~\bibnamefont {Paganelli}}, \bibinfo
  {author} {\bibfnamefont {H.}~\bibnamefont {Rieger}}, \ and\ \bibinfo {author}
  {\bibfnamefont {G.}~\bibnamefont {Morigi}},\ }\href {\doibase
  10.1103/PhysRevLett.110.075304} {\bibfield  {journal} {\bibinfo  {journal}
  {Phys. Rev. Lett.}\ }\textbf {\bibinfo {volume} {110}},\ \bibinfo {pages}
  {075304} (\bibinfo {year} {2013})}\BibitemShut {NoStop}%
\bibitem [{\citenamefont {Caballero-Benitez}\ and\ \citenamefont
  {Mekhov}(2015)}]{Caballero15}%
  \BibitemOpen
  \bibfield  {author} {\bibinfo {author} {\bibfnamefont {S.~F.}\ \bibnamefont
  {Caballero-Benitez}}\ and\ \bibinfo {author} {\bibfnamefont {I.~B.}\
  \bibnamefont {Mekhov}},\ }\href {\doibase 10.1103/PhysRevLett.115.243604}
  {\bibfield  {journal} {\bibinfo  {journal} {Phys. Rev. Lett.}\ }\textbf
  {\bibinfo {volume} {115}},\ \bibinfo {pages} {243604} (\bibinfo {year}
  {2015})}\BibitemShut {NoStop}%
\bibitem [{\citenamefont {Caballero-Benitez}\ and\ \citenamefont
  {Mekhov}(2016)}]{Caballero16}%
  \BibitemOpen
  \bibfield  {author} {\bibinfo {author} {\bibfnamefont {S.~F.}\ \bibnamefont
  {Caballero-Benitez}}\ and\ \bibinfo {author} {\bibfnamefont {I.~B.}\
  \bibnamefont {Mekhov}},\ }\href {\doibase 10.1088/1367-2630/18/11/113010}
  {\bibfield  {journal} {\bibinfo  {journal} {New Journal of Physics}\ }\textbf
  {\bibinfo {volume} {18}},\ \bibinfo {pages} {113010} (\bibinfo {year}
  {2016})}\BibitemShut {NoStop}%
\bibitem [{\citenamefont {Elliott}\ and\ \citenamefont
  {Mekhov}(2016)}]{Elliott16}%
  \BibitemOpen
  \bibfield  {author} {\bibinfo {author} {\bibfnamefont {T.~J.}\ \bibnamefont
  {Elliott}}\ and\ \bibinfo {author} {\bibfnamefont {I.~B.}\ \bibnamefont
  {Mekhov}},\ }\href {\doibase 10.1103/PhysRevA.94.013614} {\bibfield
  {journal} {\bibinfo  {journal} {Phys. Rev. A}\ }\textbf {\bibinfo {volume}
  {94}},\ \bibinfo {pages} {013614} (\bibinfo {year} {2016})}\BibitemShut
  {NoStop}%
\bibitem [{\citenamefont {Gr\"uner}(1988)}]{CDW:Grunwe1988}%
  \BibitemOpen
  \bibfield  {author} {\bibinfo {author} {\bibfnamefont {G.}~\bibnamefont
  {Gr\"uner}},\ }\href {\doibase 10.1103/RevModPhys.60.1129} {\bibfield
  {journal} {\bibinfo  {journal} {Rev. Mod. Phys.}\ }\textbf {\bibinfo {volume}
  {60}},\ \bibinfo {pages} {1129} (\bibinfo {year} {1988})}\BibitemShut
  {NoStop}%
\bibitem [{\citenamefont {Gor'kov}\ and\ \citenamefont
  {Gr{\"u}ner}(2012)}]{CDW:gorkov2012}%
  \BibitemOpen
  \bibfield  {author} {\bibinfo {author} {\bibfnamefont {L.~P.}\ \bibnamefont
  {Gor'kov}}\ and\ \bibinfo {author} {\bibfnamefont {G.}~\bibnamefont
  {Gr{\"u}ner}},\ }\href@noop {} {\emph {\bibinfo {title} {{Charge Density
  Waves in Solids}}}}\ (\bibinfo  {publisher} {Elsevier},\ \bibinfo {year}
  {2012})\BibitemShut {NoStop}%
\bibitem [{\citenamefont {Monceau}(1985)}]{CDW:monceau1985electronic}%
  \BibitemOpen
  \bibfield  {author} {\bibinfo {author} {\bibfnamefont {P.}~\bibnamefont
  {Monceau}},\ }\href {\doibase 10.1007/978-94-015-6923-1} {\emph {\bibinfo
  {title} {Electronic Properties of Inorganic Quasi-One-Dimensional
  Compounds}}}\ (\bibinfo  {publisher} {Springer Netherlands},\ \bibinfo {year}
  {1985})\BibitemShut {NoStop}%
\bibitem [{\citenamefont {Zhu}\ \emph {et~al.}(2015)\citenamefont {Zhu},
  \citenamefont {Cao}, \citenamefont {Zhang}, \citenamefont {Plummer},\ and\
  \citenamefont {Guo}}]{CDW:Zhu2015}%
  \BibitemOpen
  \bibfield  {author} {\bibinfo {author} {\bibfnamefont {X.}~\bibnamefont
  {Zhu}}, \bibinfo {author} {\bibfnamefont {Y.}~\bibnamefont {Cao}}, \bibinfo
  {author} {\bibfnamefont {J.}~\bibnamefont {Zhang}}, \bibinfo {author}
  {\bibfnamefont {E.~W.}\ \bibnamefont {Plummer}}, \ and\ \bibinfo {author}
  {\bibfnamefont {J.}~\bibnamefont {Guo}},\ }\href {\doibase
  10.1073/pnas.1424791112} {\bibfield  {journal} {\bibinfo  {journal}
  {Proceedings of the National Academy of Sciences}\ }\textbf {\bibinfo
  {volume} {112}},\ \bibinfo {pages} {2367} (\bibinfo {year}
  {2015})}\BibitemShut {NoStop}%
\bibitem [{\citenamefont {Chiriac\`o}\ and\ \citenamefont
  {Millis}(2018)}]{CDW:Chiriaco2018}%
  \BibitemOpen
  \bibfield  {author} {\bibinfo {author} {\bibfnamefont {G.}~\bibnamefont
  {Chiriac\`o}}\ and\ \bibinfo {author} {\bibfnamefont {A.~J.}\ \bibnamefont
  {Millis}},\ }\href {\doibase 10.1103/PhysRevB.98.205152} {\bibfield
  {journal} {\bibinfo  {journal} {Phys. Rev. B}\ }\textbf {\bibinfo {volume}
  {98}},\ \bibinfo {pages} {205152} (\bibinfo {year} {2018})}\BibitemShut
  {NoStop}%
\bibitem [{\citenamefont {Chen}\ \emph {et~al.}(2014)\citenamefont {Chen},
  \citenamefont {Yu},\ and\ \citenamefont {Zhai}}]{SuperCDW:Chen2014}%
  \BibitemOpen
  \bibfield  {author} {\bibinfo {author} {\bibfnamefont {Y.}~\bibnamefont
  {Chen}}, \bibinfo {author} {\bibfnamefont {Z.}~\bibnamefont {Yu}}, \ and\
  \bibinfo {author} {\bibfnamefont {H.}~\bibnamefont {Zhai}},\ }\href {\doibase
  10.1103/PhysRevLett.112.143004} {\bibfield  {journal} {\bibinfo  {journal}
  {Phys. Rev. Lett.}\ }\textbf {\bibinfo {volume} {112}},\ \bibinfo {pages}
  {143004} (\bibinfo {year} {2014})}\BibitemShut {NoStop}%
\bibitem [{\citenamefont {Keeling}\ \emph {et~al.}(2014)\citenamefont
  {Keeling}, \citenamefont {Bhaseen},\ and\ \citenamefont
  {Simons}}]{SuperCDW:Keeling2014}%
  \BibitemOpen
  \bibfield  {author} {\bibinfo {author} {\bibfnamefont {J.}~\bibnamefont
  {Keeling}}, \bibinfo {author} {\bibfnamefont {M.~J.}\ \bibnamefont
  {Bhaseen}}, \ and\ \bibinfo {author} {\bibfnamefont {B.~D.}\ \bibnamefont
  {Simons}},\ }\href {\doibase 10.1103/PhysRevLett.112.143002} {\bibfield
  {journal} {\bibinfo  {journal} {Phys. Rev. Lett.}\ }\textbf {\bibinfo
  {volume} {112}},\ \bibinfo {pages} {143002} (\bibinfo {year}
  {2014})}\BibitemShut {NoStop}%
\bibitem [{\citenamefont {Piazza}\ and\ \citenamefont
  {Strack}(2014)}]{SuperCDW:Piazza2014}%
  \BibitemOpen
  \bibfield  {author} {\bibinfo {author} {\bibfnamefont {F.}~\bibnamefont
  {Piazza}}\ and\ \bibinfo {author} {\bibfnamefont {P.}~\bibnamefont
  {Strack}},\ }\href {\doibase 10.1103/PhysRevLett.112.143003} {\bibfield
  {journal} {\bibinfo  {journal} {Phys. Rev. Lett.}\ }\textbf {\bibinfo
  {volume} {112}},\ \bibinfo {pages} {143003} (\bibinfo {year}
  {2014})}\BibitemShut {NoStop}%
\bibitem [{\citenamefont {Rylands}\ \emph {et~al.}(2020)\citenamefont
  {Rylands}, \citenamefont {Guo}, \citenamefont {Lev}, \citenamefont
  {Keeling},\ and\ \citenamefont {Galitski}}]{SuperCDW:Rylands2020}%
  \BibitemOpen
  \bibfield  {author} {\bibinfo {author} {\bibfnamefont {C.}~\bibnamefont
  {Rylands}}, \bibinfo {author} {\bibfnamefont {Y.}~\bibnamefont {Guo}},
  \bibinfo {author} {\bibfnamefont {B.~L.}\ \bibnamefont {Lev}}, \bibinfo
  {author} {\bibfnamefont {J.}~\bibnamefont {Keeling}}, \ and\ \bibinfo
  {author} {\bibfnamefont {V.}~\bibnamefont {Galitski}},\ }\href {\doibase
  10.1103/PhysRevLett.125.010404} {\bibfield  {journal} {\bibinfo  {journal}
  {Phys. Rev. Lett.}\ }\textbf {\bibinfo {volume} {125}},\ \bibinfo {pages}
  {010404} (\bibinfo {year} {2020})}\BibitemShut {NoStop}%
\bibitem [{\citenamefont {Viehmann}\ \emph {et~al.}(2013)\citenamefont
  {Viehmann}, \citenamefont {von Delft},\ and\ \citenamefont
  {Marquardt}}]{viehmann2013observing}%
  \BibitemOpen
  \bibfield  {author} {\bibinfo {author} {\bibfnamefont {O.}~\bibnamefont
  {Viehmann}}, \bibinfo {author} {\bibfnamefont {J.}~\bibnamefont {von Delft}},
  \ and\ \bibinfo {author} {\bibfnamefont {F.}~\bibnamefont {Marquardt}},\
  }\href {\doibase 10.1103/PhysRevLett.110.030601} {\bibfield  {journal}
  {\bibinfo  {journal} {Phys. Rev. Lett.}\ }\textbf {\bibinfo {volume} {110}},\
  \bibinfo {pages} {030601} (\bibinfo {year} {2013})}\BibitemShut {NoStop}%
\bibitem [{\citenamefont {Dalmonte}\ \emph {et~al.}(2015)\citenamefont
  {Dalmonte}, \citenamefont {Mirzaei}, \citenamefont {Muppalla}, \citenamefont
  {Marcos}, \citenamefont {Zoller},\ and\ \citenamefont
  {Kirchmair}}]{dalmonte2015realizing}%
  \BibitemOpen
  \bibfield  {author} {\bibinfo {author} {\bibfnamefont {M.}~\bibnamefont
  {Dalmonte}}, \bibinfo {author} {\bibfnamefont {S.~I.}\ \bibnamefont
  {Mirzaei}}, \bibinfo {author} {\bibfnamefont {P.~R.}\ \bibnamefont
  {Muppalla}}, \bibinfo {author} {\bibfnamefont {D.}~\bibnamefont {Marcos}},
  \bibinfo {author} {\bibfnamefont {P.}~\bibnamefont {Zoller}}, \ and\ \bibinfo
  {author} {\bibfnamefont {G.}~\bibnamefont {Kirchmair}},\ }\href {\doibase
  10.1103/PhysRevB.92.174507} {\bibfield  {journal} {\bibinfo  {journal} {Phys.
  Rev. B}\ }\textbf {\bibinfo {volume} {92}},\ \bibinfo {pages} {174507}
  (\bibinfo {year} {2015})}\BibitemShut {NoStop}%
\bibitem [{\citenamefont {Sandvik}(2010)}]{Sandvik2010}%
  \BibitemOpen
  \bibfield  {author} {\bibinfo {author} {\bibfnamefont {A.~W.}\ \bibnamefont
  {Sandvik}},\ }\href {\doibase 10.1063/1.3518900} {\bibfield  {journal}
  {\bibinfo  {journal} {AIP Conference Proceedings}\ }\textbf {\bibinfo
  {volume} {1297}},\ \bibinfo {pages} {135} (\bibinfo {year}
  {2010})}\BibitemShut {NoStop}%
\bibitem [{\citenamefont {White}(1992)}]{white_prl_1992}%
  \BibitemOpen
  \bibfield  {author} {\bibinfo {author} {\bibfnamefont {S.~R.}\ \bibnamefont
  {White}},\ }\href {\doibase 10.1103/PhysRevLett.69.2863} {\bibfield
  {journal} {\bibinfo  {journal} {Phys. Rev. Lett.}\ }\textbf {\bibinfo
  {volume} {69}},\ \bibinfo {pages} {2863} (\bibinfo {year}
  {1992})}\BibitemShut {NoStop}%
\bibitem [{\citenamefont {White}(1993)}]{white_prb_1993}%
  \BibitemOpen
  \bibfield  {author} {\bibinfo {author} {\bibfnamefont {S.~R.}\ \bibnamefont
  {White}},\ }\href {\doibase 10.1103/PhysRevB.48.10345} {\bibfield  {journal}
  {\bibinfo  {journal} {Phys. Rev. B}\ }\textbf {\bibinfo {volume} {48}},\
  \bibinfo {pages} {10345} (\bibinfo {year} {1993})}\BibitemShut {NoStop}%
\bibitem [{\citenamefont {Schollw\"{o}ck}(2011)}]{schollwock_aop_2011}%
  \BibitemOpen
  \bibfield  {author} {\bibinfo {author} {\bibfnamefont {U.}~\bibnamefont
  {Schollw\"{o}ck}},\ }\href {\doibase 10.1016/j.aop.2010.09.012} {\bibfield
  {journal} {\bibinfo  {journal} {Annals of Physics}\ }\textbf {\bibinfo
  {volume} {326}},\ \bibinfo {pages} {96} (\bibinfo {year} {2011})}\BibitemShut
  {NoStop}%
\bibitem [{\citenamefont {Or{\'{u}}s}(2014)}]{Orus_aop_2014}%
  \BibitemOpen
  \bibfield  {author} {\bibinfo {author} {\bibfnamefont {R.}~\bibnamefont
  {Or{\'{u}}s}},\ }\href {\doibase 10.1016/j.aop.2014.06.013} {\bibfield
  {journal} {\bibinfo  {journal} {Annals of Physics}\ }\textbf {\bibinfo
  {volume} {349}},\ \bibinfo {pages} {117} (\bibinfo {year}
  {2014})}\BibitemShut {NoStop}%
\bibitem [{\citenamefont {Singh}\ \emph {et~al.}(2010)\citenamefont {Singh},
  \citenamefont {Pfeifer},\ and\ \citenamefont {Vidal}}]{Singh_PRA_2010}%
  \BibitemOpen
  \bibfield  {author} {\bibinfo {author} {\bibfnamefont {S.}~\bibnamefont
  {Singh}}, \bibinfo {author} {\bibfnamefont {R.~N.~C.}\ \bibnamefont
  {Pfeifer}}, \ and\ \bibinfo {author} {\bibfnamefont {G.}~\bibnamefont
  {Vidal}},\ }\href {\doibase 10.1103/PhysRevA.82.050301} {\bibfield  {journal}
  {\bibinfo  {journal} {Phys. Rev. A}\ }\textbf {\bibinfo {volume} {82}},\
  \bibinfo {pages} {050301} (\bibinfo {year} {2010})}\BibitemShut {NoStop}%
\bibitem [{\citenamefont {Singh}\ \emph {et~al.}(2011)\citenamefont {Singh},
  \citenamefont {Pfeifer},\ and\ \citenamefont {Vidal}}]{Singh_PRB_2011}%
  \BibitemOpen
  \bibfield  {author} {\bibinfo {author} {\bibfnamefont {S.}~\bibnamefont
  {Singh}}, \bibinfo {author} {\bibfnamefont {R.~N.~C.}\ \bibnamefont
  {Pfeifer}}, \ and\ \bibinfo {author} {\bibfnamefont {G.}~\bibnamefont
  {Vidal}},\ }\href {\doibase 10.1103/PhysRevB.83.115125} {\bibfield  {journal}
  {\bibinfo  {journal} {Phys. Rev. B}\ }\textbf {\bibinfo {volume} {83}},\
  \bibinfo {pages} {115125} (\bibinfo {year} {2011})}\BibitemShut {NoStop}%
\bibitem [{\citenamefont {Halati}\ \emph
  {et~al.}(2020{\natexlab{a}})\citenamefont {Halati}, \citenamefont
  {Sheikhan},\ and\ \citenamefont {Kollath}}]{Halati20a}%
  \BibitemOpen
  \bibfield  {author} {\bibinfo {author} {\bibfnamefont {C.-M.}\ \bibnamefont
  {Halati}}, \bibinfo {author} {\bibfnamefont {A.}~\bibnamefont {Sheikhan}}, \
  and\ \bibinfo {author} {\bibfnamefont {C.}~\bibnamefont {Kollath}},\ }\href
  {\doibase 10.1103/PhysRevResearch.2.043255} {\bibfield  {journal} {\bibinfo
  {journal} {Phys. Rev. Research}\ }\textbf {\bibinfo {volume} {2}},\ \bibinfo
  {pages} {043255} (\bibinfo {year} {2020}{\natexlab{a}})}\BibitemShut
  {NoStop}%
\bibitem [{\citenamefont {Halati}\ \emph
  {et~al.}(2020{\natexlab{b}})\citenamefont {Halati}, \citenamefont {Sheikhan},
  \citenamefont {Ritsch},\ and\ \citenamefont {Kollath}}]{Halati20b}%
  \BibitemOpen
  \bibfield  {author} {\bibinfo {author} {\bibfnamefont {C.-M.}\ \bibnamefont
  {Halati}}, \bibinfo {author} {\bibfnamefont {A.}~\bibnamefont {Sheikhan}},
  \bibinfo {author} {\bibfnamefont {H.}~\bibnamefont {Ritsch}}, \ and\ \bibinfo
  {author} {\bibfnamefont {C.}~\bibnamefont {Kollath}},\ }\href {\doibase
  10.1103/PhysRevLett.125.093604} {\bibfield  {journal} {\bibinfo  {journal}
  {Phys. Rev. Lett.}\ }\textbf {\bibinfo {volume} {125}},\ \bibinfo {pages}
  {093604} (\bibinfo {year} {2020}{\natexlab{b}})}\BibitemShut {NoStop}%
\bibitem [{\citenamefont {Halati}\ \emph {et~al.}(2022)\citenamefont {Halati},
  \citenamefont {Sheikhan},\ and\ \citenamefont {Kollath}}]{Halati21}%
  \BibitemOpen
  \bibfield  {author} {\bibinfo {author} {\bibfnamefont {C.-M.}\ \bibnamefont
  {Halati}}, \bibinfo {author} {\bibfnamefont {A.}~\bibnamefont {Sheikhan}}, \
  and\ \bibinfo {author} {\bibfnamefont {C.}~\bibnamefont {Kollath}},\ }\href
  {\doibase 10.1103/PhysRevResearch.4.L012015} {\bibfield  {journal} {\bibinfo
  {journal} {Phys. Rev. Research}\ }\textbf {\bibinfo {volume} {4}},\ \bibinfo
  {pages} {L012015} (\bibinfo {year} {2022})}\BibitemShut {NoStop}%
\bibitem [{\citenamefont {Baumann}\ \emph {et~al.}(2011)\citenamefont
  {Baumann}, \citenamefont {Mottl}, \citenamefont {Brennecke},\ and\
  \citenamefont {Esslinger}}]{Dicke:Baumann2011}%
  \BibitemOpen
  \bibfield  {author} {\bibinfo {author} {\bibfnamefont {K.}~\bibnamefont
  {Baumann}}, \bibinfo {author} {\bibfnamefont {R.}~\bibnamefont {Mottl}},
  \bibinfo {author} {\bibfnamefont {F.}~\bibnamefont {Brennecke}}, \ and\
  \bibinfo {author} {\bibfnamefont {T.}~\bibnamefont {Esslinger}},\ }\href
  {\doibase 10.1103/physrevlett.107.140402} {\bibfield  {journal} {\bibinfo
  {journal} {Phys. Rev. Lett.}\ }\textbf {\bibinfo {volume} {107}},\ \bibinfo
  {pages} {140402} (\bibinfo {year} {2011})}\BibitemShut {NoStop}%
\bibitem [{\citenamefont {Hruby}\ \emph {et~al.}(2018)\citenamefont {Hruby},
  \citenamefont {Dogra}, \citenamefont {Landini}, \citenamefont {Donner},\ and\
  \citenamefont {Esslinger}}]{Hruby2018}%
  \BibitemOpen
  \bibfield  {author} {\bibinfo {author} {\bibfnamefont {L.}~\bibnamefont
  {Hruby}}, \bibinfo {author} {\bibfnamefont {N.}~\bibnamefont {Dogra}},
  \bibinfo {author} {\bibfnamefont {M.}~\bibnamefont {Landini}}, \bibinfo
  {author} {\bibfnamefont {T.}~\bibnamefont {Donner}}, \ and\ \bibinfo {author}
  {\bibfnamefont {T.}~\bibnamefont {Esslinger}},\ }\href {\doibase
  10.1073/pnas.1720415115} {\bibfield  {journal} {\bibinfo  {journal}
  {Proceedings of the National Academy of Sciences}\ }\textbf {\bibinfo
  {volume} {115}},\ \bibinfo {pages} {3279} (\bibinfo {year}
  {2018})}\BibitemShut {NoStop}%
\bibitem [{\citenamefont {Callan}\ and\ \citenamefont
  {Wilczek}(1994)}]{callan_geometric_1994}%
  \BibitemOpen
  \bibfield  {author} {\bibinfo {author} {\bibfnamefont {C.}~\bibnamefont
  {Callan}}\ and\ \bibinfo {author} {\bibfnamefont {F.}~\bibnamefont
  {Wilczek}},\ }\href {\doibase 10.1016/0370-2693(94)91007-3} {\bibfield
  {journal} {\bibinfo  {journal} {Physics Letters B}\ }\textbf {\bibinfo
  {volume} {333}},\ \bibinfo {pages} {55} (\bibinfo {year} {1994})}\BibitemShut
  {NoStop}%
\bibitem [{\citenamefont {Calabrese}\ and\ \citenamefont
  {Cardy}(2004)}]{calabrese_entanglement_2004}%
  \BibitemOpen
  \bibfield  {author} {\bibinfo {author} {\bibfnamefont {P.}~\bibnamefont
  {Calabrese}}\ and\ \bibinfo {author} {\bibfnamefont {J.}~\bibnamefont
  {Cardy}},\ }\href {\doibase 10.1088/1742-5468/2004/06/p06002} {\bibfield
  {journal} {\bibinfo  {journal} {Journal of Statistical Mechanics}\ }\textbf
  {\bibinfo {volume} {2004}},\ \bibinfo {pages} {P06002} (\bibinfo {year}
  {2004})}\BibitemShut {NoStop}%
\bibitem [{\citenamefont {Calabrese}\ and\ \citenamefont
  {Cardy}(2009)}]{ENT:Calabrese2009}%
  \BibitemOpen
  \bibfield  {author} {\bibinfo {author} {\bibfnamefont {P.}~\bibnamefont
  {Calabrese}}\ and\ \bibinfo {author} {\bibfnamefont {J.}~\bibnamefont
  {Cardy}},\ }\href {\doibase 10.1088/1751-8113/42/50/504005} {\bibfield
  {journal} {\bibinfo  {journal} {Journal of Physics A: Mathematical and
  Theoretical}\ }\textbf {\bibinfo {volume} {42}},\ \bibinfo {pages} {504005}
  (\bibinfo {year} {2009})}\BibitemShut {NoStop}%
\bibitem [{\citenamefont {Ballarini}\ and\ \citenamefont
  {Liberato}(2019)}]{Exp:Ballarini2019}%
  \BibitemOpen
  \bibfield  {author} {\bibinfo {author} {\bibfnamefont {D.}~\bibnamefont
  {Ballarini}}\ and\ \bibinfo {author} {\bibfnamefont {S.~D.}\ \bibnamefont
  {Liberato}},\ }\href {\doibase 10.1515/nanoph-2018-0188} {\bibfield
  {journal} {\bibinfo  {journal} {Nanophotonics}\ }\textbf {\bibinfo {volume}
  {8}},\ \bibinfo {pages} {641} (\bibinfo {year} {2019})}\BibitemShut {NoStop}%
\bibitem [{\citenamefont {Forn-D\'{\i}az}\ \emph {et~al.}(2019)\citenamefont
  {Forn-D\'{\i}az}, \citenamefont {Lamata}, \citenamefont {Rico}, \citenamefont
  {Kono},\ and\ \citenamefont {Solano}}]{Exp:Forn-Diaz2019}%
  \BibitemOpen
  \bibfield  {author} {\bibinfo {author} {\bibfnamefont {P.}~\bibnamefont
  {Forn-D\'{\i}az}}, \bibinfo {author} {\bibfnamefont {L.}~\bibnamefont
  {Lamata}}, \bibinfo {author} {\bibfnamefont {E.}~\bibnamefont {Rico}},
  \bibinfo {author} {\bibfnamefont {J.}~\bibnamefont {Kono}}, \ and\ \bibinfo
  {author} {\bibfnamefont {E.}~\bibnamefont {Solano}},\ }\href {\doibase
  10.1103/RevModPhys.91.025005} {\bibfield  {journal} {\bibinfo  {journal}
  {Rev. Mod. Phys.}\ }\textbf {\bibinfo {volume} {91}},\ \bibinfo {pages}
  {025005} (\bibinfo {year} {2019})}\BibitemShut {NoStop}%
\bibitem [{\citenamefont {Kavokin}\ \emph {et~al.}(2007)\citenamefont
  {Kavokin}, \citenamefont {Baumberg}, \citenamefont {Malpuech},\ and\
  \citenamefont {Laussy}}]{Exp:Kavokin2008}%
  \BibitemOpen
  \bibfield  {author} {\bibinfo {author} {\bibfnamefont {A.}~\bibnamefont
  {Kavokin}}, \bibinfo {author} {\bibfnamefont {J.~J.}\ \bibnamefont
  {Baumberg}}, \bibinfo {author} {\bibfnamefont {G.}~\bibnamefont {Malpuech}},
  \ and\ \bibinfo {author} {\bibfnamefont {F.~P.}\ \bibnamefont {Laussy}},\
  }\href {\doibase 10.1093/acprof:oso/9780199228942.001.0001} {\emph {\bibinfo
  {title} {Microcavities}}}\ (\bibinfo  {publisher} {Oxford University Press},\
  \bibinfo {year} {2007})\BibitemShut {NoStop}%
\bibitem [{\citenamefont {Schneider}\ \emph {et~al.}(2018)\citenamefont
  {Schneider}, \citenamefont {Glazov}, \citenamefont {Korn}, \citenamefont
  {H\"{o}fling},\ and\ \citenamefont {Urbaszek}}]{Exp:Schneider2018}%
  \BibitemOpen
  \bibfield  {author} {\bibinfo {author} {\bibfnamefont {C.}~\bibnamefont
  {Schneider}}, \bibinfo {author} {\bibfnamefont {M.~M.}\ \bibnamefont
  {Glazov}}, \bibinfo {author} {\bibfnamefont {T.}~\bibnamefont {Korn}},
  \bibinfo {author} {\bibfnamefont {S.}~\bibnamefont {H\"{o}fling}}, \ and\
  \bibinfo {author} {\bibfnamefont {B.}~\bibnamefont {Urbaszek}},\ }\href
  {\doibase 10.1038/s41467-018-04866-6} {\bibfield  {journal} {\bibinfo
  {journal} {Nature Communications}\ }\textbf {\bibinfo {volume} {9}},\
  \bibinfo {pages} {2695} (\bibinfo {year} {2018})}\BibitemShut {NoStop}%
\bibitem [{\citenamefont {Halbhuber}\ \emph {et~al.}(2020)\citenamefont
  {Halbhuber}, \citenamefont {Mornhinweg}, \citenamefont {Zeller},
  \citenamefont {Ciuti}, \citenamefont {Bougeard}, \citenamefont {Huber},\ and\
  \citenamefont {Lange}}]{Exp:Halbhuber2020}%
  \BibitemOpen
  \bibfield  {author} {\bibinfo {author} {\bibfnamefont {M.}~\bibnamefont
  {Halbhuber}}, \bibinfo {author} {\bibfnamefont {J.}~\bibnamefont
  {Mornhinweg}}, \bibinfo {author} {\bibfnamefont {V.}~\bibnamefont {Zeller}},
  \bibinfo {author} {\bibfnamefont {C.}~\bibnamefont {Ciuti}}, \bibinfo
  {author} {\bibfnamefont {D.}~\bibnamefont {Bougeard}}, \bibinfo {author}
  {\bibfnamefont {R.}~\bibnamefont {Huber}}, \ and\ \bibinfo {author}
  {\bibfnamefont {C.}~\bibnamefont {Lange}},\ }\href {\doibase
  10.1038/s41566-020-0673-2} {\bibfield  {journal} {\bibinfo  {journal} {Nature
  Photonics}\ }\textbf {\bibinfo {volume} {14}},\ \bibinfo {pages} {675}
  (\bibinfo {year} {2020})}\BibitemShut {NoStop}%
\bibitem [{\citenamefont {Maissen}\ \emph {et~al.}(2014)\citenamefont
  {Maissen}, \citenamefont {Scalari}, \citenamefont {Valmorra}, \citenamefont
  {Beck}, \citenamefont {Faist}, \citenamefont {Cibella}, \citenamefont
  {Leoni}, \citenamefont {Reichl}, \citenamefont {Charpentier},\ and\
  \citenamefont {Wegscheider}}]{Exp:Maissen2014}%
  \BibitemOpen
  \bibfield  {author} {\bibinfo {author} {\bibfnamefont {C.}~\bibnamefont
  {Maissen}}, \bibinfo {author} {\bibfnamefont {G.}~\bibnamefont {Scalari}},
  \bibinfo {author} {\bibfnamefont {F.}~\bibnamefont {Valmorra}}, \bibinfo
  {author} {\bibfnamefont {M.}~\bibnamefont {Beck}}, \bibinfo {author}
  {\bibfnamefont {J.}~\bibnamefont {Faist}}, \bibinfo {author} {\bibfnamefont
  {S.}~\bibnamefont {Cibella}}, \bibinfo {author} {\bibfnamefont
  {R.}~\bibnamefont {Leoni}}, \bibinfo {author} {\bibfnamefont
  {C.}~\bibnamefont {Reichl}}, \bibinfo {author} {\bibfnamefont
  {C.}~\bibnamefont {Charpentier}}, \ and\ \bibinfo {author} {\bibfnamefont
  {W.}~\bibnamefont {Wegscheider}},\ }\href {\doibase
  10.1103/PhysRevB.90.205309} {\bibfield  {journal} {\bibinfo  {journal} {Phys.
  Rev. B}\ }\textbf {\bibinfo {volume} {90}},\ \bibinfo {pages} {205309}
  (\bibinfo {year} {2014})}\BibitemShut {NoStop}%
\bibitem [{\citenamefont {Scalari}\ \emph {et~al.}(2012)\citenamefont
  {Scalari}, \citenamefont {Maissen}, \citenamefont {Tur{\v{c}}inkov{\'{a}}},
  \citenamefont {Hagenm\"{u}ller}, \citenamefont {Liberato}, \citenamefont
  {Ciuti}, \citenamefont {Reichl}, \citenamefont {Schuh}, \citenamefont
  {Wegscheider}, \citenamefont {Beck},\ and\ \citenamefont
  {Faist}}]{Exp:Scalari2012}%
  \BibitemOpen
  \bibfield  {author} {\bibinfo {author} {\bibfnamefont {G.}~\bibnamefont
  {Scalari}}, \bibinfo {author} {\bibfnamefont {C.}~\bibnamefont {Maissen}},
  \bibinfo {author} {\bibfnamefont {D.}~\bibnamefont {Tur{\v{c}}inkov{\'{a}}}},
  \bibinfo {author} {\bibfnamefont {D.}~\bibnamefont {Hagenm\"{u}ller}},
  \bibinfo {author} {\bibfnamefont {S.~D.}\ \bibnamefont {Liberato}}, \bibinfo
  {author} {\bibfnamefont {C.}~\bibnamefont {Ciuti}}, \bibinfo {author}
  {\bibfnamefont {C.}~\bibnamefont {Reichl}}, \bibinfo {author} {\bibfnamefont
  {D.}~\bibnamefont {Schuh}}, \bibinfo {author} {\bibfnamefont
  {W.}~\bibnamefont {Wegscheider}}, \bibinfo {author} {\bibfnamefont
  {M.}~\bibnamefont {Beck}}, \ and\ \bibinfo {author} {\bibfnamefont
  {J.}~\bibnamefont {Faist}},\ }\href {\doibase 10.1126/science.1216022}
  {\bibfield  {journal} {\bibinfo  {journal} {Science}\ }\textbf {\bibinfo
  {volume} {335}},\ \bibinfo {pages} {1323} (\bibinfo {year}
  {2012})}\BibitemShut {NoStop}%
\bibitem [{\citenamefont {Li}\ \emph {et~al.}(2018)\citenamefont {Li},
  \citenamefont {Bamba}, \citenamefont {Zhang}, \citenamefont {Fallahi},
  \citenamefont {Gardner}, \citenamefont {Gao}, \citenamefont {Lou},
  \citenamefont {Yoshioka}, \citenamefont {Manfra},\ and\ \citenamefont
  {Kono}}]{Exp:Li2018}%
  \BibitemOpen
  \bibfield  {author} {\bibinfo {author} {\bibfnamefont {X.}~\bibnamefont
  {Li}}, \bibinfo {author} {\bibfnamefont {M.}~\bibnamefont {Bamba}}, \bibinfo
  {author} {\bibfnamefont {Q.}~\bibnamefont {Zhang}}, \bibinfo {author}
  {\bibfnamefont {S.}~\bibnamefont {Fallahi}}, \bibinfo {author} {\bibfnamefont
  {G.~C.}\ \bibnamefont {Gardner}}, \bibinfo {author} {\bibfnamefont
  {W.}~\bibnamefont {Gao}}, \bibinfo {author} {\bibfnamefont {M.}~\bibnamefont
  {Lou}}, \bibinfo {author} {\bibfnamefont {K.}~\bibnamefont {Yoshioka}},
  \bibinfo {author} {\bibfnamefont {M.~J.}\ \bibnamefont {Manfra}}, \ and\
  \bibinfo {author} {\bibfnamefont {J.}~\bibnamefont {Kono}},\ }\href {\doibase
  10.1038/s41566-018-0153-0} {\bibfield  {journal} {\bibinfo  {journal} {Nature
  Photonics}\ }\textbf {\bibinfo {volume} {12}},\ \bibinfo {pages} {324}
  (\bibinfo {year} {2018})}\BibitemShut {NoStop}%
\bibitem [{\citenamefont {Zhang}\ \emph {et~al.}(2016)\citenamefont {Zhang},
  \citenamefont {Lou}, \citenamefont {Li}, \citenamefont {Reno}, \citenamefont
  {Pan}, \citenamefont {Watson}, \citenamefont {Manfra},\ and\ \citenamefont
  {Kono}}]{Exp:Zhang2016}%
  \BibitemOpen
  \bibfield  {author} {\bibinfo {author} {\bibfnamefont {Q.}~\bibnamefont
  {Zhang}}, \bibinfo {author} {\bibfnamefont {M.}~\bibnamefont {Lou}}, \bibinfo
  {author} {\bibfnamefont {X.}~\bibnamefont {Li}}, \bibinfo {author}
  {\bibfnamefont {J.~L.}\ \bibnamefont {Reno}}, \bibinfo {author}
  {\bibfnamefont {W.}~\bibnamefont {Pan}}, \bibinfo {author} {\bibfnamefont
  {J.~D.}\ \bibnamefont {Watson}}, \bibinfo {author} {\bibfnamefont {M.~J.}\
  \bibnamefont {Manfra}}, \ and\ \bibinfo {author} {\bibfnamefont
  {J.}~\bibnamefont {Kono}},\ }\href {\doibase 10.1038/nphys3850} {\bibfield
  {journal} {\bibinfo  {journal} {Nature Physics 2016 12:11}\ }\textbf
  {\bibinfo {volume} {12}},\ \bibinfo {pages} {1005} (\bibinfo {year}
  {2016})}\BibitemShut {NoStop}%
\bibitem [{\citenamefont {Jarc}\ \emph {et~al.}(2022)\citenamefont {Jarc},
  \citenamefont {Mathengattil}, \citenamefont {Giusti}, \citenamefont
  {Barnaba}, \citenamefont {Singh}, \citenamefont {Montanaro}, \citenamefont
  {Glerean}, \citenamefont {Rigoni}, \citenamefont {Zilio}, \citenamefont
  {Winnerl},\ and\ \citenamefont {Fausti}}]{Exp:Jarc2022}%
  \BibitemOpen
  \bibfield  {author} {\bibinfo {author} {\bibfnamefont {G.}~\bibnamefont
  {Jarc}}, \bibinfo {author} {\bibfnamefont {S.~Y.}\ \bibnamefont
  {Mathengattil}}, \bibinfo {author} {\bibfnamefont {F.}~\bibnamefont
  {Giusti}}, \bibinfo {author} {\bibfnamefont {M.}~\bibnamefont {Barnaba}},
  \bibinfo {author} {\bibfnamefont {A.}~\bibnamefont {Singh}}, \bibinfo
  {author} {\bibfnamefont {A.}~\bibnamefont {Montanaro}}, \bibinfo {author}
  {\bibfnamefont {F.}~\bibnamefont {Glerean}}, \bibinfo {author} {\bibfnamefont
  {E.~M.}\ \bibnamefont {Rigoni}}, \bibinfo {author} {\bibfnamefont {S.~D.}\
  \bibnamefont {Zilio}}, \bibinfo {author} {\bibfnamefont {S.}~\bibnamefont
  {Winnerl}}, \ and\ \bibinfo {author} {\bibfnamefont {D.}~\bibnamefont
  {Fausti}},\ }\href {\doibase 10.1063/5.0080045} {\bibfield  {journal}
  {\bibinfo  {journal} {Review of Scientific Instruments}\ }\textbf {\bibinfo
  {volume} {93}},\ \bibinfo {pages} {033102} (\bibinfo {year}
  {2022})}\BibitemShut {NoStop}%
\bibitem [{\citenamefont {Bylinkin}\ \emph {et~al.}(2020)\citenamefont
  {Bylinkin}, \citenamefont {Schnell}, \citenamefont {Autore}, \citenamefont
  {Calavalle}, \citenamefont {Li}, \citenamefont {Taboada-Guti{\`{e}}rrez},
  \citenamefont {Liu}, \citenamefont {Edgar}, \citenamefont {Casanova},
  \citenamefont {Hueso}, \citenamefont {Alonso-Gonzalez}, \citenamefont
  {Nikitin},\ and\ \citenamefont {Hillenbrand}}]{Exp:Bylinkin2020}%
  \BibitemOpen
  \bibfield  {author} {\bibinfo {author} {\bibfnamefont {A.}~\bibnamefont
  {Bylinkin}}, \bibinfo {author} {\bibfnamefont {M.}~\bibnamefont {Schnell}},
  \bibinfo {author} {\bibfnamefont {M.}~\bibnamefont {Autore}}, \bibinfo
  {author} {\bibfnamefont {F.}~\bibnamefont {Calavalle}}, \bibinfo {author}
  {\bibfnamefont {P.}~\bibnamefont {Li}}, \bibinfo {author} {\bibfnamefont
  {J.}~\bibnamefont {Taboada-Guti{\`{e}}rrez}}, \bibinfo {author}
  {\bibfnamefont {S.}~\bibnamefont {Liu}}, \bibinfo {author} {\bibfnamefont
  {J.~H.}\ \bibnamefont {Edgar}}, \bibinfo {author} {\bibfnamefont
  {F.}~\bibnamefont {Casanova}}, \bibinfo {author} {\bibfnamefont {L.~E.}\
  \bibnamefont {Hueso}}, \bibinfo {author} {\bibfnamefont {P.}~\bibnamefont
  {Alonso-Gonzalez}}, \bibinfo {author} {\bibfnamefont {A.~Y.}\ \bibnamefont
  {Nikitin}}, \ and\ \bibinfo {author} {\bibfnamefont {R.}~\bibnamefont
  {Hillenbrand}},\ }\href {\doibase 10.1038/s41566-020-00725-3} {\bibfield
  {journal} {\bibinfo  {journal} {Nature Photonics}\ }\textbf {\bibinfo
  {volume} {15}},\ \bibinfo {pages} {197} (\bibinfo {year} {2020})}\BibitemShut
  {NoStop}%
\bibitem [{\citenamefont {Epstein}\ \emph {et~al.}(2020)\citenamefont
  {Epstein}, \citenamefont {Alcaraz}, \citenamefont {Huang}, \citenamefont
  {Pusapati}, \citenamefont {Hugonin}, \citenamefont {Kumar}, \citenamefont
  {Deputy}, \citenamefont {Khodkov}, \citenamefont {Rappoport}, \citenamefont
  {Hong}, \citenamefont {Peres}, \citenamefont {Kong}, \citenamefont {Smith},\
  and\ \citenamefont {Koppens}}]{Exp:Epstein2020}%
  \BibitemOpen
  \bibfield  {author} {\bibinfo {author} {\bibfnamefont {I.}~\bibnamefont
  {Epstein}}, \bibinfo {author} {\bibfnamefont {D.}~\bibnamefont {Alcaraz}},
  \bibinfo {author} {\bibfnamefont {Z.}~\bibnamefont {Huang}}, \bibinfo
  {author} {\bibfnamefont {V.-V.}\ \bibnamefont {Pusapati}}, \bibinfo {author}
  {\bibfnamefont {J.-P.}\ \bibnamefont {Hugonin}}, \bibinfo {author}
  {\bibfnamefont {A.}~\bibnamefont {Kumar}}, \bibinfo {author} {\bibfnamefont
  {X.~M.}\ \bibnamefont {Deputy}}, \bibinfo {author} {\bibfnamefont
  {T.}~\bibnamefont {Khodkov}}, \bibinfo {author} {\bibfnamefont {T.~G.}\
  \bibnamefont {Rappoport}}, \bibinfo {author} {\bibfnamefont {J.-Y.}\
  \bibnamefont {Hong}}, \bibinfo {author} {\bibfnamefont {N.~M.~R.}\
  \bibnamefont {Peres}}, \bibinfo {author} {\bibfnamefont {J.}~\bibnamefont
  {Kong}}, \bibinfo {author} {\bibfnamefont {D.~R.}\ \bibnamefont {Smith}}, \
  and\ \bibinfo {author} {\bibfnamefont {F.~H.~L.}\ \bibnamefont {Koppens}},\
  }\href {\doibase 10.1126/science.abb1570} {\bibfield  {journal} {\bibinfo
  {journal} {Science}\ }\textbf {\bibinfo {volume} {368}},\ \bibinfo {pages}
  {1219} (\bibinfo {year} {2020})}\BibitemShut {NoStop}%
\bibitem [{\citenamefont {Sivarajah}\ \emph {et~al.}(2019)\citenamefont
  {Sivarajah}, \citenamefont {Steinbacher}, \citenamefont {Dastrup},
  \citenamefont {Lu}, \citenamefont {Xiang}, \citenamefont {Ren}, \citenamefont
  {Kamba}, \citenamefont {Cao},\ and\ \citenamefont
  {Nelson}}]{Exp:Sivarajah2019}%
  \BibitemOpen
  \bibfield  {author} {\bibinfo {author} {\bibfnamefont {P.}~\bibnamefont
  {Sivarajah}}, \bibinfo {author} {\bibfnamefont {A.}~\bibnamefont
  {Steinbacher}}, \bibinfo {author} {\bibfnamefont {B.}~\bibnamefont
  {Dastrup}}, \bibinfo {author} {\bibfnamefont {J.}~\bibnamefont {Lu}},
  \bibinfo {author} {\bibfnamefont {M.}~\bibnamefont {Xiang}}, \bibinfo
  {author} {\bibfnamefont {W.}~\bibnamefont {Ren}}, \bibinfo {author}
  {\bibfnamefont {S.}~\bibnamefont {Kamba}}, \bibinfo {author} {\bibfnamefont
  {S.}~\bibnamefont {Cao}}, \ and\ \bibinfo {author} {\bibfnamefont {K.~A.}\
  \bibnamefont {Nelson}},\ }\href {\doibase 10.1063/1.5083849} {\bibfield
  {journal} {\bibinfo  {journal} {Journal of Applied Physics}\ }\textbf
  {\bibinfo {volume} {125}},\ \bibinfo {pages} {213103} (\bibinfo {year}
  {2019})}\BibitemShut {NoStop}%
\bibitem [{\citenamefont {Zhang}\ \emph {et~al.}(2021)\citenamefont {Zhang},
  \citenamefont {Chen}, \citenamefont {Wu}, \citenamefont {Wang}, \citenamefont
  {Fan}, \citenamefont {Deng},\ and\ \citenamefont {Wu}}]{SuperEXP:Zhang2021}%
  \BibitemOpen
  \bibfield  {author} {\bibinfo {author} {\bibfnamefont {X.}~\bibnamefont
  {Zhang}}, \bibinfo {author} {\bibfnamefont {Y.}~\bibnamefont {Chen}},
  \bibinfo {author} {\bibfnamefont {Z.}~\bibnamefont {Wu}}, \bibinfo {author}
  {\bibfnamefont {J.}~\bibnamefont {Wang}}, \bibinfo {author} {\bibfnamefont
  {J.}~\bibnamefont {Fan}}, \bibinfo {author} {\bibfnamefont {S.}~\bibnamefont
  {Deng}}, \ and\ \bibinfo {author} {\bibfnamefont {H.}~\bibnamefont {Wu}},\
  }\href {\doibase 10.1126/science.abd4385} {\bibfield  {journal} {\bibinfo
  {journal} {Science}\ }\textbf {\bibinfo {volume} {373}},\ \bibinfo {pages}
  {1359} (\bibinfo {year} {2021})}\BibitemShut {NoStop}%
\bibitem [{\citenamefont {Elben}\ \emph {et~al.}(2018)\citenamefont {Elben},
  \citenamefont {Vermersch}, \citenamefont {Dalmonte}, \citenamefont {Cirac},\
  and\ \citenamefont {Zoller}}]{PROBING:Elben2018}%
  \BibitemOpen
  \bibfield  {author} {\bibinfo {author} {\bibfnamefont {A.}~\bibnamefont
  {Elben}}, \bibinfo {author} {\bibfnamefont {B.}~\bibnamefont {Vermersch}},
  \bibinfo {author} {\bibfnamefont {M.}~\bibnamefont {Dalmonte}}, \bibinfo
  {author} {\bibfnamefont {J.~I.}\ \bibnamefont {Cirac}}, \ and\ \bibinfo
  {author} {\bibfnamefont {P.}~\bibnamefont {Zoller}},\ }\href {\doibase
  10.1103/PhysRevLett.120.050406} {\bibfield  {journal} {\bibinfo  {journal}
  {Phys. Rev. Lett.}\ }\textbf {\bibinfo {volume} {120}},\ \bibinfo {pages}
  {050406} (\bibinfo {year} {2018})}\BibitemShut {NoStop}%
\bibitem [{\citenamefont {Brydges}\ \emph {et~al.}(2019)\citenamefont
  {Brydges}, \citenamefont {Elben}, \citenamefont {Jurcevic}, \citenamefont
  {Vermersch}, \citenamefont {Maier}, \citenamefont {Lanyon}, \citenamefont
  {Zoller}, \citenamefont {Blatt},\ and\ \citenamefont
  {Roos}}]{PROBING:Brydges2019probing}%
  \BibitemOpen
  \bibfield  {author} {\bibinfo {author} {\bibfnamefont {T.}~\bibnamefont
  {Brydges}}, \bibinfo {author} {\bibfnamefont {A.}~\bibnamefont {Elben}},
  \bibinfo {author} {\bibfnamefont {P.}~\bibnamefont {Jurcevic}}, \bibinfo
  {author} {\bibfnamefont {B.}~\bibnamefont {Vermersch}}, \bibinfo {author}
  {\bibfnamefont {C.}~\bibnamefont {Maier}}, \bibinfo {author} {\bibfnamefont
  {B.~P.}\ \bibnamefont {Lanyon}}, \bibinfo {author} {\bibfnamefont
  {P.}~\bibnamefont {Zoller}}, \bibinfo {author} {\bibfnamefont
  {R.}~\bibnamefont {Blatt}}, \ and\ \bibinfo {author} {\bibfnamefont {C.~F.}\
  \bibnamefont {Roos}},\ }\href {\doibase 10.1126/science.aau4963} {\bibfield
  {journal} {\bibinfo  {journal} {Science}\ }\textbf {\bibinfo {volume}
  {364}},\ \bibinfo {pages} {260} (\bibinfo {year} {2019})}\BibitemShut
  {NoStop}%
\bibitem [{\citenamefont {Elben}\ \emph {et~al.}(2020)\citenamefont {Elben},
  \citenamefont {Kueng}, \citenamefont {Huang}, \citenamefont {van Bijnen},
  \citenamefont {Kokail}, \citenamefont {Dalmonte}, \citenamefont {Calabrese},
  \citenamefont {Kraus}, \citenamefont {Preskill}, \citenamefont {Zoller},\
  and\ \citenamefont {Vermersch}}]{PROBING:elben2020mixed}%
  \BibitemOpen
  \bibfield  {author} {\bibinfo {author} {\bibfnamefont {A.}~\bibnamefont
  {Elben}}, \bibinfo {author} {\bibfnamefont {R.}~\bibnamefont {Kueng}},
  \bibinfo {author} {\bibfnamefont {H.-Y.~R.}\ \bibnamefont {Huang}}, \bibinfo
  {author} {\bibfnamefont {R.}~\bibnamefont {van Bijnen}}, \bibinfo {author}
  {\bibfnamefont {C.}~\bibnamefont {Kokail}}, \bibinfo {author} {\bibfnamefont
  {M.}~\bibnamefont {Dalmonte}}, \bibinfo {author} {\bibfnamefont
  {P.}~\bibnamefont {Calabrese}}, \bibinfo {author} {\bibfnamefont
  {B.}~\bibnamefont {Kraus}}, \bibinfo {author} {\bibfnamefont
  {J.}~\bibnamefont {Preskill}}, \bibinfo {author} {\bibfnamefont
  {P.}~\bibnamefont {Zoller}}, \ and\ \bibinfo {author} {\bibfnamefont
  {B.}~\bibnamefont {Vermersch}},\ }\href {\doibase
  10.1103/PhysRevLett.125.200501} {\bibfield  {journal} {\bibinfo  {journal}
  {Phys. Rev. Lett.}\ }\textbf {\bibinfo {volume} {125}},\ \bibinfo {pages}
  {200501} (\bibinfo {year} {2020})}\BibitemShut {NoStop}%
\bibitem [{\citenamefont {Neven}\ \emph {et~al.}(2021)\citenamefont {Neven},
  \citenamefont {Carrasco}, \citenamefont {Vitale}, \citenamefont {Kokail},
  \citenamefont {Elben}, \citenamefont {Dalmonte}, \citenamefont {Calabrese},
  \citenamefont {Zoller}, \citenamefont {Vermersch}, \citenamefont {Kueng},\
  and\ \citenamefont {Kraus}}]{PROBING:neven2021symmetry}%
  \BibitemOpen
  \bibfield  {author} {\bibinfo {author} {\bibfnamefont {A.}~\bibnamefont
  {Neven}}, \bibinfo {author} {\bibfnamefont {J.}~\bibnamefont {Carrasco}},
  \bibinfo {author} {\bibfnamefont {V.}~\bibnamefont {Vitale}}, \bibinfo
  {author} {\bibfnamefont {C.}~\bibnamefont {Kokail}}, \bibinfo {author}
  {\bibfnamefont {A.}~\bibnamefont {Elben}}, \bibinfo {author} {\bibfnamefont
  {M.}~\bibnamefont {Dalmonte}}, \bibinfo {author} {\bibfnamefont
  {P.}~\bibnamefont {Calabrese}}, \bibinfo {author} {\bibfnamefont
  {P.}~\bibnamefont {Zoller}}, \bibinfo {author} {\bibfnamefont
  {B.}~\bibnamefont {Vermersch}}, \bibinfo {author} {\bibfnamefont
  {R.}~\bibnamefont {Kueng}}, \ and\ \bibinfo {author} {\bibfnamefont
  {B.}~\bibnamefont {Kraus}},\ }\href {\doibase 10.1038/s41534-021-00487-y}
  {\bibfield  {journal} {\bibinfo  {journal} {npj Quantum Information}\
  }\textbf {\bibinfo {volume} {7}},\ \bibinfo {pages} {152} (\bibinfo {year}
  {2021})}\BibitemShut {NoStop}%
\bibitem [{\citenamefont {Vitale}\ \emph {et~al.}(2022)\citenamefont {Vitale},
  \citenamefont {Elben}, \citenamefont {Kueng}, \citenamefont {Neven},
  \citenamefont {Carrasco}, \citenamefont {Kraus}, \citenamefont {Zoller},
  \citenamefont {Calabrese}, \citenamefont {Vermersch},\ and\ \citenamefont
  {Dalmonte}}]{PROBING:Vitale2022}%
  \BibitemOpen
  \bibfield  {author} {\bibinfo {author} {\bibfnamefont {V.}~\bibnamefont
  {Vitale}}, \bibinfo {author} {\bibfnamefont {A.}~\bibnamefont {Elben}},
  \bibinfo {author} {\bibfnamefont {R.}~\bibnamefont {Kueng}}, \bibinfo
  {author} {\bibfnamefont {A.}~\bibnamefont {Neven}}, \bibinfo {author}
  {\bibfnamefont {J.}~\bibnamefont {Carrasco}}, \bibinfo {author}
  {\bibfnamefont {B.}~\bibnamefont {Kraus}}, \bibinfo {author} {\bibfnamefont
  {P.}~\bibnamefont {Zoller}}, \bibinfo {author} {\bibfnamefont
  {P.}~\bibnamefont {Calabrese}}, \bibinfo {author} {\bibfnamefont
  {B.}~\bibnamefont {Vermersch}}, \ and\ \bibinfo {author} {\bibfnamefont
  {M.}~\bibnamefont {Dalmonte}},\ }\href {\doibase
  10.21468/SciPostPhys.12.3.106} {\bibfield  {journal} {\bibinfo  {journal}
  {SciPost Phys.}\ }\textbf {\bibinfo {volume} {12}},\ \bibinfo {pages} {106}
  (\bibinfo {year} {2022})}\BibitemShut {NoStop}%
\bibitem [{\citenamefont {Wang}\ \emph {et~al.}(2011)\citenamefont {Wang},
  \citenamefont {Mariantoni}, \citenamefont {Bialczak}, \citenamefont
  {Lenander}, \citenamefont {Lucero}, \citenamefont {Neeley}, \citenamefont
  {O'Connell}, \citenamefont {Sank}, \citenamefont {Weides}, \citenamefont
  {Wenner}, \citenamefont {Yamamoto}, \citenamefont {Yin}, \citenamefont
  {Zhao}, \citenamefont {Martinis},\ and\ \citenamefont
  {Cleland}}]{EXP:Wang2011}%
  \BibitemOpen
  \bibfield  {author} {\bibinfo {author} {\bibfnamefont {H.}~\bibnamefont
  {Wang}}, \bibinfo {author} {\bibfnamefont {M.}~\bibnamefont {Mariantoni}},
  \bibinfo {author} {\bibfnamefont {R.~C.}\ \bibnamefont {Bialczak}}, \bibinfo
  {author} {\bibfnamefont {M.}~\bibnamefont {Lenander}}, \bibinfo {author}
  {\bibfnamefont {E.}~\bibnamefont {Lucero}}, \bibinfo {author} {\bibfnamefont
  {M.}~\bibnamefont {Neeley}}, \bibinfo {author} {\bibfnamefont {A.~D.}\
  \bibnamefont {O'Connell}}, \bibinfo {author} {\bibfnamefont {D.}~\bibnamefont
  {Sank}}, \bibinfo {author} {\bibfnamefont {M.}~\bibnamefont {Weides}},
  \bibinfo {author} {\bibfnamefont {J.}~\bibnamefont {Wenner}}, \bibinfo
  {author} {\bibfnamefont {T.}~\bibnamefont {Yamamoto}}, \bibinfo {author}
  {\bibfnamefont {Y.}~\bibnamefont {Yin}}, \bibinfo {author} {\bibfnamefont
  {J.}~\bibnamefont {Zhao}}, \bibinfo {author} {\bibfnamefont {J.~M.}\
  \bibnamefont {Martinis}}, \ and\ \bibinfo {author} {\bibfnamefont {A.~N.}\
  \bibnamefont {Cleland}},\ }\href {\doibase 10.1103/PhysRevLett.106.060401}
  {\bibfield  {journal} {\bibinfo  {journal} {Phys. Rev. Lett.}\ }\textbf
  {\bibinfo {volume} {106}},\ \bibinfo {pages} {060401} (\bibinfo {year}
  {2011})}\BibitemShut {NoStop}%
\bibitem [{\citenamefont {Eichler}\ \emph {et~al.}(2011)\citenamefont
  {Eichler}, \citenamefont {Bozyigit}, \citenamefont {Lang}, \citenamefont
  {Baur}, \citenamefont {Steffen}, \citenamefont {Fink}, \citenamefont
  {Filipp},\ and\ \citenamefont {Wallraff}}]{EXP:eichler2011observation}%
  \BibitemOpen
  \bibfield  {author} {\bibinfo {author} {\bibfnamefont {C.}~\bibnamefont
  {Eichler}}, \bibinfo {author} {\bibfnamefont {D.}~\bibnamefont {Bozyigit}},
  \bibinfo {author} {\bibfnamefont {C.}~\bibnamefont {Lang}}, \bibinfo {author}
  {\bibfnamefont {M.}~\bibnamefont {Baur}}, \bibinfo {author} {\bibfnamefont
  {L.}~\bibnamefont {Steffen}}, \bibinfo {author} {\bibfnamefont {J.~M.}\
  \bibnamefont {Fink}}, \bibinfo {author} {\bibfnamefont {S.}~\bibnamefont
  {Filipp}}, \ and\ \bibinfo {author} {\bibfnamefont {A.}~\bibnamefont
  {Wallraff}},\ }\href {\doibase 10.1103/PhysRevLett.107.113601} {\bibfield
  {journal} {\bibinfo  {journal} {Phys. Rev. Lett.}\ }\textbf {\bibinfo
  {volume} {107}},\ \bibinfo {pages} {113601} (\bibinfo {year}
  {2011})}\BibitemShut {NoStop}%
\bibitem [{\citenamefont {Brennecke}\ \emph {et~al.}(2013)\citenamefont
  {Brennecke}, \citenamefont {Mottl}, \citenamefont {Baumann}, \citenamefont
  {Landig}, \citenamefont {Donner},\ and\ \citenamefont
  {Esslinger}}]{EXP:brennecke2013real}%
  \BibitemOpen
  \bibfield  {author} {\bibinfo {author} {\bibfnamefont {F.}~\bibnamefont
  {Brennecke}}, \bibinfo {author} {\bibfnamefont {R.}~\bibnamefont {Mottl}},
  \bibinfo {author} {\bibfnamefont {K.}~\bibnamefont {Baumann}}, \bibinfo
  {author} {\bibfnamefont {R.}~\bibnamefont {Landig}}, \bibinfo {author}
  {\bibfnamefont {T.}~\bibnamefont {Donner}}, \ and\ \bibinfo {author}
  {\bibfnamefont {T.}~\bibnamefont {Esslinger}},\ }\href {\doibase
  10.1073/pnas.1306993110} {\bibfield  {journal} {\bibinfo  {journal}
  {Proceedings of the National Academy of Sciences}\ }\textbf {\bibinfo
  {volume} {110}},\ \bibinfo {pages} {11763} (\bibinfo {year}
  {2013})}\BibitemShut {NoStop}%
\bibitem [{\citenamefont {Mlynek}\ \emph {et~al.}(2012)\citenamefont {Mlynek},
  \citenamefont {Abdumalikov}, \citenamefont {Fink}, \citenamefont {Steffen},
  \citenamefont {Baur}, \citenamefont {Lang}, \citenamefont {van Loo},\ and\
  \citenamefont {Wallraff}}]{ENT:Mlynek2012Dicke}%
  \BibitemOpen
  \bibfield  {author} {\bibinfo {author} {\bibfnamefont {J.~A.}\ \bibnamefont
  {Mlynek}}, \bibinfo {author} {\bibfnamefont {A.~A.}\ \bibnamefont
  {Abdumalikov}}, \bibinfo {author} {\bibfnamefont {J.~M.}\ \bibnamefont
  {Fink}}, \bibinfo {author} {\bibfnamefont {L.}~\bibnamefont {Steffen}},
  \bibinfo {author} {\bibfnamefont {M.}~\bibnamefont {Baur}}, \bibinfo {author}
  {\bibfnamefont {C.}~\bibnamefont {Lang}}, \bibinfo {author} {\bibfnamefont
  {A.~F.}\ \bibnamefont {van Loo}}, \ and\ \bibinfo {author} {\bibfnamefont
  {A.}~\bibnamefont {Wallraff}},\ }\href {\doibase 10.1103/PhysRevA.86.053838}
  {\bibfield  {journal} {\bibinfo  {journal} {Phys. Rev. A}\ }\textbf {\bibinfo
  {volume} {86}},\ \bibinfo {pages} {053838} (\bibinfo {year}
  {2012})}\BibitemShut {NoStop}%
\end{thebibliography}%

\end{document}